\documentclass[12pt,preprint2]{aastex62}

\usepackage{natbib}
\usepackage{longtable}
\usepackage{enumerate}
\usepackage{amsmath}
\usepackage{hyperref}
\usepackage{CJK}
\usepackage[title]{appendix}
\usepackage{rotating}

\def\msun{\ifmmode {\rm M}_{\mathord\odot}\else $M_{\mathord\odot}$\fi}
\def\rsun{\ifmmode {\rm R}_{\mathord\odot}\else $R_{\mathord\odot}$\fi}
\def\lsun{\ifmmode {\rm L}_{\mathord\odot}\else $L_{\mathord\odot}$\fi}

\def\co{$^{12}$CO}
\def\c18o{C$^{18}$O}
\def\h2{H$_{2}$}
\def\13co{$^{13}$CO}
\def\n2hp{$_{2}$H$^{+}$}

\def\radmc{{\sc radmc-3d}}
\def\cm2{cm$^{-2}$}
\def\cmc{cm$^{-3}$}

\newcommand{\xmark}{\text{\sffamily X}}
\newcommand{\cmark}{\text{\sffamily \checkmark}}
\newcommand{\kms}{km s$^{-1}$}
\newcommand{\hii}{H\,{\sc ii}}
\newcommand{\Brut}{\textit{Brut}}

\newcommand{\CASI}{{\sc casi}}
\newcommand{\CASItD}{{\sc casi-3d}}
\def\turbustat{{\sc turbustat}}

\shorttitle{}
\shortauthors{}

\begin{document}
\begin{CJK*}{UTF8}{gbsn}

\title{Application of Convolutional Neural Networks to Identify Stellar Feedback Bubbles in CO Emission}

\author{Duo Xu}
\affil{Department of Astronomy, The University of Texas at Austin, Austin, TX 78712, USA;}
\email{xuduo117@utexas.edu}

\author{Stella S. R. Offner}
\affil{Department of Astronomy, The University of Texas at Austin, Austin, TX 78712, USA;}
\email{soffner@astro.as.utexas.edu}

\author{Robert Gutermuth}
\affil{Department of Astronomy, University of Massachusetts, Amherst, MA 01003, USA;}

\author{Colin Van Oort}
\affil{University of Vermont, Burlington, VT 05405, USA;}

\begin{abstract}

We adopt a deep learning method \CASI\ (Convolutional Approach to Shell Identification) and extend it to 3D (\CASItD) to identify signatures of stellar feedback in molecular line spectra, such as \13co. We adopt magneto-hydrodynamics simulations that study the impact of stellar winds in a turbulent molecular cloud as an input to generate synthetic observations. We apply the 3D radiation transfer code \radmc\ to model \13co\ ($J$=1-0) line emission from the simulated clouds. We train two \CASItD\ models: ME1 is trained to predict only the position of feedback, while MF is trained to predict the fraction of the mass coming from feedback in each voxel. We adopt 75\% of the synthetic observations as the training set and assess the accuracy of the two models  with the remaining data. We demonstrate that model ME1 identifies bubbles in simulated data with 95\% accuracy, and model MF predicts the bubble mass within 4\% of the true value. We test the two models on bubbles that were previously visually identified in Taurus in \13co. We show our models perform well on the highest confidence bubbles that have a clear ring morphology and contain one or more sources. We apply our two models on the full 98 deg$^2$ FCRAO \13co\ survey of the Taurus cloud. Models ME1 and MF predict feedback gas mass of 2894 \msun\ and 302 \msun, respectively. When including a correction factor for missing energy due to the limited velocity range of the \13co\ data cube, model ME1 predicts feedback kinetic energies of 4.0$\times10^{46}$ ergs and 1.5$\times10^{47}$ ergs with/without subtracting the cloud velocity gradient. Model MF predicts feedback kinetic energy of 9.6$\times10^{45}$ ergs and 2.8$\times10^{46}$ ergs with/without subtracting the cloud velocity gradient. Model ME1 predicts bubble locations and properties consistent with previous visually identified bubbles. However, model MF demonstrates that feedback properties computed based on visual identifications are significantly over-estimated due to line of sight confusion and contamination from background and foreground gas. 

\end{abstract}


\section{Introduction} 

Stellar winds driven by young stars create distinct features in molecular clouds. The ejected mass compresses and heats the ambient gas, producing shocks \citep{1999RvMP...71..173H}. In the case of more massive stars, stellar winds combined with radiation create bubbles containing luminous \hii\ regions. Observational surveys find that signatures of such bubbles are ubiquitous. For example, \cite{2006ApJ...649..759C, 2007ApJ...670..428C} visually identified numerous \hii\ regions in the Spitzer Galactic Legacy Infrared Mid-Plane Survey Extraordinaire (GLIMPSE) data. They concluded these bubbles have a significant impact on the dynamics and star formation of molecular clouds, and they found 12\% of the shells are associated with young sources, which may have been triggered by shell expansion. Over 50\% of the bubbles identified are not spherically symmetric due to fluctuations in local gas density and/or anisotropic stellar winds and radiation fields. These complications make bubble identification more challenging. 

A variety of groups have investigated the impact of stellar feedback bubbles due to stellar winds within molecular clouds. \citet{2012ApJ...746...25N} found several parsec-scale bubbles expanding and compressing the ambient gas, which they proposed has contributed to the formation of several dense filaments. They suggested one of these dense filaments is converging with another filament, triggering recent star formation in the cloud. \citet{2005ApJ...632..941Q} mapped the expanding cavities in NGC 1333, a sub-region in the Perseus molecular cloud. They found the kinetic energy of these cavities is sufficient to power the turbulence in this region. \citet{2011ApJ...742..105A} identified stellar feedback bubbles in a full map of the Perseus molecular cloud and drew similar conclusions about the energy budget. However, \citet{2015ApJS..219...20L} found the energy injected from bubbles in the Taurus molecular cloud is only 29\% of the turbulent energy of Taurus. 

A variety of theoretical work has also investigated the impact and signatures of stellar feedback. \citet{2016ApJ...833..233B} studied statistical signatures of stellar winds in synthetic observations. They found that the covariance matrices of the velocity channels are sensitive to the existence of stellar feedback. \citet{2018NatAs...2..896O} found that the slope of the velocity power spectrum becomes steeper in simulations with feedback compared to those without feedback. However, stellar feedback might also indirectly add energy to the ambient gas where there is no injected feedback mass \citep{2018NatAs...2..896O}, making quantitative statistical study of feedback more challenging in observational data. 
 
Historically, bubbles, such as those found in the above studies, have been identified ``by eye‘’. However, given the exponentially increasing amount of data, visual identification is not scalable, i.e. it is almost impossible for humans to look through all the data by eye \citep{2010PASP..122..314M, 2011ApJS..194...20P}. Moreover, to study the dynamics of bubbles, it is necessary to switch from two dimensional images \citep{2014ApJS..214....3B} to three dimensional data cubes \citep{2011ApJ...742..105A,2015ApJS..219...20L} those contain information about the gas motion. An extra dimension makes it much more time intensive to identify the bubble features. Moreover, it is impossible for humans to consistently identify or classify without bias. However, systematic and repeatable identification is possible with the aid of machine learning approaches \citep{2011ApJ...741...14B, 2014ApJS..214....3B,Van Oort}. 

Several machine learning algorithms have been applied to identify stellar feedback features \citep{2011ApJ...741...14B, 2014ApJS..214....3B}. \citet{2011ApJ...741...14B} applied Support Vector Machines (SVM) to distinguish a supernova remnant from the ambient gas in $^{12}$CO $J=3-2$ emission.  \citet{2014ApJS..214....3B} developed the Brut algorithm, based on “Random Forests,” to identify bubbles in dust emission. To train Brut, they adopted bubble identification results from over 35,000 participants in the Milky Way Project, a citizen science project based on GLIMPSE data. \citet{2017ApJ...851..149X} expanded on this work by supplementing the \Brut\ training set with synthetic observations of bubbles in simulated clouds. After retraining on the enhanced training set, \Brut\ more efficiently identified ultra-compact and compact \hii\ regions generated by B-type stars. Leveraging both observational data \citep[e.g.,][]{2012MNRAS.424.2442S,2019MNRAS.488.1141J} and synthetic observations can significantly enhance the performance of machine learning algorithms. However, \Brut\ requires the bubble to be centered in the image for it to be accurately identified. This makes it computationally expensive to identify bubbles in a large sky survey map because the data must be cropped into small chunks centered at different positions with different sizes to ensure the target bubbles are centered in at least one image.

Due to the evolution of high performance computing and the power of GPUs (Graphics Processing Units), deep learning is gaining popularity thanks to its general applicability and high accuracy. Recently developed deep learning methods are more powerful in image recognition than earlier methods, like \Brut. \citet{2019ApJ...876...82N} developed a deep machine learning tool based on Convolutional Neural Networks (CNNs) to estimate the mass of galaxy clusters in X-ray emission. The CNN is not sensitive to the position of galaxy clusters, making it straightforward to apply to large sky survey maps. \citet{Van Oort} developed an ``Encoder-Decoder'' Convolutional Approach to Shell Identification (\CASI) to identify stellar wind bubbles in density slices and 2D CO emission. Once trained, \CASI\ can identify structures in minutes and achieves a 98\% pixel-level accuracy.  However, one caveat of these CNN models is that they are limited to 2D integrated intensity maps or individual velocity channels. These algorithms do not take radial velocity information into consideration, which may lead to a high false detection rate when applied to 3D data. In other words, this technique may identify a clear ring structure as a bubble even though this structure is caused by a turbulent pattern without any evidence of expansion in the spectra. Alternatively, it may miss structures with coherent expansion across a range of channels but which do not have a bubble morphology in most channels.

In this paper, we adopt the deep learning method \CASI\  \citep{Van Oort} and extend it to 3D (\CASItD) in order to exploit the full 3D CO spectral information to identify bubbles. We develop two deep machine learning tasks. Task I predicts the position of feedback. Task II predicts the fraction of the mass in the voxels that is coming from feedback. We describe our deep machine learning algorithm architecture and  how we generate the training set from synthetic observations in Section~\ref{Method}. In Section~\ref{Result}, we present the performance of the CNN model in identifying bubbles in both synthetic data and observational data. We
summarize our results and conclusions in Section~\ref{Conclusion}.

\section{Method }
\label{Method}

\subsection{\CASItD\ Architecture}
\label{CASI-3D Architecture}  

In this section we give a brief overview of the \CASI\ architecture and describe our implementation in 3D. \CASItD\ is available on GitLab\footnote{https://gitlab.com/casi-project/casi-3d}.

\citet{Van Oort} developed the \CASI\ architecture with residual networks \citep{He2016} and ``U-net'' \citep{Ronneberger2015}. The residual networks exponentially increase the complexity of the networks to prevent over-fitting \citep{He2016,Veit2016}. ``U-net'' adds cross connections between different layers, which enhance the performance in constructing the output image \citep{Ronneberger2015}. \CASI\ utilizes a widely-used optimization method, stochastic gradient descent (SGD) with momentum, in the training. Momentum indicates that the SGD takes the past time step into consideration when conducting the optimization. It helps accelerate SGD during training.  \CASI\ is trained on simulated molecular cloud density 2D slices or CO integrated intensity maps. It learns to predict the ``tracer field,'' which is an {\sc orion} field \citep{2012ApJ...745..139L} that tracks the fraction of gas in each cell that is launched in the stellar winds. 
A more detailed description of \CASI\ can be found in \citet{Van Oort}.

\begin{figure*}[hbt!]
\centering
\includegraphics[width=0.95\linewidth]{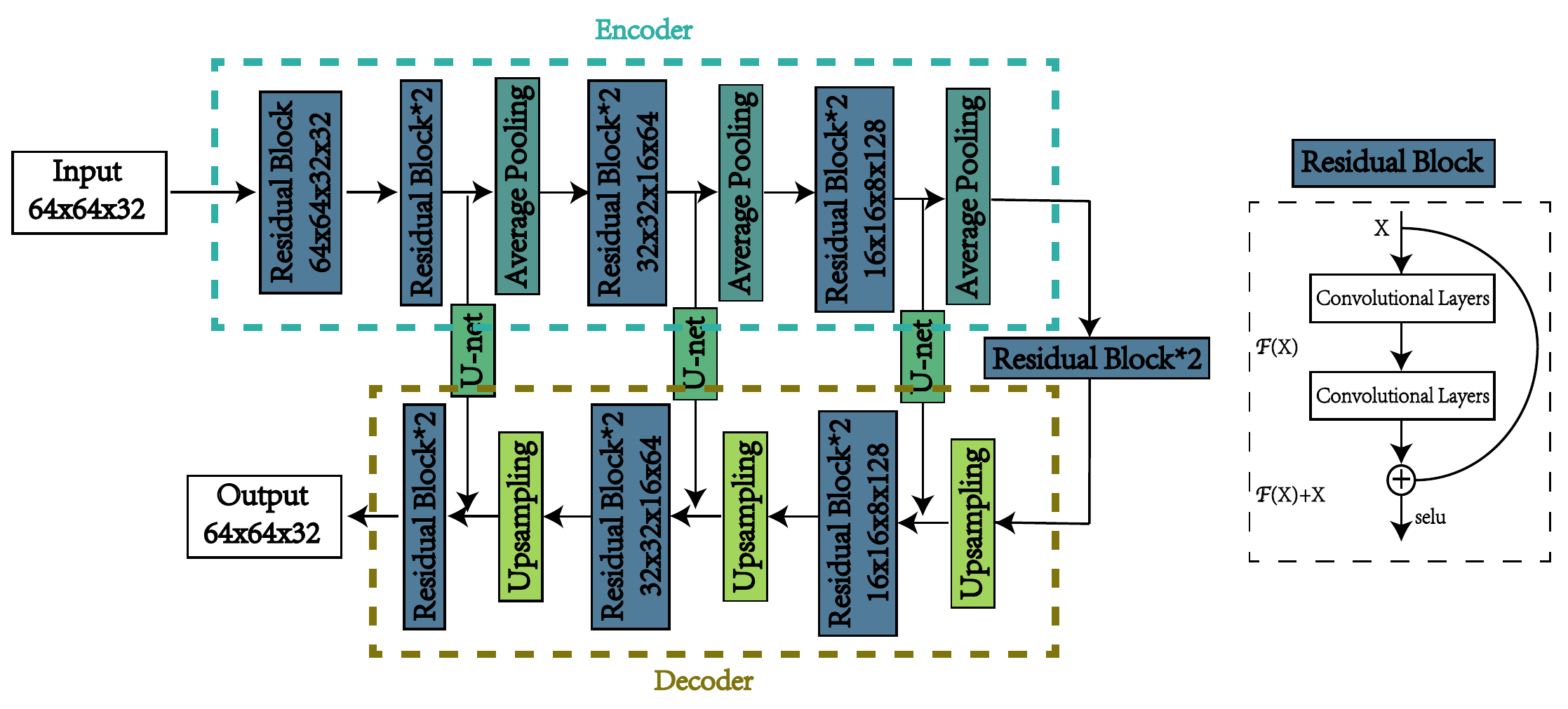}
\caption{The architecture of the U-net CNN model with residual functions. }
\label{fig.block3}
\end{figure*}

We modify the \CASI\ architecture and replace the 2D convolutions with 3D convolutions, as shown in Figure~\ref{fig.block3}. The ``encoder'' part extracts the features from the input data, then the ``decoder'' part translates these features into another image, e.g., the tracer field. In \citet{Van Oort}, the input data is a $256\times 256$ array. The image goes through four down-sampling (max pooling) layers. Given the extra dimension in our model, we cannot maintain the same spatial resolution in our 3D model due to the limited memory on GPU. We reform the input data to an array shape of $64\times 64 \times 32$ in position-position-velocity (PPV). The data cube undergoes three down-sampling layers. We apply $7\times 7\times 7$ filters in the first layer to represent the large scale structure in the training set. We then apply $3\times 3\times 3$ filters in the following layers, which capture the details of bubble morphology and velocity information. We begin with 32 filters and double them after each down-sampling layer. We adopt the average pooling method to down sample the data. See Appendix~\ref{Down-sampling Methods} for more discussion of the down-sampling methods. 

Owing to the limitation of the GPU memory, we set the batch size to be 8. The batch size indicates the number of samples that propagates through the network at one time. We explore different numbers of epochs during the training. An epoch is one iteration of the prediction, error estimation, and weight update cycle over all the training data. After some number of epochs, the model error converges and does not decrease much more. The learning rate is a hyper-parameter that controls the magnitude of filter weight updates with respect to the loss gradient. The learning rate affects how quickly our model converges to an acceptable minimum squared error value. If the learning rate is too high, the model diverges and can not reach the minimum squared error value. If the learning rates are too low, it will take a long time for the model to converge {as it makes very tiny updates to 
the weights in the network}. 

We test two strategies to set the learning rate during training: a fixed learning rate during the whole training process or an adaptive learning rate which changes with the epoch. The fixed learning rate is set to 0.01. The adaptive learning rate is initially set to 0.02 and halved every 19 epochs. Considering the computing time {limits} during the training, we set the maximum epoch number to be 277. 
{Meanwhile, we set an early-stopping criterion for models ME1, ME2, ME5, ME6 and ME7. The early-stopping criterion provides guidance on how many epochs can be run before the model begins to over-fit. The model stops the training if the validation loss decreases less than 0.0001 after 100 epochs. In practice, most models (ME2, ME5, ME6, ME7) reach the early-stopping criterion before reaching the epoch limit.}


\subsection{Training Task}
\label{Training Task}
 We develop two training tasks to identify stellar feedback bubbles. Task I involves training to predict the position of feedback, 
 which reproduces the morphology of bubbles. Task II involves training to predict the fraction of the mass that comes from stellar feedback, which gives a better mass estimation of the bubbles.
 
\subsubsection{Training Task I: Predicting the Position of Feedback} 
\label{Predicting on Pixel by Pixel Position of Feedback}
In Task I, we aim to predict the position of feedback on a pixel by pixel basis. We use the mean squared error (MSE) as the loss function in the training. The loss function is a metric to quantify the performance of the model predictions. The mean square error is defined as:
\begin{equation}
\label{eq-1}
MSE=\frac{1}{n}\sum^{n}_{i=1} (Y_{\rm pred}-Y_{\rm tracer})^{2},
\end{equation}
where $Y_{\rm pred}$ represents the prediction from the model, $Y_{\rm tracer}$ represents the ``ground truth" as described by the tracer field, and $n$ indicates the number of samples. In this model, $Y_{\rm tracer}$ is the emission of \13co\ at the specific positions where there is feedback gas. The \13co\ intensity is proportional to the mass of the gas in the optically thin regime. 

Observational data includes noise so we must adopt a prescription to define feedback that accounts for noise limits. We adopt MSE in part because it is less affected by the noise in the data cube than other loss functions. We set a threshold based on the input data noise level, 0.2 K, to binarize the prediction map, i.e., a pixel is assigned a 1 if the predicted emission is above 0.2 K, and a pixel is assigned a 0 if the predicted emission is below 0.2 K. The binarized prediction map indicates the location of the bubbles in PPV space. The final loss indicates the error of the prediction, which is related to the mass estimation uncertainty. 

To explore the performance of different models with different hyper-parameters, we train models with different learning rates, different epochs, different optimization methods and different loss functions. We easily rule out other loss functions 
and optimization methods based on the performance on the training set, and adopt MSE as the loss function and SGD as the optimization method. 
We list different models with different learning rates and with different epochs in Table~\ref{Training Model Parameter}. We additionally explore the completeness of the training set on the performance. More discussion can be found in Section~\ref{Data Augmentation}.

\subsubsection{Training Task II: Predictions for the Fraction of Feedback Mass}
\label{Predictions on The Fraction of Feedback Mass}
Task I basically classifies pixels individually as belonging to feedback or not belonging to feedback. However, it does not take into account that many pixels contain some feedback and some non-feedback gas. To address this, we train another model to predict the fraction of the mass that comes from stellar feedback. We adopt the same learning rate, epoch and optimization method as the best model in Section~\ref{Model Selection}. We define the new training data to be PPV cubes containing the fraction of mass, rather than the emission, that comes from the feedback. We describe how these cubes are constructed in more detail in Section~\ref{Training Target: Tracer Field for Task I}.

We test the MSE, Intersection over Union (IoU)  and a combination of MSE and IoU as the loss function and compare their performance in Appendix~\ref{Loss Function}. The IoU is a metric to evaluate the overlap fraction between the prediction and the tracer field, which is defined as:
\begin{equation}
\label{eq-2}
IoU=\frac{Y_{\rm pred} \cap Y_{\rm tracer}}{Y_{\rm pred} \cup Y_{\rm tracer}}.
\end{equation}
The combination of MSE and IoU is defined as:
\begin{equation}
\label{eq-3}
L_{\rm new}=\omega \times MSE+IoU,
\end{equation}
where $\omega$ is the weight of the MSE in the new loss function. Here we set $\omega=100$. Since the fraction is between 0 and 1, the MSE is not sensitive to small values, while the IoU is strongly sensitive to small values. We find the combination loss function performs the best in predicting the fraction of the mass that comes from stellar feedback.

\begin{table*}[]
\begin{center}
\caption{Training Model Parameter\label{Training Model Parameter}}
\begin{tabular}{cccccccc}
\hline
Model &Task& \multicolumn{3}{c}{Training Set}& Learning Rate  &  Epoch   \\
  & &Fiducial Resolution & Negative Set & High Resolution  & &    \\
 & &(5 pc$\times$5 pc) & & (2.5 pc$\times$2.5 pc)   & &    \\
 \hline   
   ME1 & I & \cmark & \cmark & \cmark & adaptive & 277 \\
   ME2 & I & \cmark & \cmark & \xmark & adaptive & 223 \\
   ME3 & I & \cmark & \cmark & \cmark & fixed & 60 \\
   ME4 & I & \cmark & \cmark & \cmark & adaptive & 60 \\
   ME5 & I & \cmark & \xmark & \cmark & adaptive & 189 \\
   ME6 & I & \cmark & \xmark & \cmark & adaptive & 260 \\
   ME7 & I & \xmark & \cmark & \cmark & adaptive & 260 \\
   MF$^{a}$ & II & \cmark & \cmark & \cmark & adaptive & 277 \\
\hline
\multicolumn{7}{p{0.95\linewidth}}{$^{a}$: The training set and the hyper-parameters of model MF is the same as those of ME1. The only difference is the training set. Model MF adopts the fraction of the feedback mass as the target. The other models adopt the intensity of the feedback emission in the training. 
}\\
\end{tabular}
\end{center}
\end{table*}

\subsection{Training Sets}
\label{Training Sets}

\subsubsection{Synthetic \13co\ Observations} 
\label{Synthetic Observations}

To train \CASItD, we adopt magneto-hydrodynamic (MHD) simulations that model sources launching stellar winds in a piece of a turbulent molecular cloud \citep{2015ApJ...811..146O}. The simulation box is $5\times 5\times 5$ pc$^3$, with a total mass of $3762~\msun$ and mean density of $\sim500 $ cm$ ^{-3} $. \citet{2015ApJ...811..146O} conduct different simulation runs with different mass loss rates, different magnetic fields, different turbulent patterns and different evolutionary stages to study the impact of stellar winds on the ambient gas. More details about the simulations can be found in \citet{2015ApJ...811..146O}.

We apply the publicly available radiation transfer code \radmc\ \citep{2012ascl.soft02015D} to model the 
\13co\ ($J$=1-0) line emission of the simulation gas. We use \13co\ emission rather than \co\ since \co\ is optically thick at the average simulation column density. 
To construct the synthetic observations, we adopt the simulation density, temperature and velocity distribution for the \radmc\ inputs. In the radiative transfer, \h2 is assumed to be the only collisional partner with \13co. In general, we assume the \13co\ abundance is constant where [\13co/\h2]=$1.5\times10^{-6}$. However, when gas conditions are likely to result in full dissociation of \13co\ ($T>1000$~K or $n(\rm H_{2})<$~50~\cmc), we set the abundance to zero. In addition, we also set the \13co\ abundance to zero in conditions where it would freeze out onto dust grains ($n(\rm H_{2})<10^{4}$~\cmc) or where it would be dissociated by strong shocks ($|v| > 10$~km/s).



We increase the training set by also considering thin clouds. \citet{2015ApJ...811...71Q} study the thickness of molecular clouds from the core velocity dispersion (CVD) and find the line-of-sight thickness of Taurus, Perseus and Ophiuchus molecular clouds are $\leq \frac{1}{8}$ of their length. To recreate the distinct circular structures of the observed stellar feedback, we crop the data cube into thinner slices, including widths of 2 pc and 0.9 pc. We show the emission from \13co\ with different cloud thicknesses in Appendix~\ref{12co and 13co Comparison} and discuss why we adopt \13co\ in favor of \co\ in our analysis.

\subsubsection{Training Target: Tracer Field for Task I}
\label{Training Target: Tracer Field for Task I}

To construct the training set for task I, we first define the position of the bubbles using the tracer field that indicates the fraction of gas coming from stellar feedback at each position. A given voxel is assigned to be part of a bubble structure where more than 2\% of the mass comes from stellar feedback. Conversely, a voxel is assigned to be pristine gas where less than 2\% of the mass comes from stellar feedback. 
To better capture the morphology of the bubbles, we define the position of a bubble by the gas temperature, where $T\ge 12$ K. We discuss different definitions of bubbles in Appendix~\ref{Different Definitions for the Bubble Extends}, including a criterion using the gas velocity.

We mask all the positions of pristine gas. We set the \13co\ abundance to be 0 in the masked region and compute the radiative transfer to obtain the \13co\ emission that is only coming from the stellar feedback, which we refer to as the \13co\ feedback map. Figure~\ref{fig.bubbleco-inte-all} shows an example of synthetic \13co\ observations and the \13co\ feedback map. In PPV space, the feedback map is the \13co\ emission that comes from stellar feedback, which allows us to distinguish between the feedback and diffuse emission from the host molecular cloud. The wind tracer is the positive signal that \CASItD\ learns to pick out from the messy \13co\ emission of the molecular cloud. 

In PPV space, the \13co\ feedback map emission is only a fraction of the total emission in each voxel, due to the foreground and background emission along the line of sight. As a result, we built the target in Task I by filling the voxels where there is feedback with the corresponding \13co\ emission in PPV space. This closely emulates what astronomers do in observational data \citep{2011ApJ...742..105A,2015ApJS..219...20L}. The model in Task I adopts the raw \13co\ data cube and returns a data cube of the same shape that only has the \13co\ intensity in the feedback regions.

In the final step, we compute the bubble properties from the output prediction. The prediction is the reconstructed \13co\ emission in voxels associated with feedback in PPV space. We apply a threshold based on the noise level of the input data to mask out noise-induced false bubble identifications. We then sum over all the voxels to calculate the mass, momentum and energy of the identified bubbles.  

\subsubsection{Training Target: Mass Fraction for Task II}
\label{Training Target: Mass Fraction for Task II}

To attempt to improve mass, momentum and energy estimation in the feedback identification, we adopt the fraction of feedback mass as the target in Task II. To build the target, we first convert the raw density from position-position-position (PPP) space to PPV space, just like the \13co\ data cube. Next, we convert the density cubes that only include the feedback gas, to PPV space. We take the ratio of the two converted cubes to get the fraction of feedback mass in PPV space. The fraction ranges from 0 to 1 and the fraction value is not necessarily proportional to the actual \13co\ intensity. If the \13co\ emission is optically thin, its column density is proportional to the emission intensity. Knowing the fraction of the feedback in each position allows us to calculate the actual feedback mass. 

Note that the emission predicted by Task I does not exclusively come from the feedback gas. Pristine gas that is not associated with feedback also contributes to the emission. Thus, Task I overestimates the mass coming from feedback. Consequently, Task II is a more advanced approach to estimate the physical properties of the feedback bubbles.

\subsubsection{Data Augmentation}
\label{Data Augmentation}

\begin{figure*}[hbt!]
\centering
\includegraphics[width=0.95\linewidth]{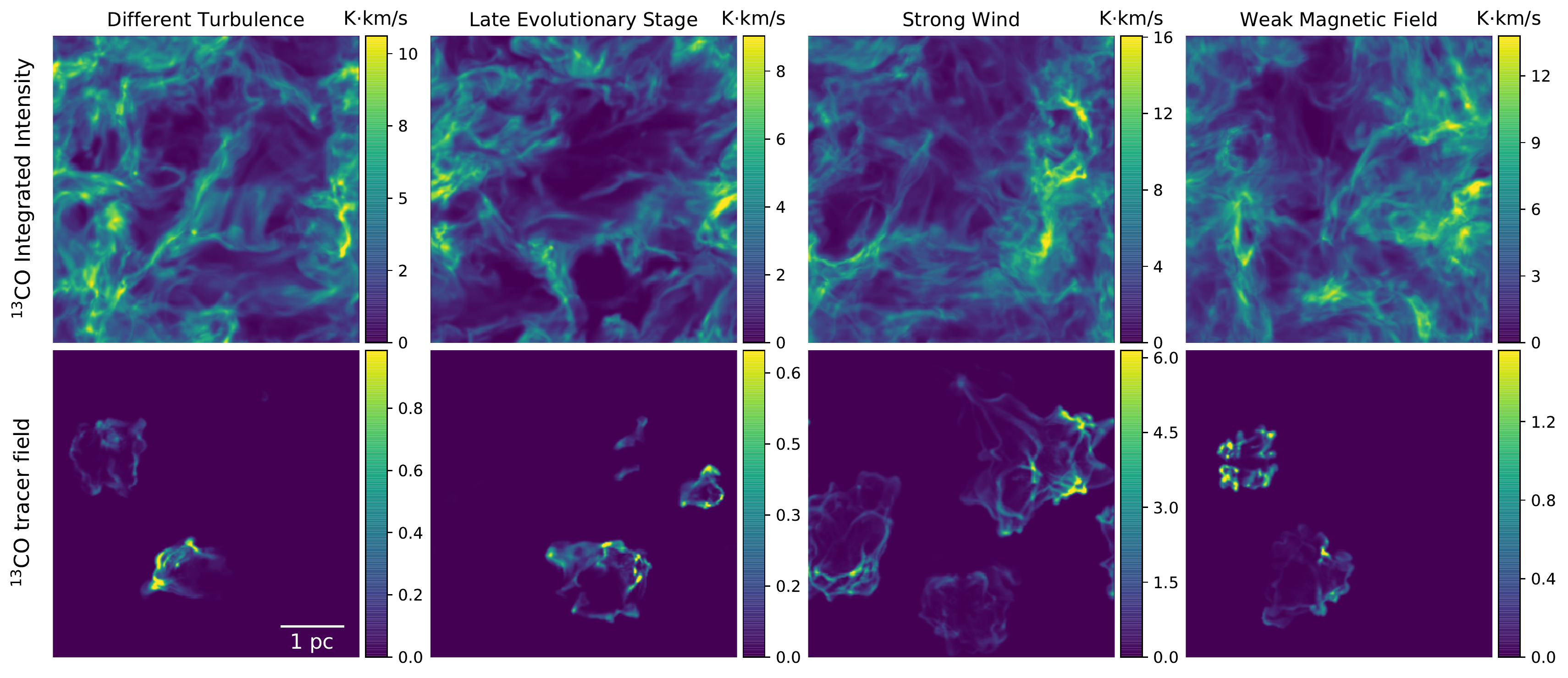}
\caption{Upper row: \13co\ integrated intensity including emission from the full cloud. Bottom row: integrated intensity of the full \13co (upper row) masked by a \13co\ synthetic observation of the tracer field to obtain the pixel locations of the feedback in PPV space (see Section~\ref{Training Target: Tracer Field for Task I}). First column: synthetic observations corresponding to model W2\_T2\_t0 in \citet{2015ApJ...811..146O}. Second Column: synthetic observations corresponding to model W2\_T2\_t1. Third column: synthetic observations corresponding to model W1\_T2\_t0. Fourth column: synthetic observations corresponding to model W2\_T3\_t0. }
\label{fig.bubbleco-inte-all}
\end{figure*}

We adopt simulations with different mass loss rates, different magnetic fields, different turbulent patterns and different evolutionary stages to create synthetic observations. The simulations have a physical scale of 5 pc on a side. To enhance the diversity of the training set, we conduct radiative transfer from three different angular views and rotate the images every 15$^{\circ}$ from 0$^{\circ}$ to 360$^{\circ}$. Figure~\ref{fig.bubbleco-inte-all} shows an example of the integrated intensity and the tracer field of \13co\ with different model outputs. We also construct a ``zoomed in" synthetic observation on bubbles with an image size of 2.5 pc $\times$2.5 pc. Different scales of bubbles in the training set reinforce the ability of the model to detect bubbles on different scales. 

To help \CASI\ distinguish feedback bubbles from shell-like structures produced from supersonic turbulence in the molecular cloud, we also conduct synthetic observations on purely turbulent simulations {including noise}, which do not contain feedback sources. We adopt the non-feedback cloud emission data as a negative training set. {This negative training set is essential because it trains the algorithm to ignore large (e.g., 0.5-2 pc) arc-like and shell-like features that ambient turbulent motions may generate despite no recent stellar feedback. In the Taurus survey, for instance, there are areas that are considerably larger than our adopted postage-stamp view-port size that lack any feedback sources while curving, filamentary structures are visually apparent. The model must perform well in those areas or else may provide false detections.} Table~\ref{Training Model Parameter} lists the properties of the training sets adopted by seven different models.

\begin{figure*}[hbt!]
\centering
\includegraphics[width=0.95\linewidth]{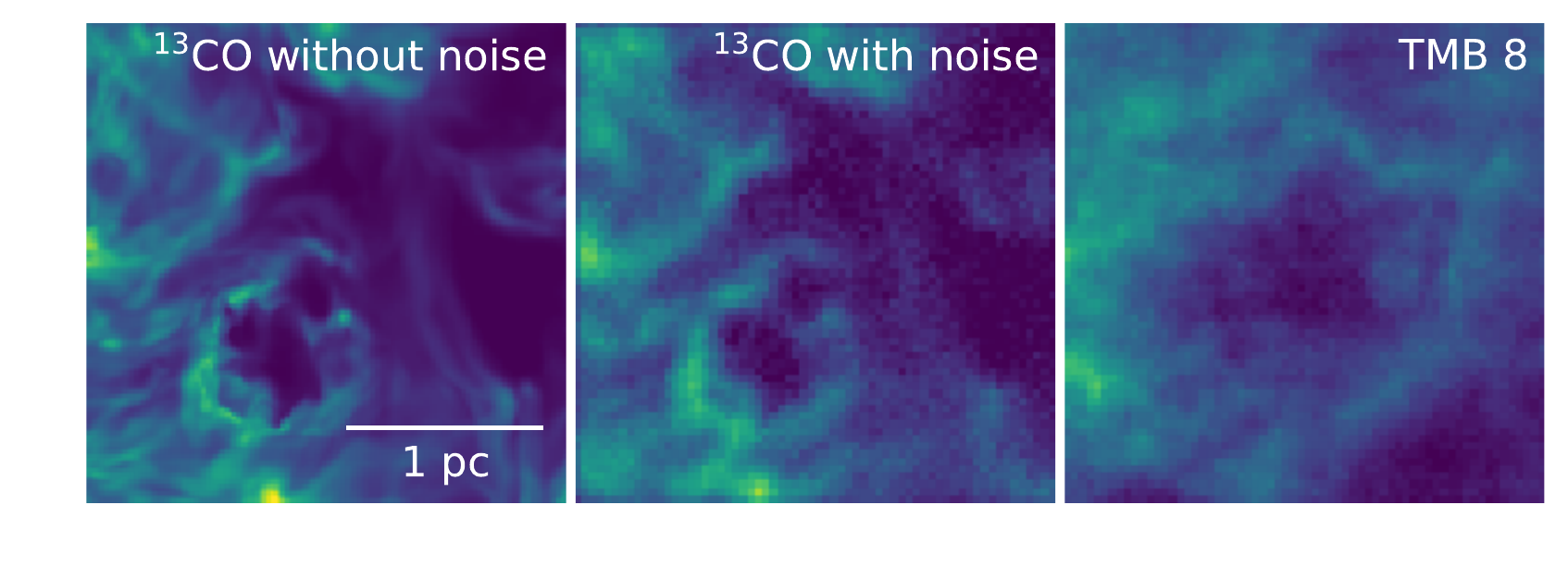}
\caption{Integrated intensity of \13co\ ($J$=1-0). Left: \13co\ integrated intensity of an simulated bubble without noise. Middle: \13co\ integrated intensity of a simulated bubble convolved with a beam of size 50'' and with 0.125\,K noise.
Right: \13co\ integrated intensity of an observed bubble, TMB 8, identified by \citet{2015ApJS..219...20L}.}
\label{fig.bubbleco-inte-all-noise}
\end{figure*}

To make the synthetic cubes closer to the real observational data in Section~\ref{Taurus Data}, we convolve them with a telescope beam of 50'' and add noise. We assume the synthetic images are at a distance of either 140 pc or 250 pc and are observed by Five College Radio Astronomy Observatory (FCRAO) \citep{2006AJ....131.2921R}. Figure~\ref{fig.bubbleco-inte-all-noise} shows a bubble before and after we convolve the image with the beam and add noise. The noise level, 0.125 K, is the same as the RMS noise in the Taurus \13co\ observational data. 
Moreover, we randomly shift the central velocity of the cubes between -1 to 1 \kms\, to increase the diversity of the training set. 

In total, we generate 7821 synthetic data cubes: 3910 have a field of view (FoV) of 5~pc~$\times$~5~pc, 3648 have a FoV of 2.5~pc~$\times$~2.5~pc  and 509 contain no feedback sources. We adopt 4693 of the data cubes as a training set, 1564 data cubes as a test set and 1564 data cubes as a validation set.  The validation set 
allows us to estimate how well the model has been trained. The test set assesses the accuracy of the final model.

\subsection{Taurus Data}
\label{Taurus Data}

The Taurus \13co\ $J=1-0$ map was observed between 2003 and 2005 using the 13.7 m FCRAO Telescope \citep{2008ApJS..177..341N}. The map covers an area of $\sim$98 deg$^2$ with a beam size of 50$^{\prime\prime}$. The data have a mean RMS antenna temperature of 0.125 K. We combine this with the Class III YSO catalog of \citet{2017ApJ...838..150K} as a reference to determine the potential driving sources of the bubbles. \citet{2017ApJ...838..150K} reexamined 396 candidate members from previous surveys in the literature covering $3^{h}50^{m}<\alpha<5^{h}40^{m}$ and $14^{\circ}<\delta<34^{\circ}$. They concluded 218 YSOs are confirmed or likely Taurus members, but 160 candidates are confirmed or likely interlopers, and the remaining 18 objects are uncertain. 

\citet{2015ApJS..219...20L} visually identified 37 bubbles in the Taurus molecular cloud from the \13co\ emission, and we adopt these as an observational test sample for our models. 
We divide these 37 bubbles into three categories based on their morphology and likely driving source. The three ranked categories of bubbles are:
\begin{enumerate}[\null]

\item  A: An A bubble contains at least one YSO inside the bubble and has clear circle/arc morphology. 

\item B: An B bubble contains no YSOs inside but contains at least one YSO on the bubble rim or near the bubble boundary.

\item C: An C bubble contains no known YSOs in/around the bubbles.

\end{enumerate}
Among the 37 bubbles, 7 are Rank A, 16 are Rank B and 14 are Rank C. 

To make the observational data suitable for the algorithm, we first down-sample the Taurus \13co\ data cube by a factor of 3, so that it has a similar resolution ($\sim 1^{\prime}$) to that of the training set. We shift the down-sampled cube's mean velocity to 0, and then crop the velocity range from -4 \kms\ to 4 \kms. We crop the cube to 2.7 pc $\times$ 2.7 pc after centering on each bubble identified by \citet{2015ApJS..219...20L}. This procedure generates a stack of data cubes with a shape of $64\times 64 \times 32$. See Appendix~\ref{Assessing the Sensitivity of the Data Window} for more detail.

However, investigating only the 37 previously identified bubbles is limiting. \CASItD\ does not require the bubbles to be centered in the image, and the algorithm is able to rapidly search the entire Taurus map if it is divided into smaller data cubes. Furthermore, comparing the \CASItD\ identification of the previously defined, cropped bubbles to those identified in the full map in an unbiased search allows us to verify that the algorithm is translation invariant and insensitive to the position of the bubble.

\CASItD\ requires input data that has the same dimensions as the training data. When applying the CNN model to the full map, we decompose the Taurus data into a series of $64\times 64 \times 32$ cubes. Each cube is offset by 5 pixels ($\sim$5\arcmin) resulting in 92\% overlap with adjacent steps. We begin cropping from the northeast corner of the full map. Then we move to the next position along the RA direction with a step size of 5 pixels. When we finish the sampling procedure along the RA direction at fixed declination, we move 5 pixels in declination and then repeat the process again.

 \subsection{Model Selection} 
\label{Model Selection}

\subsubsection{Validation}
\label{validation}

\begin{figure}[hbt!]
\centering
\includegraphics[width=0.98\linewidth]{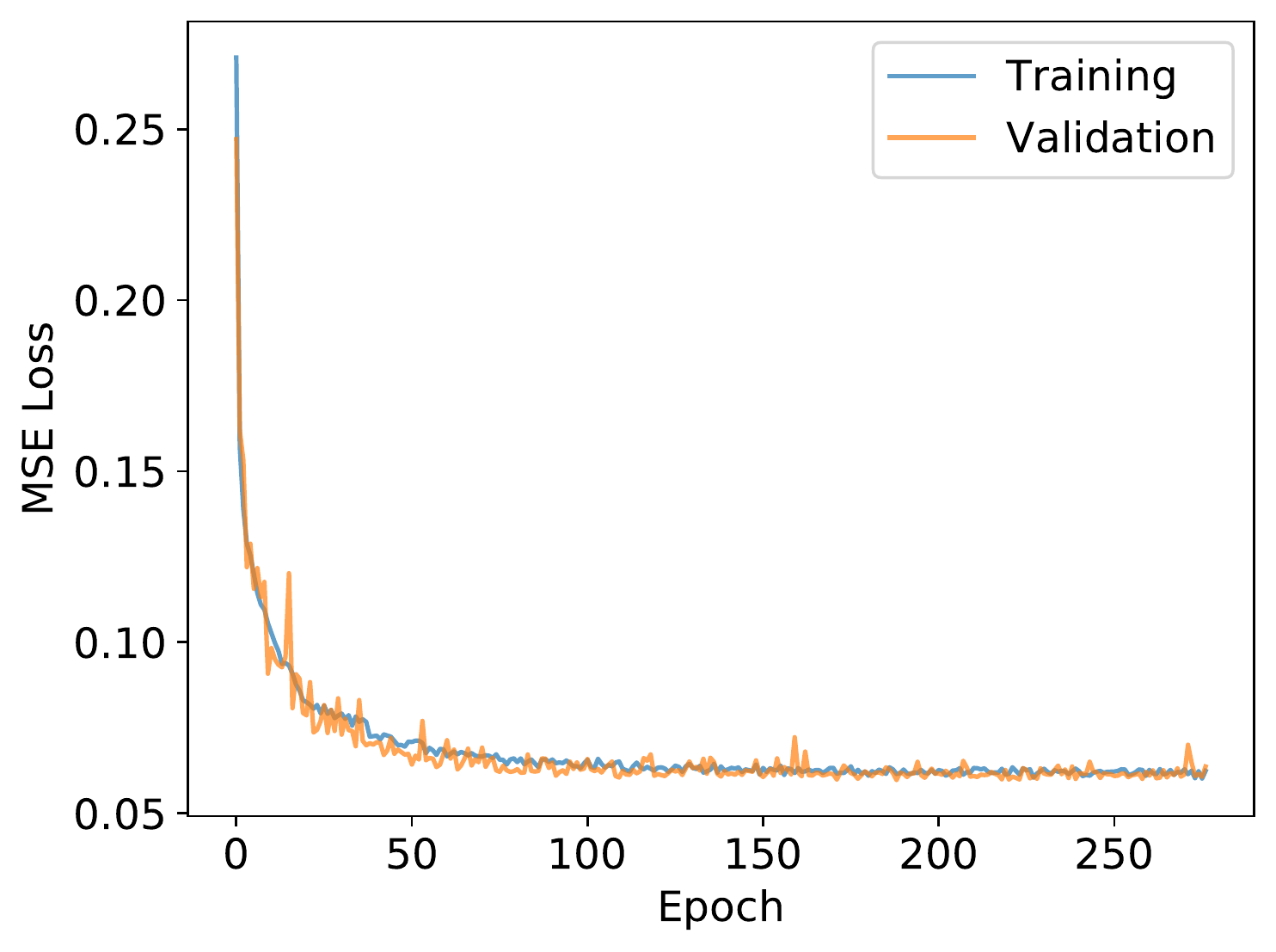}
\caption{Training and validation errors of model ME1 during training.}
\label{fig.bubbleco-lossfunction}
\end{figure}

\begin{figure*}[hbt!]
\centering
\includegraphics[width=0.98\linewidth]{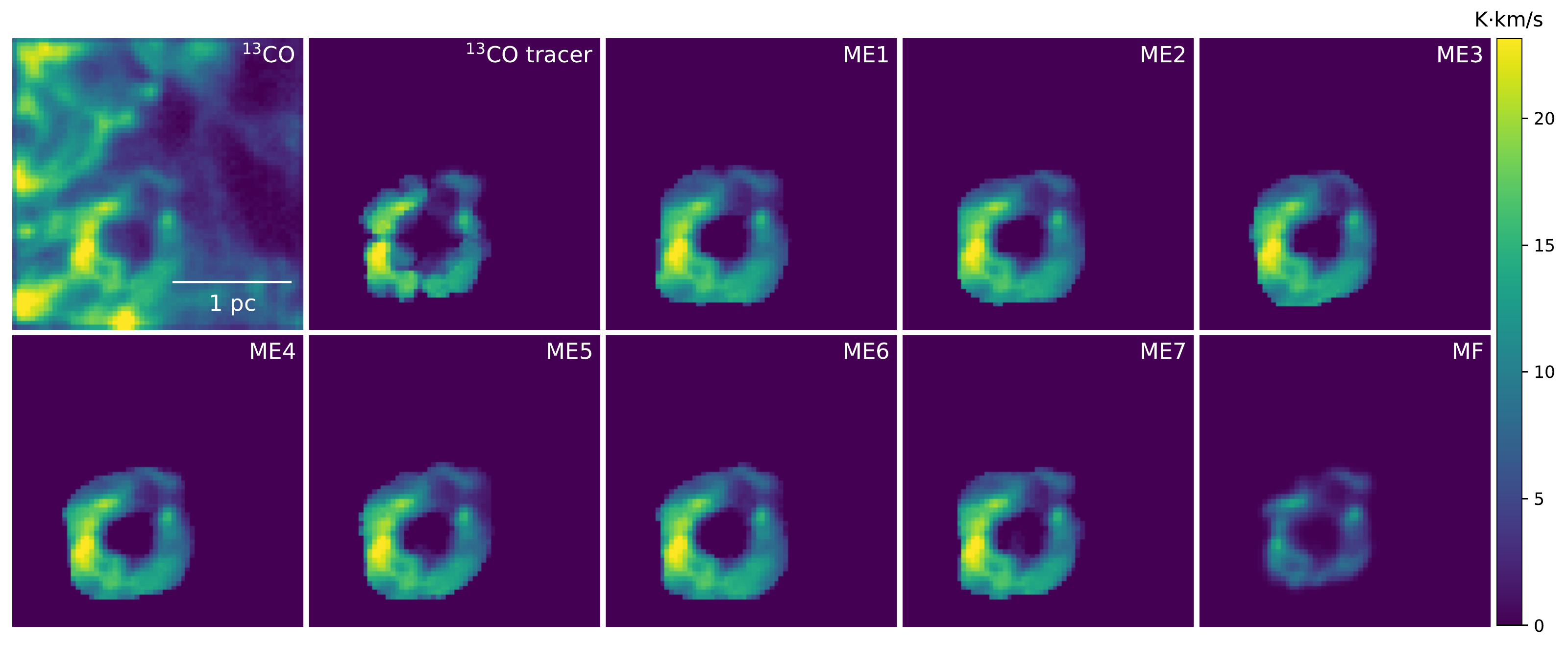}
\caption{ The integrated intensity of the \13co\ (top left), the integrated intensity of the full \13co masked by a \13co synthetic observation of the tracer field to obtain the pixel locations of the feedback in PPV space (second from top left),  and the prediction of eight CNN models for the bubble emission.
}
\label{fig.bubbleco-test-syn}
\end{figure*}

After training, we find all models in Task I converge to a MSE below 0.1, and a combination of MSE and IoU in Task II (model MF) converges to unity. Figure~\ref{fig.bubbleco-lossfunction} shows the training and validation errors of model ME1. After 277 epochs, this model converges to a MSE of 0.06. The  number of epochs used in the training, 277, is set by the maximum job run-time permitted on our computing resources. We show the performance of seven CNN models on a test set of synthetic observations in Figure~\ref{fig.bubbleco-test-syn}. All models in Task I and II capture the shell features produced by stellar feedback clearly. 

\subsubsection{Mean Opinion Score: Visually Assessing the Model Performance}
\label{Mean Opinion Score: Visually Assessing the Model Performance}

To visually assess the performance of the CNN models on observational data, we apply eight CNN models as listed in Table~\ref{Training Model Parameter} to the Taurus \13co\ bubble data. All test bubbles have a clear circular or arc-like structure across a range of velocity channels, which provides an appropriate test sample.
However, we do not quantitatively know the true bubble boundaries, so we use visual identification and assessment to evaluate the performance of the models. 
We introduce the mean opinion score (MOS) to visually quantify the performance of the Task I models on the Rank A and B bubbles in Taurus. The MOS is expressed as an integer ranging from 1 to 5, where 1 is poor performance, and 5 is excellent performance. We create a rubric outlining the characteristics of each score to ensure visual ranking between assessors is as uniform as possible.  The rubric is as follows:

\begin{enumerate}[]

\item 5: excellent performance. The prediction covers the full rim structure of the bubble, with less than 10\% extra emission, i.e., a minimal amount of false positive pixels. 

\item 4: good performance. The prediction covers 75\% of the rim structure of the bubble, with less than 25\% extra emission. 

\item 3: average performance. The prediction covers half the rim structure of the bubble, with less than 50\% extra prediction. 

\item 2: fair performance. The prediction covers one third of the rim structure of the bubble, with less than 70\% extra prediction. 

\item 1: poor performance. The prediction covers less than one third of the rim structure of the bubble, with more than 70\% extra prediction. 

\end{enumerate}

\begin{figure*}[hbt!]
\centering
\includegraphics[width=0.98\linewidth]{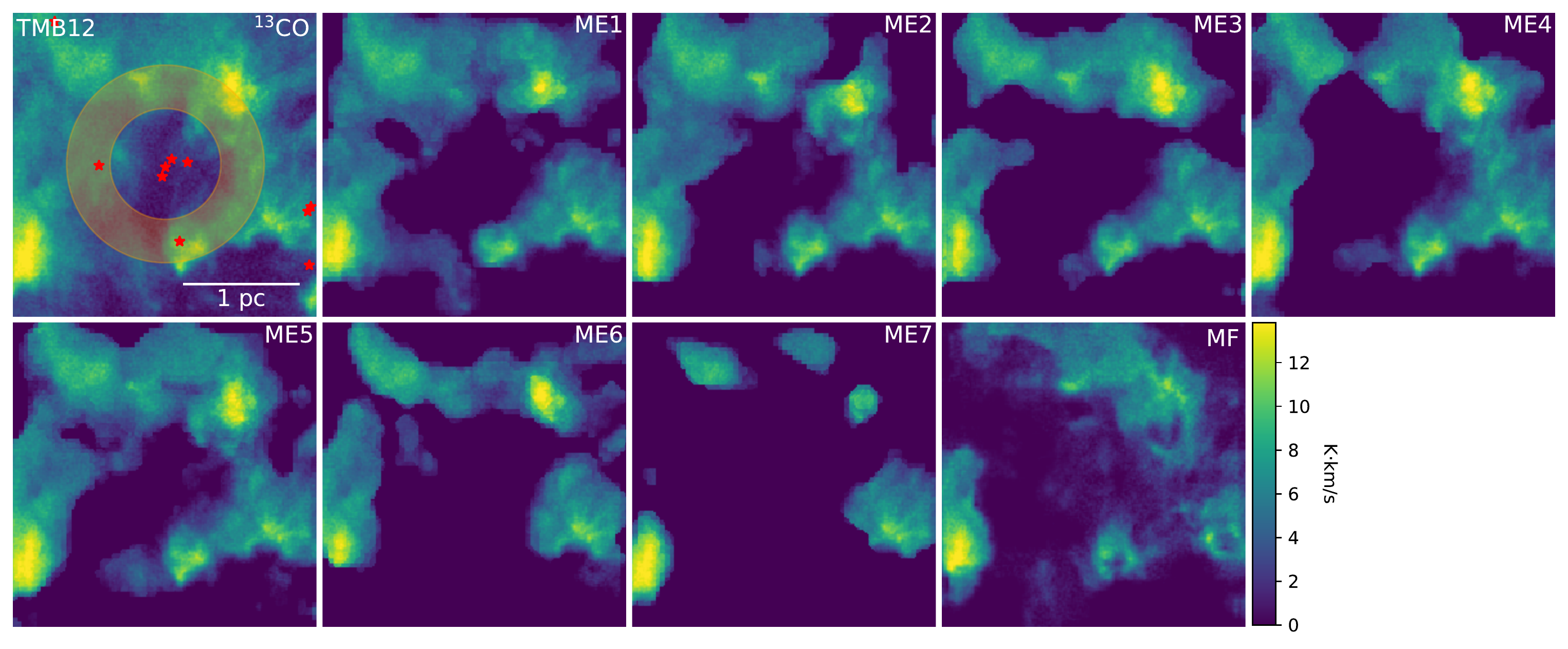}
\caption{The results of the eight models applied to Rank A bubble TMB12. First row, first column panel: integrated intensity of \13co\ overlaid with a yellow ring indicating the position and thickness of the bubble. Star symbols show the location of the Class III YSOs from \citet{2017ApJ...838..150K}. The remaining panels: the predicted intensity integrated along the velocity axis for the eight CNN models.}
\label{fig.bubbleco-taurus1}
\end{figure*}

\begin{figure*}[hbt!]
\centering
\includegraphics[width=0.98\linewidth]{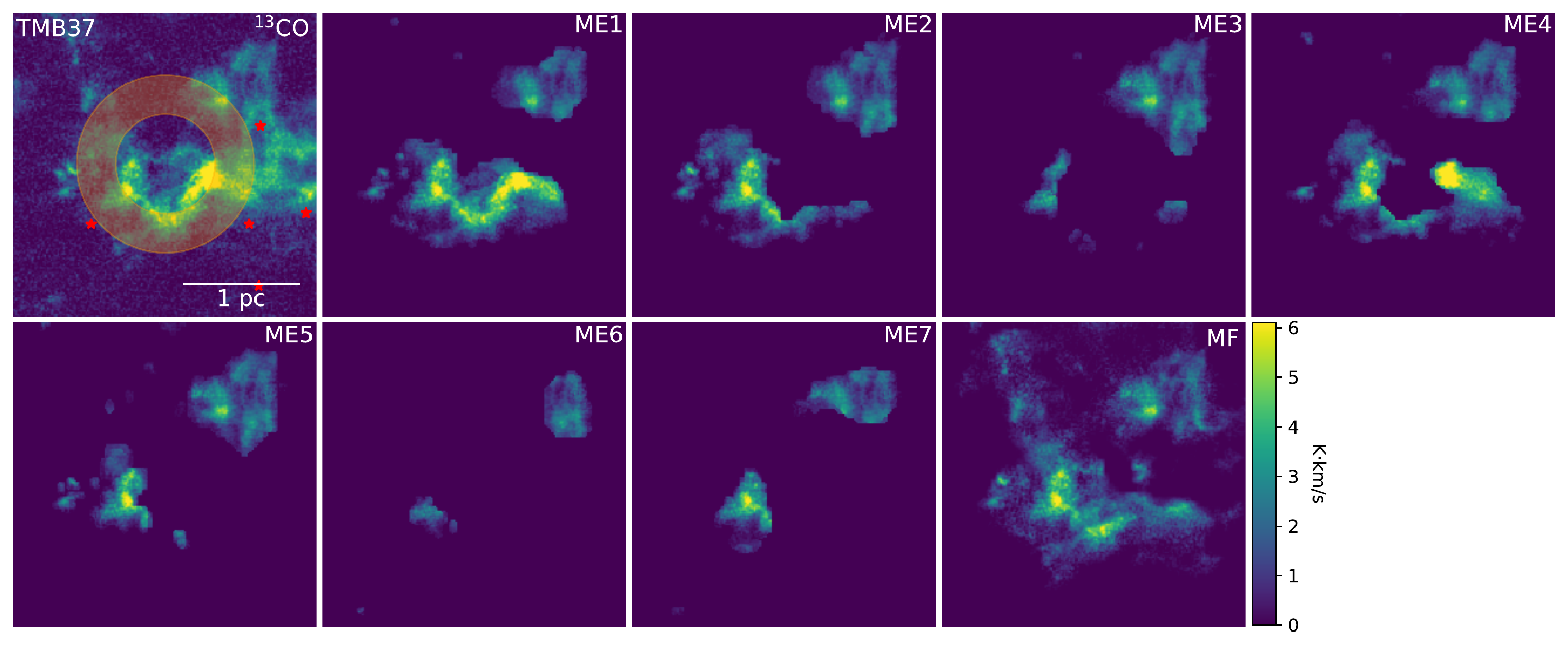}
\caption{The same as Figure~\ref{fig.bubbleco-taurus1} but depicting the Rank B bubble TMB37.}  
\label{fig.bubbleco-taurus2}

\end{figure*}

Figure~\ref{fig.bubbleco-taurus1} and \ref{fig.bubbleco-taurus2} show the integrated intensity of example Rank A and B bubbles and compare the predictions from the CNN models. 

\begin{figure}[hbt!]
\centering
\includegraphics[width=0.98\linewidth]{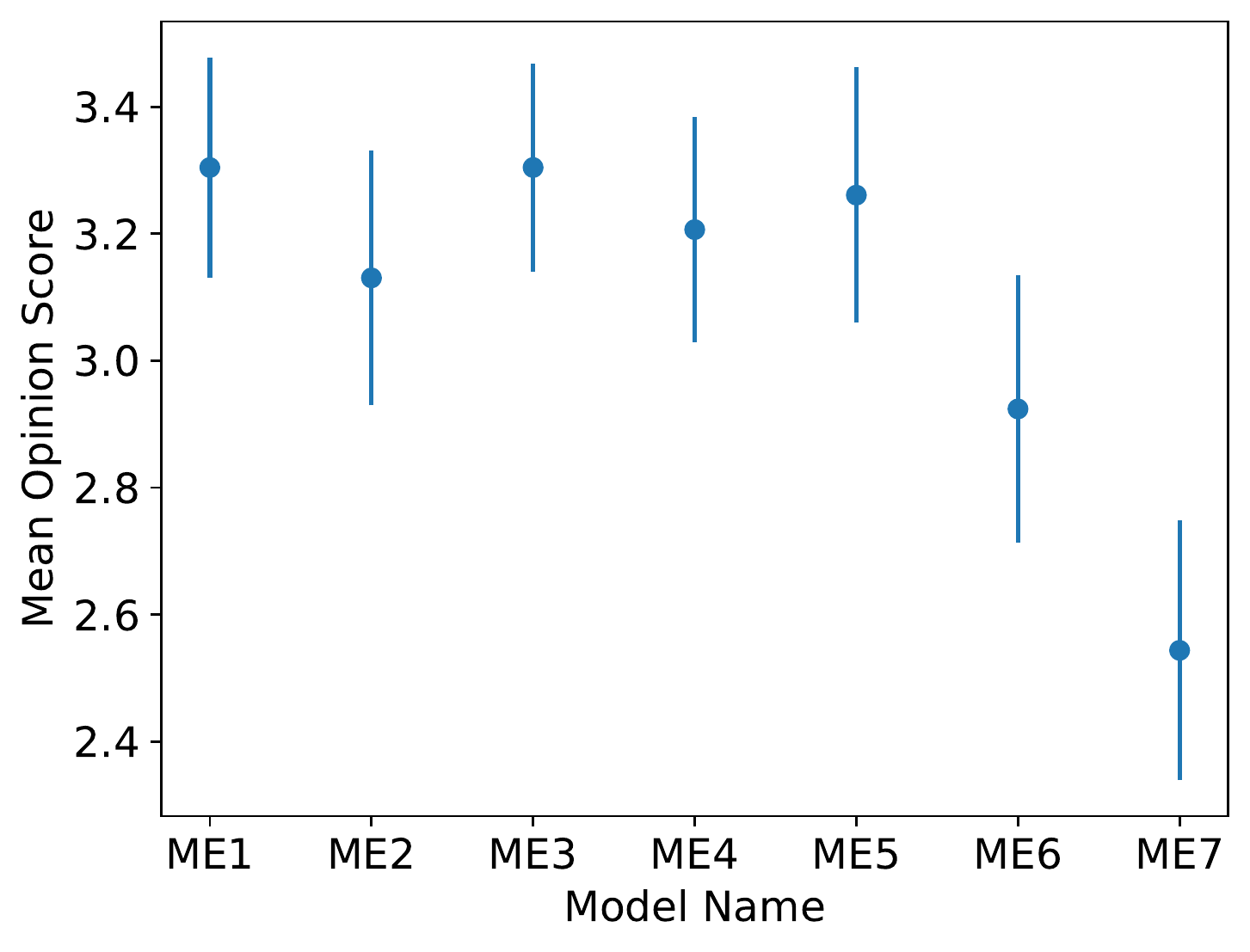}
\caption{The overall mean opinion score (MOS for each model, averaged over all 23 Rank A and Rank B bubbles and the visual rankings by the four human judges. }  
\label{fig.MOS-model}
\end{figure}

We conduct a blind rating of the performance of the seven models in Task I, where each co-author assessed the quality of each bubble prediction. To judge the prediction, the co-authors looked at the integrated intensity map and the channel by channel prediction of each model for each bubble. Figure~\ref{fig.MOS-model} shows the MOS of each model as rated by the four co-authors. Five of the seven models 
have an overall MOS that is above average. We easily rule out models ME6 and ME7, which have lower MOSs. ME1 exhibits decent performance on the Taurus bubbles, and it is trained using the most complete training set that includes both negative examples (no bubbles) and higher resolution bubbles. We adopt ME1 as the fiducial model in the following analysis. 

For Task II, we maintain the same training set and hyper-parameters as those adopted for ME1 and simply replace the training target data 
with the fraction of mass coming from feedback in the training set 
as discussed in Section~\ref{CASI-3D Architecture}.

\section{Results} 
\label{Result}

\subsection{Assessing Model Accuracy Using Synthetic Observations}
\label{Assessing Model Accuracy Using Synthetic Observations}

In this section we use the synthetic images to assess how accurately physical properties can be determined from the identified bubbles. We apply all the models to the synthetic observations of the bubbles in the test set as shown in Figure~\ref{fig.bubbleco-test-syn}.  We mask the prediction cubes at 0.2 K, which is consistent with the input cubes' noise level. We then calculate the mass of the bubbles by assuming that the \13co\ emission line is optically thin and has an excitation temperature of 25 K \citep{2012MNRAS.425.2641N,2015ApJS..219...20L}. {We examine the uncertainty of the bubble mass estimation in terms of the choice of excitation temperatures in Appendix~\ref{Excitation Temperature Selection and Impact}.} We take $1.5\times10^{-6}$ as the abundance ratio between \13co\ and \h2 \citep{2012MNRAS.425.2641N,2015ApJS..219...20L}. Finally, we compute the total mass by summing over the bubble volume.

\begin{figure}[hbt!]
\centering
\includegraphics[width=0.98\linewidth]{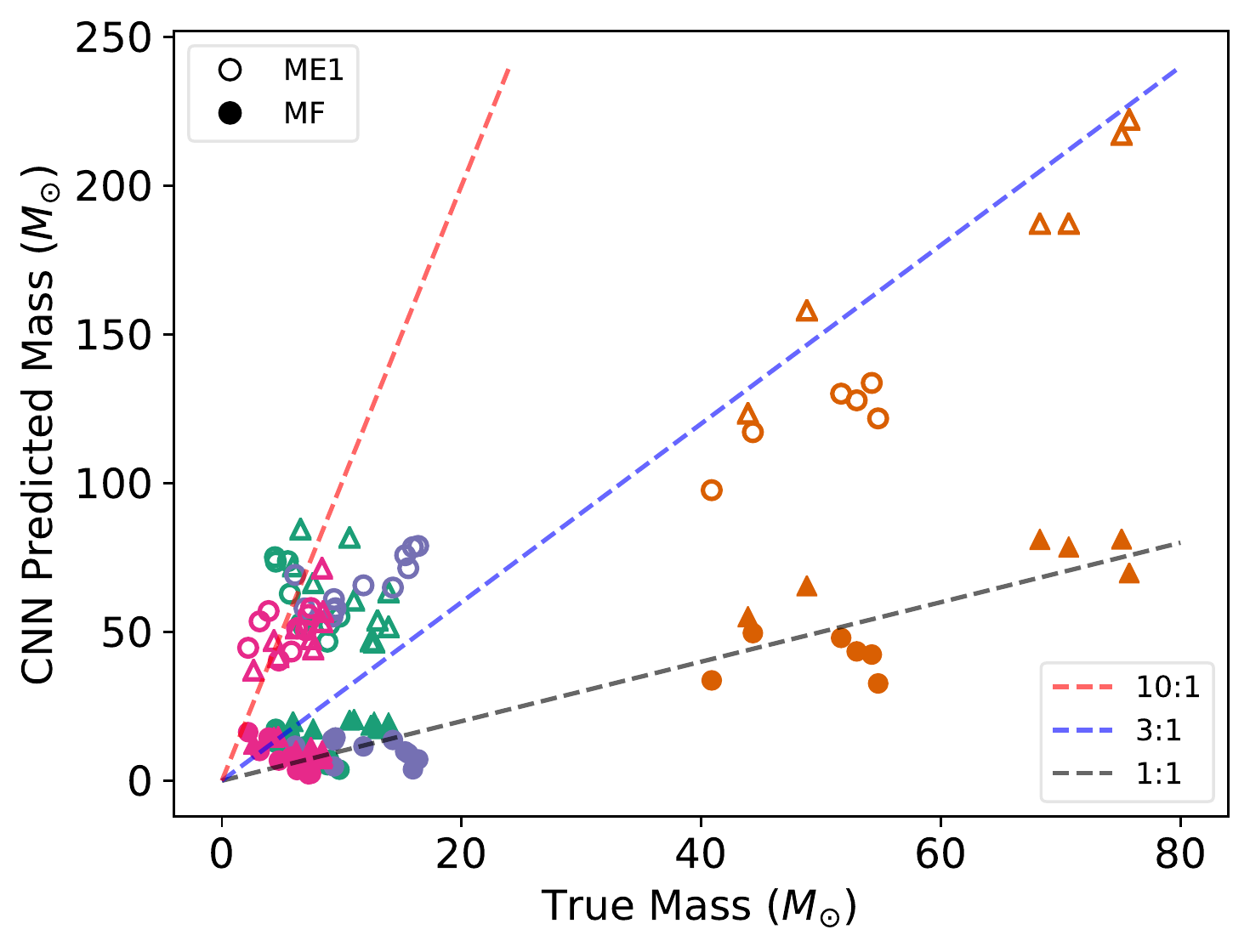}
\caption{The relation between the CNN predicted bubble mass and the true feedback mass for different bubbles. The filled symbols indicate the mass calculated from model MF. The open symbols represent the mass calculated from model ME1. The {black} dashed line indicates where the CNN correctly predicts the true mass as determined by the tracer field and gas temperature. {The blue dashed line has a slope of 3 and the red dashed line has a slope of 10.} The physical parameters of the simulations with different labels are listed in \citet{2015ApJ...811..146O}. }
\label{fig.tacer-comp-cnn2}
\end{figure}

Figure~\ref{fig.tacer-comp-cnn2} shows the mass estimated from the two models, ME1 and MF. We also plot the true feedback mass, which we estimate directly by adding the mass contained in all cells with $T\ge12$\,K and a tracer fraction $\ge$2\% (see \S2.3.2). We find ME1 overestimates the bubble mass by a factor of 3 {or more}, while MF correctly predicts the bubble mass within 4\% error. 
{The low-mass bubbles are overestimated by a factor of ten by model ME1, while the high-mass bubbles are overestimated by a factor of three by model ME1. Since the low-mass bubbles are usually small, and they do not expand enough to break out of the cloud 
such that more gas 
along the line of sight contributes to the \13co\ emission. 
Their velocities are also small, yielding more surrounding gas in the velocity channels where the feedback is. On the other hand, high-mass bubbles are usually large with more gas coming from the driving YSOs and have larger expanding velocities. The gas 
along the line of sight of the high-mass bubbles occupies a smaller fraction (but still a large portion) in each velocity channel compared to that of low-mass bubbles.}

\begin{figure}[hbt!]
\centering
\includegraphics[width=0.98\linewidth]{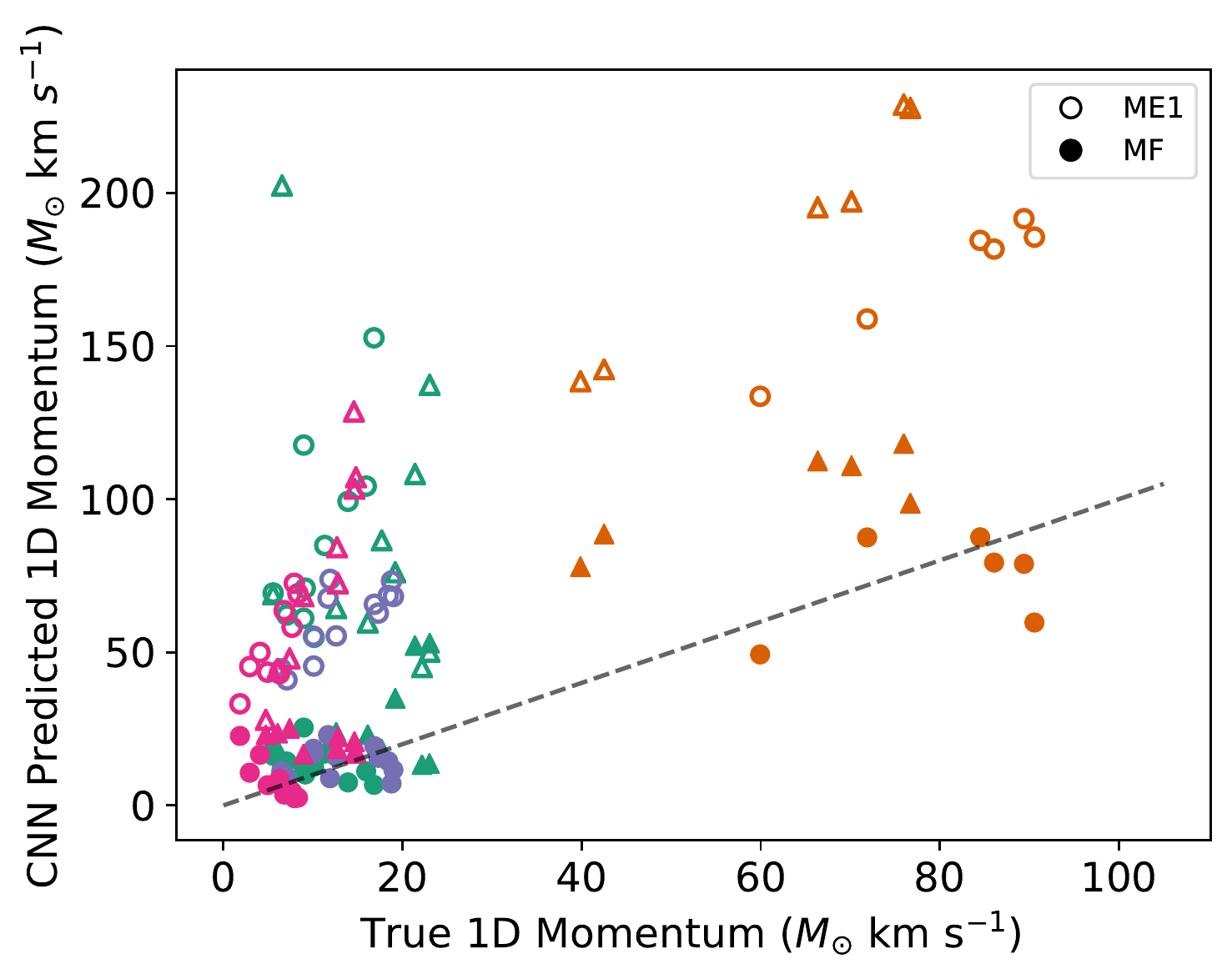}
\caption{The relation between the CNN predicted bubble momentum and the true feedback 1D momentum from different bubbles. The filled symbols indicate the momentum calculation from model MF. The open symbols represent the momentum calculation from model ME1. The dashed line indicates where the CNN correctly predicts the true momentum. }
\label{fig.tacer-comp-cnn2-momentum}
\end{figure}

We compare the 1D line of sight momentum between the model prediction and the true simulation feedback in Figure~\ref{fig.tacer-comp-cnn2-momentum}.  We define the 1D momentum as the sum of the gas mass in each channel multiplied by the channel velocity, where we have shifted the mean cloud velocity to zero. Model ME1 overestimates the 1D momentum by a factor of 2.8. In contrast, model MF is able to correctly predict the 1D momentum within 10\% error. 

\begin{figure}[hbt!]
\centering
\includegraphics[width=0.98\linewidth]{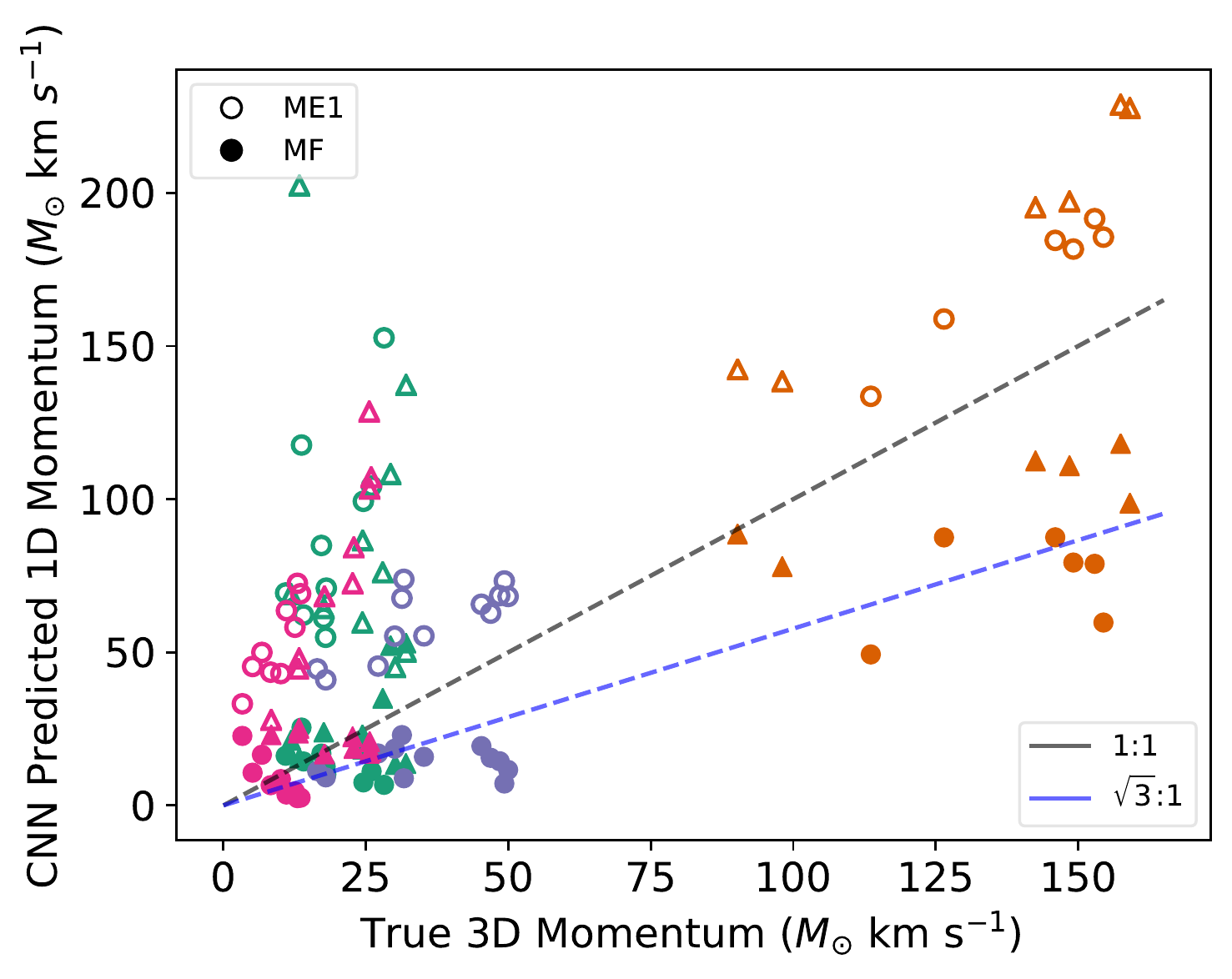}
\caption{The relation between the CNN predicted bubble momentum and the true feedback 3D momentum from different bubbles. The filled symbols indicate the momentum calculation from model ME1. The open symbols represent the momentum calculation from  model ME1. The black dashed line has a slope of 1 and the blue dashed line has a slope of 1/$\sqrt{3}$, which indicates that velocity symmetry is a reasonable assumption to estimate the true 3D momentum. }
\label{fig.tacer-comp-cnn2-momentum-3d}
\end{figure}

\begin{figure}[hbt!]
\centering
\includegraphics[width=0.98\linewidth]{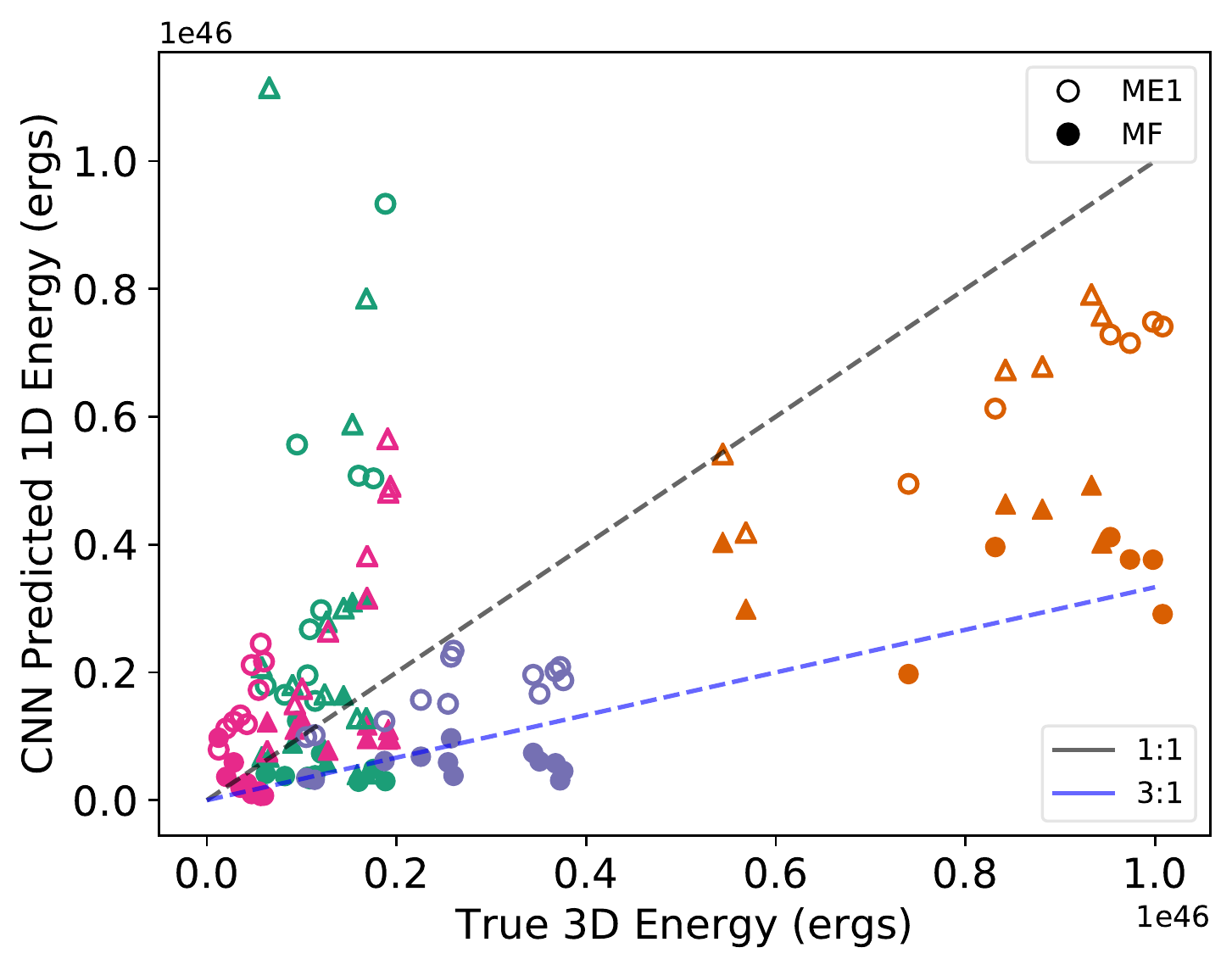}
\caption{The relation between the CNN predicted bubble energy and the true feedback energy from different bubbles. The filled symbols indicate the energy calculation from MF. The open symbols represent the mass calculation from ME1. The black dashed line has a slope of 1 and the blue dashed line has a slope of 1/3. }
\label{fig.tacer-comp-cnn2-energy}
\end{figure}

Under the assumption of isotropic expansion, the 3D momentum would be expected to be a factor of $\sqrt{3}$ larger compared to the 1D momentum, while the 3D kinetic energy would be a factor of 3 larger compared to the 1D kinetic energy. Figures~\ref{fig.tacer-comp-cnn2-momentum-3d} and \ref{fig.tacer-comp-cnn2-energy} show the 1D momentum and energy, respectively, predicted by the two models compared to the respective 3D quantities calculated from the simulation. 
Again, we find the momentum and energy predicted by model MF are comparable to the true simulation values. 

One caveat here is that we limit the velocity range of the synthetic observations in the training set to match the Taurus observation. To assess how much mass, momentum and energy are missed by applying this cutoff, we calculate the total mass, momentum and energy associated with velocities that exceed the CO spectrum velocity range. We find that 9\% of the mass is in gas with $|v|>4$ \kms. Meanwhile, 20\% of the momentum and 44\% of the energy are missed due to the limited CO velocity range.  In observations, the amount of missing mass, momentum and energy will depend both on the observation spectral range and the source masses, which are often not well constrained. In the following sections, we correct the totals for the missing mass, momentum and energy.

The different accuracies achieved by the ME1 and MF models can be understood as follows.
Model ME1 is trained using the tracer intensity and thus can only predict the feedback position. We find the training set for ME1, namely the full \13co\ emission masked by the position of the tracer field in PPV space, is not a good indicator of the fraction of feedback mass in each voxel, because the emission at the predicted position does not exclusively come from feedback gas. Instead, gas along the line of sight that is not associated with feedback contributes to the emission and can dominate the total. This matches the current state of the art in human identification, since visual identification of feedback cannot disentangle the feedback from the non-feedback gas within a given velocity range and voxel. This is why our model ME1 predicts total mass, momentum and energy similar to that estimated by \citet{2015ApJS..219...20L} as we will show in Section~\ref{Physical Properties of the Individual Taurus Bubbles}.
However, model MF adopts the fraction of mass coming from feedback as the training target, which allows the model not only to predict the position of feedback but also to predict the fraction of the mass coming from feedback in each voxel. This allows a significantly more accurate determination of the mass, momentum and energy.

\subsection{Physical Properties of the Individual Taurus Bubbles}
\label{Physical Properties of the Individual Taurus Bubbles}

Next, we estimate the masses of the bubbles in Taurus identified by models ME1 and MF and compare them with the previous observational estimates.  For the purpose of comparison with the prior visual identifications, we analyze postage stamps centered on each of the 37 Rank A, B and C bubbles.
We calculate the bubble mass and momentum for each model as described in Section~\ref{Assessing Model Accuracy Using Synthetic Observations}.
The observational approach to calculate the observed bubble mass, momentum and energy is as follows.
\citet{2015ApJS..219...20L} adopts an annulus as a mask for each bubble rim region, where the inner and outer radii of the annuli are determined by visual inspection, and then adds up the emission of \13co\ in the masked region to calculate the mass. The bubble  velocity extent along the line of sight is also determined by eye. 

\begin{figure}[hbt!]
\centering
\includegraphics[width=0.98\linewidth]{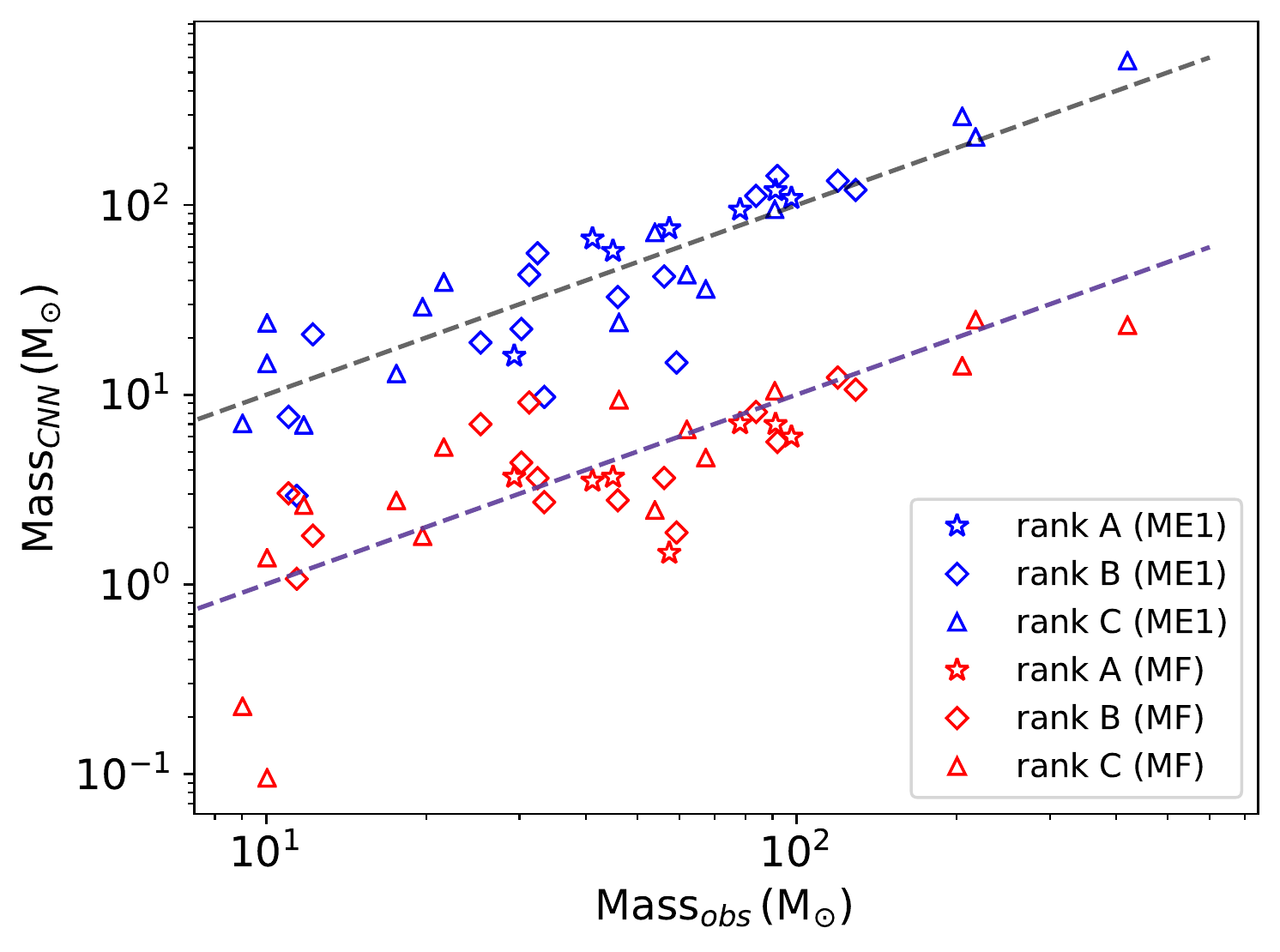}
\caption{Bubble mass estimated from the {\sc casi-3D} model  predictions and the observational mass estimate from \citet{2015ApJS..219...20L}. The grey dashed line indicates the trend for equal mass, while the purple dashed line is scaled down by 10. The blue symbols indicate the mass calculated from model ME1. The red symbols represent the mass calculated from model MF.}
\label{fig.bubbleco-mass-taurus}
\end{figure}

Figure~\ref{fig.bubbleco-mass-taurus} compares the bubble mass calculated from the two CNN models and that from the observational approach for all the bubbles identified by \citet{2015ApJS..219...20L}. The mass estimated from ME1 shows an approximately linear trend with that from the observational approach within a factor of 2. Figure~\ref{fig.bubbleco-mass-taurus} also compares the mass calculated from MF with the mass estimated from the observational approach. We find the observational approach overestimates the bubble mass by an order of magnitude.

\begin{figure}[hbt!]
\centering
\includegraphics[width=0.98\linewidth]{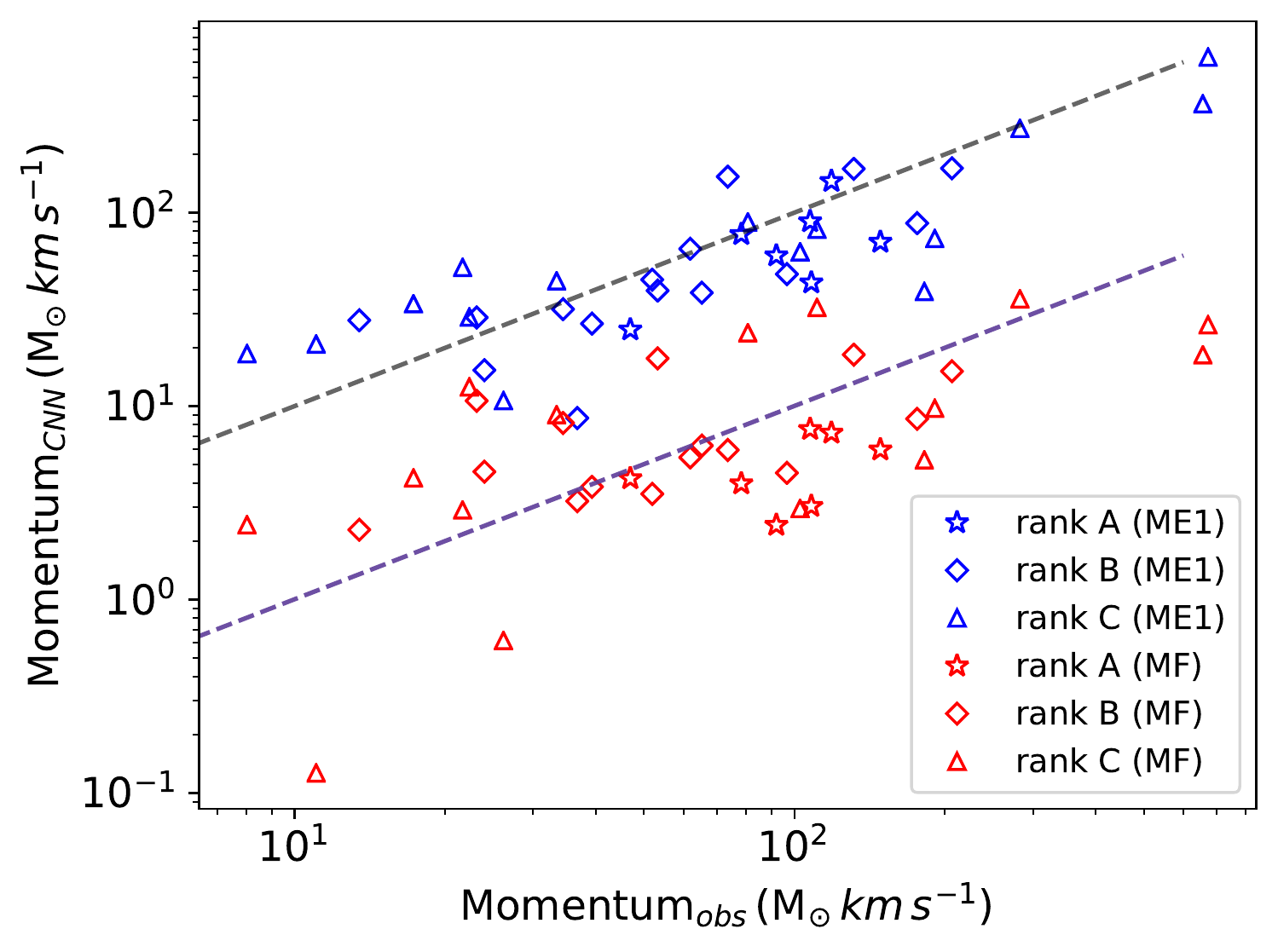}
\caption{Bubble momentum estimated from  the {\sc casi-3D} model predictions and the observational  momentum estimate from \citet{2015ApJS..219...20L}. The grey dashed line indicates the trend for equal momentum, while the purple dashed line is scaled down by 10. The blue symbols indicate the momentum calculated from model ME1. The red symbols represent the momentum calculated from model MF. }
\label{fig.bubbleco-momentum-taurus}
\end{figure}
\begin{figure}[hbt!]
\centering
\includegraphics[width=0.98\linewidth]{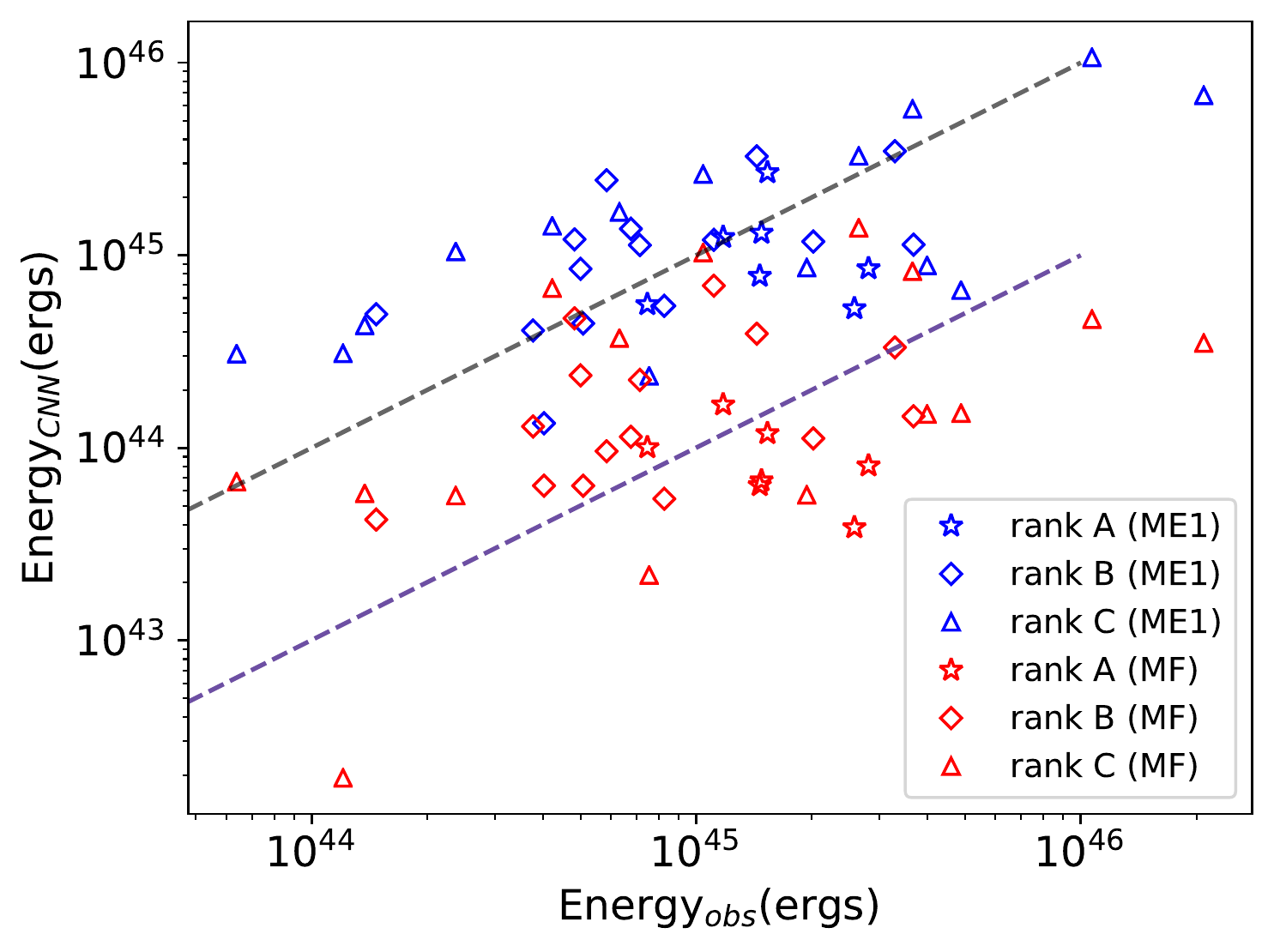}
\caption{Bubble energy estimated from the {\sc casi-3D} model predictions and the  observational  energy estimate from \citet{2015ApJS..219...20L}.  The grey dashed line indicates the trend for equal energy, while the purple dashed line is scaled down by 10. The blue symbols indicate the energy calculated from model ME1. The red symbols represent the energy calculated from model MF.}
\label{fig.bubbleco-energy-taurus}
\end{figure}

Figures~\ref{fig.bubbleco-momentum-taurus} and ~\ref{fig.bubbleco-energy-taurus} compare the momentum and energy calculated from the CNN models with those from the observational approach. The momentum and energy estimated from ME1 both show approximately linear trends compared to those from the observational approach and are within a factor of 2. However, when considering the fraction of mass coming from feedback, both model ME1 and the observational approach overestimate the momentum and energy by an order of magnitude. 

The differences between the observational approach and model ME1, are not too surprising, since
the observational approach to calculate the bubble mass, momentum and energy is fairly simple. For example,  \citet{2015ApJS..219...20L} may overestimate or underestimate the mass depending on the ring mask they draw. In addition, bubbles are not always a closed, symmetric circle. They are likely to be discontinuous on the rim as depicted in Figure~\ref{fig.bubbleco-taurus1}. As we showed in Section~\ref{Assessing Model Accuracy Using Synthetic Observations}, the higher values obtained by model ME1 and the observations compared to model MF are also not surprising since the former approaches include excess material along the line of sight that is not part of the feedback.
However, the {\sc casi-3d} models may also overestimate the total bubble properties since there may be more than one bubble identified in each postage stamp. To address this, we set the postage stamp size to minimize this effect.

{Several different effects may cause errors in estimating the mass, momentum and energy from the CO emission. We find that the choice of excitation temperature could cause a factor of two error in mass estimation, but it cannot account for a factor of ten (see Appendix~\ref{Excitation Temperature Selection and Impact} for more detail). Likewise, the assumption of LTE has a small affect on the mass estimation. We conclude the line of sight gas contamination is the main uncertainty in mass estimation. As we discussed in Section~\ref{Assessing Model Accuracy Using Synthetic Observations}, low-mass bubbles are overestimated by a larger factor (a factor of ten) compared to high-mass bubbles (a factor of three) due to the line of sight gas contamination. For low-mass bubbles, the line of sight contamination is the dominant factor overestimating the mass, but for high-mass bubbles, the uncertainty that comes from line of sight contamination is similar to the uncertainty that comes from assuming a fixed excitation temperature.

 It is also worth considering the estimates from a physical perspective. The mass associated with feedback based on our understanding of the launching velocities of feedback is usually comparable to the young stars' mass. It is physically impossible for a $\sim$ few \msun\ young star to drive a 50-60 \msun\ bubble. \citet{2011ApJ...742..105A} pointed out that these high mass bubbles could be produced if $v_{wind}$=200 km/s with $\dot{m}_{\rm wind}=10^{-7}-10^{-6}$~\msun /yr. However, these mass-loss rates are orders of magnitude higher than those that can be explained by stellar winds, outflows \citep[considering  outflows are collimated,][]{2016ARA&A..54..491B}, ionization or radiation pressure from the stars observed in Taurus \citep{2014ARA&A..52..487S}. The mass directly launched by young stars in both theoretical work and observations is small, $\sim 10^{-9}$~\msun/yr \citep[e.g.,][]{1994ApJ...429..781S,1995ApJ...452..736H}. Numerical simulations suggest the entrained gas can contribute three times more mass than the direct mass loss from young stars \citep{2017ApJ...847..104O}. In observations, the mass associated with feedback is included in the estimate of the entrained gas. Model ME1 does the same thing, i.e., it predicts the gas associated with feedback (including the entrained gas) but cannot disentangle the line of sight contamination. Model MF goes one step further to predict the fraction of gas mass associated with feedback (including the entrained gas). Although we include the entrained gas in the bubble mass estimation, the result is significantly less than 10-100 \msun. Reducing the bubble mass by excluding extra gas along the line of sight brings the estimates closer in line with both empirical and theoretical models for feedback.}

Table~\ref{Physical Parameters of Taurus Bubbles} lists the physical parameters of all the Taurus bubbles. It includes the estimates from  \citet{2015ApJS..219...20L} and our models ME1 and MF. Since \citet{2015ApJS..219...20L} do not consider the correction factors for bubble mass, momentum and energy due to the limited \13co\ velocity range, to make a fair comparison, we do not apply correction factors to the predictions from models ME1 and MF.

\begin{table*}[]
\begin{center}
\caption{Physical Parameters of Taurus Bubbles\label{Physical Parameters of Taurus Bubbles}}
\begin{tabular}{ccccccccccc}
\hline
Bubble & Rank & \multicolumn{3}{c}{Li+ (2015)}& \multicolumn{3}{c}{ME1}  &  \multicolumn{3}{c}{MF}   \\
\hline
ID &  & Mass& Momentum&Energy & Mass& Momentum&Energy & Mass & Momentum & Energy \\
 & & (M$_{\odot}$) &(M$_{\odot}$ km/s) & (10$^{44}$ ergs)& (M$_{\odot}$) &(M$_{\odot}$ km/s) & (10$^{44}$ ergs) & (M$_{\odot}$) &(M$_{\odot}$ km/s) & (10$^{44}$ ergs)   \\
\hline

    TMS\_1 & B     & 59    & 65    & 7     & 15    & 39    & 11    & 1.9   & 6.2   & 2.25 \\
    TMS\_2 & B     & 31    & 34    & 4     & 43    & 32    & 4     & 9.1   & 8.2   & 1.29 \\
    TMS\_3 & B     & 129   & 207   & 33    & 120   & 169   & 35    & 10.7  & 15.1  & 3.33 \\
    TMS\_4 & B     & 56    & 62    & 7     & 42    & 65    & 14    & 3.6   & 5.4   & 1.14 \\
    TMS\_5 & B     & 32    & 52    & 8     & 56    & 45    & 5     & 3.6   & 3.5   & 0.54 \\
    TMS\_6 & B     & 33    & 37    & 4     & 10    & 9     & 1     & 2.7   & 3.2   & 0.64 \\
    TMS\_7 & C     & 91    & 191   & 40    & 95    & 73    & 9     & 10.5  & 9.7   & 1.50 \\
    TMS\_8 & B     & 46    & 97    & 20    & 33    & 48    & 12    & 2.8   & 4.5   & 1.12 \\
    TMS\_9 & C     & 22    & 17    & 1     & 39    & 34    & 4     & 5.3   & 4.3   & 0.58 \\
    TMS\_10 & C     & 217   & 282   & 37    & 228   & 272   & 57    & 24.8  & 35.8  & 8.26 \\
    TMS\_11 & A     & 78    & 149   & 28    & 94    & 70    & 9     & 7.1   & 5.9   & 0.81 \\
    TMS\_12 & A     & 45    & 108   & 26    & 57    & 43    & 5     & 3.7   & 3.0   & 0.39 \\
    TMS\_13 & B     & 84    & 176   & 37    & 112   & 88    & 11    & 8.1   & 8.6   & 1.46 \\
    TMS\_14 & C     & 18    & 33    & 6     & 13    & 44    & 17    & 2.8   & 9.0   & 3.71 \\
    TMS\_15 & A     & 98    & 107   & 12    & 109   & 90    & 12    & 6.0   & 7.6   & 1.68 \\
    TMS\_16 & A     & 57    & 92    & 15    & 75    & 60    & 8     & 1.5   & 2.4   & 0.64 \\
    TMS\_17 & C     & 10    & 8     & 1     & 15    & 19    & 3     & 1.4   & 2.4   & 0.67 \\
    TMS\_18 & C     & 54    & 103   & 19    & 72    & 63    & 9     & 2.5   & 3.0   & 0.57 \\
    TMS\_19 & A     & 41    & 78    & 15    & 66    & 77    & 13    & 3.5   & 4.0   & 0.68 \\
    TMS\_20 & B     & 12    & 13    & 1     & 21    & 28    & 5     & 1.8   & 2.3   & 0.42 \\
    TMS\_21 & B     & 11    & 24    & 5     & 3     & 15    & 9     & 1.1   & 4.6   & 2.38 \\
    TMS\_22 & B     & 119   & 131   & 14    & 134   & 168   & 33    & 12.3  & 18.4  & 3.92 \\
    TMS\_23 & C     & 10    & 11    & 1     & 24    & 21    & 3     & 0.1   & 0.1   & 0.02 \\
    TMS\_24 & C     & 20    & 22    & 2     & 29    & 52    & 10    & 1.8   & 2.9   & 0.56 \\
    TMS\_25 & C     & 9     & 26    & 8     & 7     & 11    & 2     & 0.2   & 0.6   & 0.22 \\
    TMS\_26 & B     & 11    & 23    & 5     & 8     & 29    & 12    & 3.0   & 10.6  & 4.71 \\
    TMS\_27 & C     & 46    & 111   & 26    & 24    & 82    & 33    & 9.4   & 32.3  & 13.87 \\
    TMS\_28 & C     & 205   & 656   & 209   & 293   & 365   & 68    & 14.2  & 18.4  & 3.50 \\
    TMS\_29 & A     & 91    & 119   & 15    & 120   & 145   & 27    & 7.0   & 7.3   & 1.19 \\
    TMS\_30 & C     & 420   & 672   & 107   & 576   & 636   & 106   & 23.2  & 26.3  & 4.65 \\
    TMS\_31 & C     & 62    & 81    & 10    & 43    & 89    & 26    & 6.6   & 23.9  & 10.33 \\
    TMS\_32 & C     & 12    & 22    & 4     & 7     & 29    & 14    & 2.6   & 12.6  & 6.74 \\
    TMS\_33 & B     & 92    & 74    & 6     & 143   & 153   & 25    & 5.6   & 5.9   & 0.96 \\
    TMS\_34 & C     & 67    & 182   & 49    & 36    & 39    & 7     & 4.7   & 5.3   & 1.51 \\
    TMS\_35 & B     & 30    & 39    & 5     & 22    & 27    & 4     & 4.4   & 3.8   & 0.64 \\
    TMS\_36 & A     & 29    & 47    & 7     & 16    & 25    & 6     & 3.7   & 4.2   & 1.00 \\
    TMS\_37 & B    & 25    & 53    & 11    & 19    & 40    & 12    & 7.0   & 17.6  & 6.96 \\
\hline
\end{tabular}
\end{center}
\end{table*}

\subsection{Assessing the Global Impact of Feedback: Full Taurus Map }
\label{Assessing the Global Impact of Feedback}

\subsubsection{Feedback Features Identified in the Full Map}
\label{Feedback Features Identified in the Full Map}

We apply the {\sc casi-3d} models to the complete Taurus map to predict all the emission associated with feedback. We divide the Taurus map into smaller cubes as discussed in Section~\ref{Taurus Data}. To create the full prediction map, 
we adopt the largest value from the overlapping predictions at each pixel. Note that the 5x5 pixel regions in the map corners have only one cube prediction for each pixel. 

\begin{figure*}[hbt!]
\centering
\includegraphics[width=0.99\linewidth]{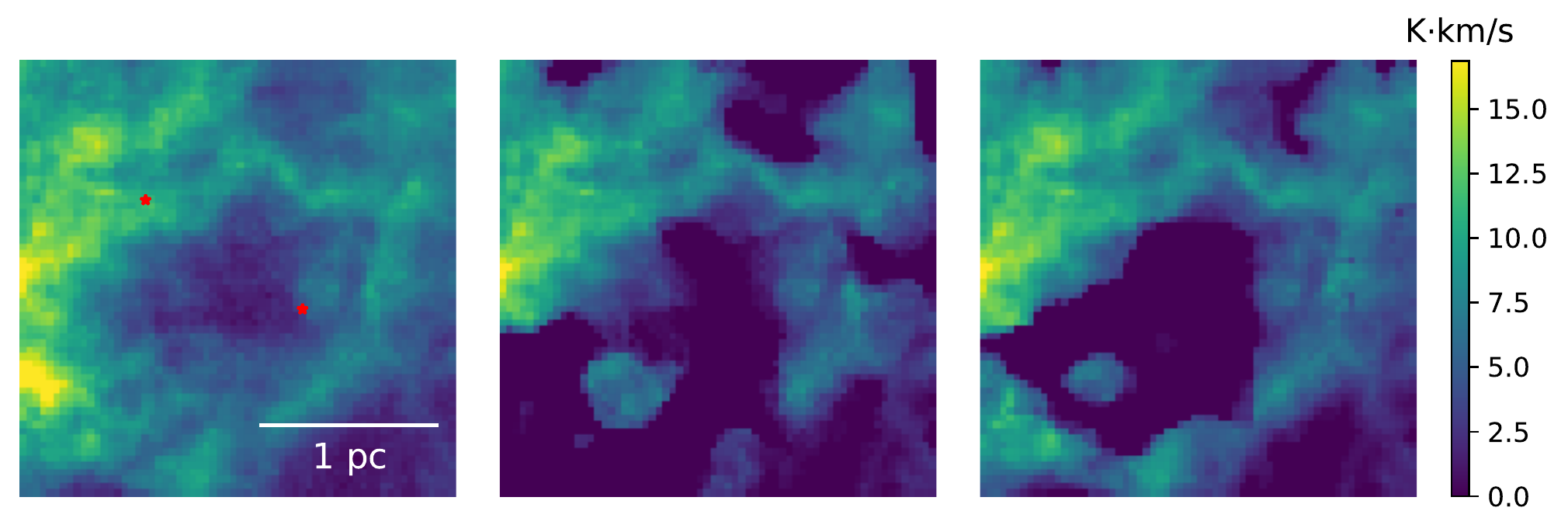}
\caption{Comparison of the prediction of a postage stamp and that from the large map on bubble TMB29. Left: integrated \13co\ intensity. The red stars indicate Class III YSOs from the \citet{2017ApJ...838..150K} catalog. Middle: integrated prediction from model ME1 run on the postage stamp shown in the left panel. Right: integrated ME1 prediction from the full map prediction, reconstructed from overlapping postage stamps.}
\label{fig.comp-ps-lm-TMB29}
\end{figure*}

To check the accuracy of this method, we compare the model predictions of the postage stamps and those from the large map. Figure~\ref{fig.comp-ps-lm-TMB29} shows that the large map prediction captures the bubble rims better than the single postage stamp predictions. 

\begin{figure*}[hbt!]
\centering
\includegraphics[width=0.95\linewidth]{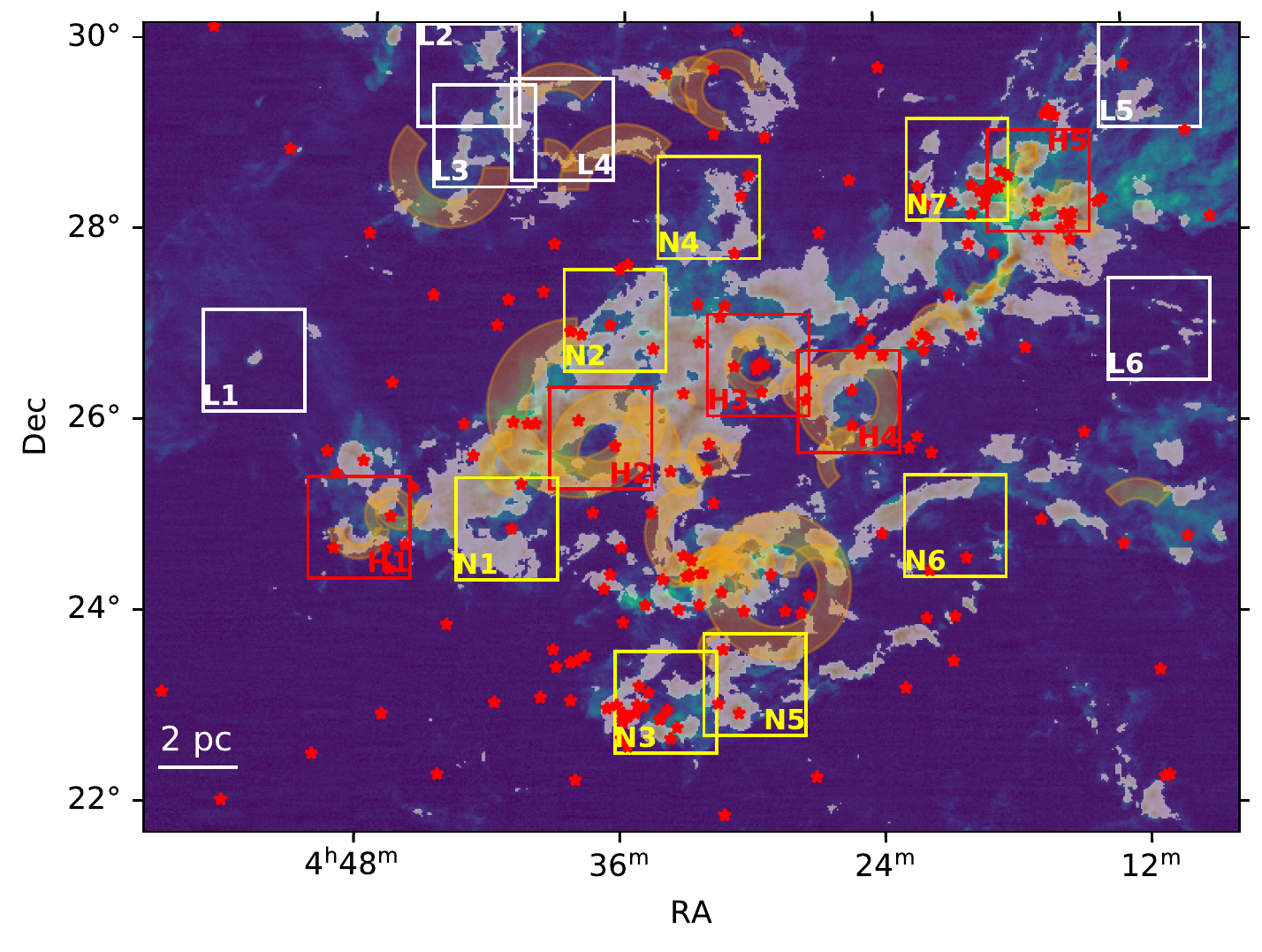}
\caption{The \13co\ integrated intensity of Taurus molecular cloud overlaid with the integrated prediction of feedback position from ME1 along velocity channels in red color. The arcs in yellow indicate the position of previously identified bubbles in \citet{2015ApJS..219...20L}. The star symbol demonstrates the location of the Class III YSOs from \citet{2017ApJ...838..150K}. }
\label{fig.large-map-ME1}
\end{figure*}

\begin{figure*}[hbt!]
\centering
\includegraphics[width=0.95\linewidth]{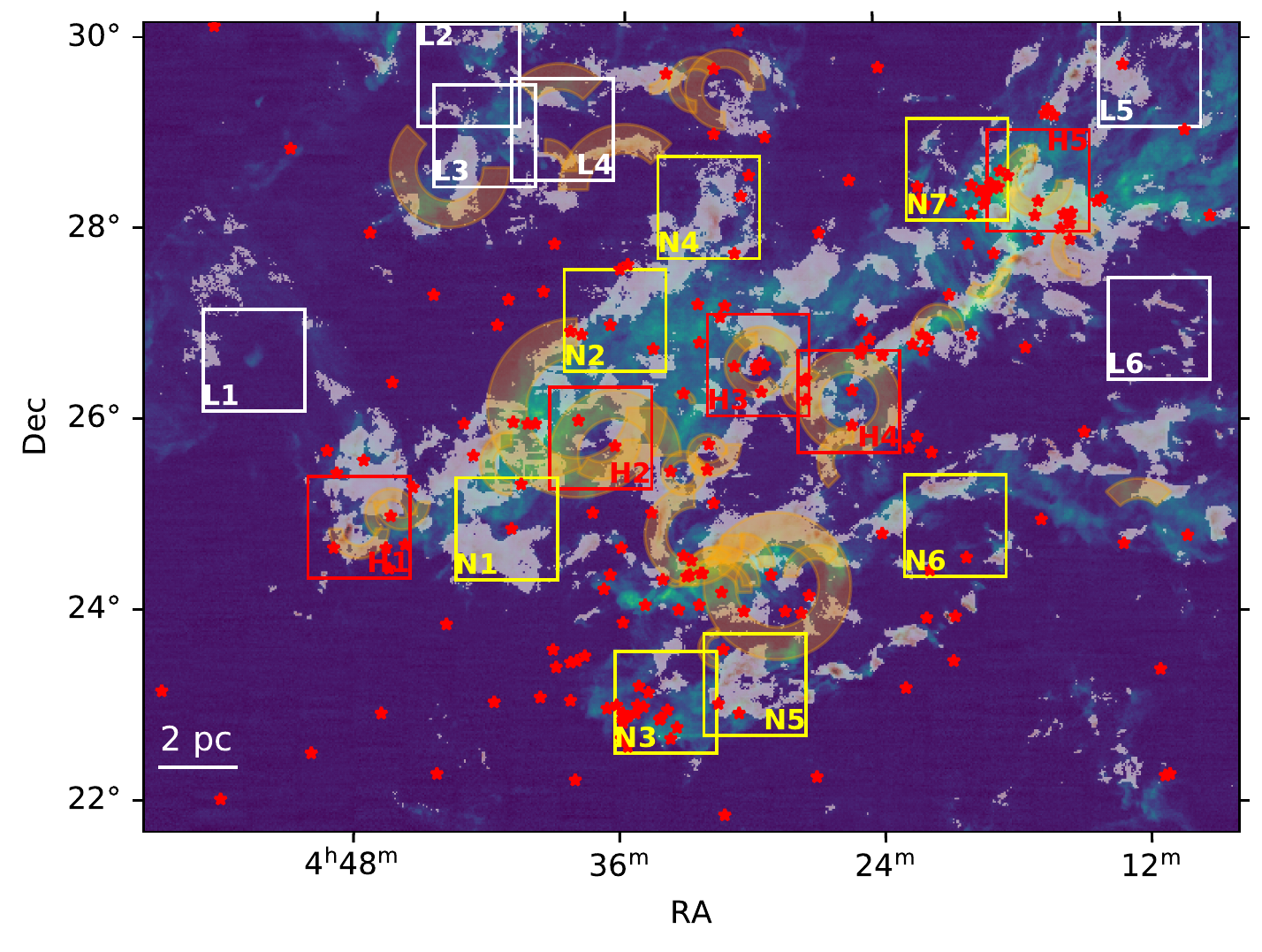}
\caption{The same as Figure~\ref{fig.large-map-ME1} but predicted by MF. }
\label{fig.large-map-MF}
\end{figure*}

Figure~\ref{fig.large-map-ME1} and \ref{fig.large-map-MF} show the predictions from models ME1 and MF for the whole Taurus map. The {\sc casi-3d} model predictions cover almost all the previously identified bubble regions and predict additional feedback regions in the Taurus map. The new predictions are correlated with the locations of Class III YSOs as shown on the map. Figure~\ref{fig.large-map-ME1} shows that most predictions are close to several groups of YSOs. For example, new bubble N3, which was not previously identified, seems to enclose a large group of YSOs. 
This suggests that the YSOs are shaping the surrounding clouds through their feedback and creating a wind signature in the \13co\ spectra.  We discuss the newly detected feedback regions further below.

We identify three types of bubbles in our model predictions: high-confidence bubbles that were identified by the previous observational survey (red boxes), high-confidence bubbles that we believe are real bubbles that were missed in the previous survey (yellow boxes), and low-confidence bubbles that are new bubbles found by our models but we believe are less certain (white boxes). The first category of  high-confidence bubbles correspond to ``true positives." The second category of missing high-confidence bubbles corresponds to  ``true negatives", and the final category of low-confidence bubbles may represent ``false positives." 

First, we discuss the high-confidence bubbles (true positives) that are consistent with the previous human identifications. These bubbles have a clear ring or arc-like structure and have at least one YSO inside. Bubbles H1, H2, H3 and H4 correspond to TMB 37, TMB 29, TMB12 and TMB7 in \citet{2015ApJS..219...20L}, respectively. These bubbles are identified by both model ME1 and MF, although the extent of the emission in model MF may be smaller if the fraction of the mass coming from feedback is predicted to be low. 

Next, we discuss the high-confidence bubbles that were not included in \citet{2015ApJS..219...20L}. These bubbles have a clear bubble rim morphology and have YSOs nearby if not directly within the bubble center. For example, in the yellow box N1 we see the bubble rim and the cavity. Moreover, one Class III YSO is centered in the cavity, which is likely to be the driving source of the bubble. The predictions from both ME1 and MF for N1 highlight the bubble rim. In another example, N3, we can easily identify the bubble rim in Figure~\ref{fig.large-map-ME1}. Supporting the bubble's existence is a group of Class III YSOs inside its rim. However, when we look at the prediction from model MF for N3 in Figure~\ref{fig.large-map-MF}, we cannot see the bubble rim prediction. This suggests that there is likely a small amount of mass coming from the feedback. 

Finally, we discuss the low-confidence bubbles. These bubbles tend not to be associated with any YSOs. In addition to the Class III YSOs identified in \citet{2017ApJ...838..150K}, we check all types of YSOs in Taurus that were identified by \citet{2010ApJS..186..259R}. These are plotted in Figures~\ref{fig.large-map-ME1-rebull} and \ref{fig.large-map-MF-rebull} in Appendix~\ref{YSOs in the Taurus molecular cloud}. We highlight four such bubbles in white boxes in Figure~\ref{fig.large-map-ME1}.
We note a number of the bubbles identified by \citet{2015ApJS..219...20L} do not contain any Class III YSOs, such as L4. In these cases, the driving source may have moved out of the bubble, or the YSO census may be incomplete. Another possible explanation is that although the morphology is circular or arc-like, they are caused by cloud turbulence, which causes coherent motion across several velocity channels. It is difficult to distinguish the bubble structure from turbulent patterns when a circular or arc-like pattern shows across multiple channels. However, we believe this last explanation is unlikely, since we include pure turbulence snapshots in the training set as negative training images. Thus, {\sc casi-3d} should not be prone to misidentify turbulent patterns as bubbles. 

Overall, we conclude that the two CNN models perform as well or better than ``by-eye” visual identifications of bubbles. They appear to reasonably predict both the bubble position and the fraction of mass coming from feedback.

\subsubsection{Mass, Momentum and Energy of the Feedback Identified in the Full Taurus Cloud }
\label{Mass Momentum and Energy analysis in the Full Taurus Map}

We now calculate the feedback mass, momentum and energy in Taurus based on the predictions from models ME1 and MF. Table~\ref{Feedback in Taurus Molecular Cloud} lists the feedback properties calculated in this work and those calculated in \citet{2015ApJS..219...20L}. 

Model ME1 predicts 2630 \msun\ of gas associated with feedback, which is consistent within a factor of two with the feedback mass calculated in \citep{2015ApJS..219...20L}. However, model MF predicts that only 275 \msun\ of gas is associated with feedback, which is an order of magnitude smaller than the previous calculations. The smaller amount of feedback mass is also consistent with the total stellar mass in Taurus, which is estimated to be on the order of 200 \msun\ in \citet{2017ApJ...838..150K}. The feedback mass predicted by model ME1 and that calculated from \citet{2015ApJS..219...20L} are 10 times the stellar mass, which is inconsistent with the expect amount of gas entrained by feedback \citep{2017ApJ...847..104O}.

Despite the detailed machine learning identification, we must still confront the challenge of how to disentangle feedback from the bulk cloud motion. For example, Taurus has a velocity gradient that stretches from the south-east to north-west. Not accounting for this gradient may artificially enhance the feedback total. The most accurate way to account for the bulk motion is not clear; thus,
we present two approaches to calculate the feedback momentum and energy. The first way treats the molecular cloud as a whole, with the same fixed central velocity. We shift the central velocity to zero and calculate the 1D momentum and 1D energy channel by channel as described in Section~\ref{Assessing Model Accuracy Using Synthetic Observations}.  The second approach is similar but treats the molecular cloud locally, which means the cloud does not have a fixed central velocity but has a central velocity gradient across the entire cloud. We subtract the central velocity pixel by pixel and then calculate the momentum and energy channel by channel.
To convert the 1D line of sight estimates to 3D, we make the assumption of isotropic expansion to calculate the 3D momentum and 3D energy.
Finally, in Section~\ref{Assessing Model Accuracy Using Synthetic Observations}, we assessed the model accuracy using synthetic observations and found that 20\% of the momentum and 44\% of the energy are missed due to the limited CO velocity coverage. Considering the limited velocity range of the \13co\ data cube, we apply these correction factors for the missing momentum and energy here. 

The momentum estimate without the velocity gradient treatment from model ME1 is close to that from \citet{2015ApJS..219...20L}. The momentum estimate with the velocity gradient treatment from model ME1 is 38\% smaller than the calculation in \citet{2015ApJS..219...20L}. Once corrected for the extra \13co\ emission from the foreground or background, the momentum (with and without the velocity gradient treatment) predicted by model MF is an order of magnitude smaller than the calculation in \citet{2015ApJS..219...20L}.

Both energy estimates (with and without the velocity gradient treatment) from model ME1 are within a factor of two compared to the energy calculated in \citet{2015ApJS..219...20L}. In contrast,  model MF implicitly corrects for the extra \13co\ emission  coming from the foreground or background, such that the predicted energy is an order of magnitude smaller than that calculated in 
\citet{2015ApJS..219...20L}. We discuss the implications in the following section.


\begin{sidewaystable*}[]

\begin{center}
\caption{Properties of Feedback in the Taurus Molecular Cloud$^{*}$\label{Feedback in Taurus Molecular Cloud}}
\begin{tabular}{ccccccccc}
\hline
Model & \multicolumn{4}{c}{Without subtracting the velocity gradient} & \multicolumn{4}{c}{Subtracting the velocity gradient}\\
  & $M$ $^{a}$ & $P_{\rm 3D}$ & $E_{3D}$ $^{b}$  &$\dot{E}$ $^{c}$ & $M$ $^{a}$ & $P_{\rm 3D}$ & $E_{3D}$ $^{b}$  &$\dot{E}$ $^{c}$ \\
 &(\msun) &  (\msun\, km/s)& ($\times 10 ^{46}$ ergs) &($\times 10 ^{33}$ ergs/s) &(\msun) & (\msun\, km/s)& ($\times 10 ^{46}$ ergs) &($\times 10 ^{33}$ ergs/s)  \\
 \hline   
   Li+ (2015) & 1707 (11.4\%) & 3780 & 9.2 (28.8\%)&6.4 (94.1\%) & - & -&- &-\\
   ME1 & 2894 (19.3\%) & 4366  & 15 (46.2\%) &10 (153\%)& 2894 (19.3\%) & 2339  & 4.0 (12.6\%) &2.8 (40.9\%)\\
   MF & 302 (2.0\%) & 609 & 2.8 (8.6\%) & 2.0 (28.6\%)& 302 (2.0\%) & 366 & 0.96 (3.0\%) & 0.67 (9.8\%) \\   
\hline
\multicolumn{9}{p{0.99\linewidth}}{$^{*}$: Model name, feedback bubble mass, 3D feedback momentum, 3D feedback energy, energy injection rate from feedback bubbles. The numbers in the table consider the correction factors due to the limited velocity range of the \13co\ data cube.} \\
\multicolumn{9}{p{0.99\linewidth}}{$^{a}$: The number in the parentheses indicates the percentage of feedback mass compared to the whole molecular cloud mass \citep{2010ApJ...721..686P}.} \\
\multicolumn{9}{p{0.99\linewidth}}{$^{b}$:  The number in the parentheses indicates the percentage of feedback energy compared to the whole molecular cloud turbulent energy \citep{2015ApJS..219...20L}.}\\
\multicolumn{9}{p{0.99\linewidth}}{$^{c}$:  The number in the parentheses indicates the percentage of energy injection rate from feedback bubbles compared to the turbulent dissipation rate of the cloud. The turbulent dissipation rate adopted here is $L_{\rm turb}=6.8\times 10^{33}$ erg s$^{-1}$, which assumes a mean cloud density of $n=100$ \cmc. This turbulent dissipation rate is about two times higher than that from \citet{2015ApJS..219...20L}, which assumes a lower mean cloud density of $n=20$ \cmc.}\\
\end{tabular}
\end{center}
\end{sidewaystable*}

\subsubsection{Assessing the Relative Energies of Turbulence and Feedback }
\label{Energy Budget Analysis in the Full Taurus Map}


In this section we compare the total energy associated with feedback and the total cloud turbulent energy. The relative magnitude of these energies impacts the cloud lifetime and whether turbulence can slow collapse by providing pressure support against self-gravity. Often, the impact of feedback is weighed against the rate of turbulence dissipation. We follow \citet{2015ApJS..219...20L} and define the turbulent dissipation rate as:
\begin{equation}
\label{eq-turb-diss}
L_{\rm turb}=\frac{E_{\rm turb}}{t_{\rm diss}},
\end{equation}
where $t_{diss}$ is the turbulent dissipation time. 
The method to estimate the turbulent dissipation time in \citet{2015ApJS..219...20L} is from \citet{1999ApJ...524..169M},
\begin{equation}
\label{eq-turb-diss-time-2}
t_{\rm diss}\sim (\frac{0.39\kappa}{\mathcal{M}_{\rm rms}})t_{\rm ff},
\end{equation}
where $t_{\rm ff}$ is the free-fall timescale, $\mathcal{M}_{rms}$ is the Mach number of the turbulence, and $\kappa$ is the ratio of the driving length to the Jean's length of the cloud. For $\mathcal{M}_{rms}=5$ and a free-fall time  $t_{\rm ff}=7\times 10^{6}$~yr, which assumes a mean cloud number density of $n=20$ \cmc, the turbulent dissipation rate is $3.1\times 10^{33}$ erg s$^{-1}$. However, Taurus is not a uniform sphere, the mean number density of $n=20$ \cmc\ adopted by \citet{2015ApJS..219...20L} is too low. The typical mean number density of a molecular cloud is around $n=100$ \cmc, which gives $t_{\rm ff}=3.3\times10^{6}$~yr and $L_{\rm turb}=6.8\times 10^{33}$ erg s$^{-1}$. \citet{2010ApJ...715.1170A} and \citet{2012MNRAS.425.2641N} also adopt this method to calculate the turbulent dissipation rates in Perseus and Taurus, respectively. 

One caveat here is that the equation to calculate the turbulent dissipation rate is obtained from simulations, which depend on the initial conditions and the way turbulence is driven.


The energy injection rate is defined as $L_{\rm bubble}=E_{\rm bubble}/t_{\rm kinetic}$, where $E_{\rm bubble}$ is the kinetic energy of the bubble and $t_{\rm kinetic}$ is the kinetic timescale of the bubble. The kinetic timescale of the bubble can be calculated as $t_{\rm kinetic}=R/V_{\rm exp}$, where R is the radius of the bubble and $V_{\rm exp}$ is the expansion velocity of the bubble. We find the energy injection rate from bubbles in ME1 is $L_{\rm turb,ME1}=1.0\times 10^{34}$~erg~s$^{-1}$, which is slightly larger that the turbulent dissipation rate of the cloud. If we subtract the velocity gradient, the energy injection rate from bubbles is $L_{\rm turb,ME1,G}=2.8\times 10^{33}$~erg~s$^{-1}$, which is about half of the turbulent dissipation rate of the cloud. In summary, like \citet{2015ApJS..219...20L}, we conclude that feedback is sufficient to maintain the current level of cloud turbulence. 

However, we have shown that model ME1 overestimates the energy because excess foreground and background material is included in the calculation.
Consequently, we find that after recalculating the feedback energy using the more accurate model MF prediction, the kinetic energy from the feedback decreases by an order of magnitude, which means the energy injection rate from stars is smaller by an order of magnitude: $L_{\rm turb,MF}=2.0\times 10^{33}$ erg s$^{-1}$. Under this circumstance, the energy injection rate from feedback is 29\% of the turbulent dissipation rate of the cloud. If we subtract the velocity gradient, the energy injection rate from feedback is $L_{\rm turb,MF,G}=6.7\times 10^{32}$ erg s$^{-1}$, which is an order of magnitude smaller than the turbulent dissipation rate. This indicates that 
some additional energy is needed to drive turbulence in the Taurus molecular cloud, which could be provided by outflows for example. Feedback from bubbles may not be sufficient to maintain the cloud turbulence over long timescales. 



The Taurus molecular cloud is host to an older population of stars ($\tau \sim 10-20$ Myr), which indicates the lifetime of the cloud is at least 10-20 million years \citep{2017ApJ...838..150K}. However, this life time is much longer than the gravitational collapse free-fall time of the Taurus molecular cloud estimated from \co, which is 3.3 million years. This suggests that there must be energy injected to support the cloud against gravitational collapse, which suggest feedback is playing some role in driving turbulence but is not dominant.




\subsubsection{Quantifying the Impact of Feedback with Turbulent Statistics}
\label{Quantifying the Impact of Feedback with Turbulent Statistics}

With an accurate prediction of the position of feedback in hand, we compute multiple astrostatistics to study the different properties between regions with and without feedback in Taurus. 
We adopt the statistical analysis package, \turbustat, to conduct the statistical analysis \citep{2017MNRAS.471.1506K,2019AJ....158....1K}. \turbustat\ contains 15 different statistics, but here we consider only the spatial power spectrum (SPS) and the covariance matrix used to compute principle component analysis (PCA). We adopt these statistics since they have previously been shown to be sensitive to feedback as discussed in the introduction. 
The SPS is defined as the square of the 2D Fourier transform of an image:
\begin{equation}
\label{sps-eq1}
\begin{split}
\mathcal{P}(k) & =\sum_{|\vec{k}|=k}|\mathcal{M}_{0}({ \vec{k}})|^{2} \\
 & =|\int_{-\infty}^{\infty}\int_{-\infty}^{\infty}M_{0}({\vec{x}})e^{-2\pi j{\vec{k}\vec{x}}}d{\vec{x}}|^2.
\end{split}
\end{equation}
It is applied to the integrated intensity map.
The covariance matrix is defined as:
\begin{equation}
\label{cov-eq1}
C_{jk}=\frac{1}{n}\sum^{n}_{i=1}X_{ij}X_{ik},
\end{equation}
where 
\begin{equation}
\label{cov-eq2}
X_{ij}=T(r_{i},v_{j})-[\sum^{n}_{k=1}T(r_{k},v_{j})]/n,
\end{equation}
in which $T(r_{i},v_{j})$ is the spectral cube, where $r_{i}=(x_{i},y_{i})$ is the position on the sky and $v_{j}$ indicates the spectral velocity channel. It provides information about velocity correlations. In addition to these two statistics, we also consider the distribution of linewidths of the feedback and non-feedback gas as well as the distance between YSOs and pixels associated with feedback.

\begin{figure*}[hbt!]
\centering
\includegraphics[width=0.47\linewidth]{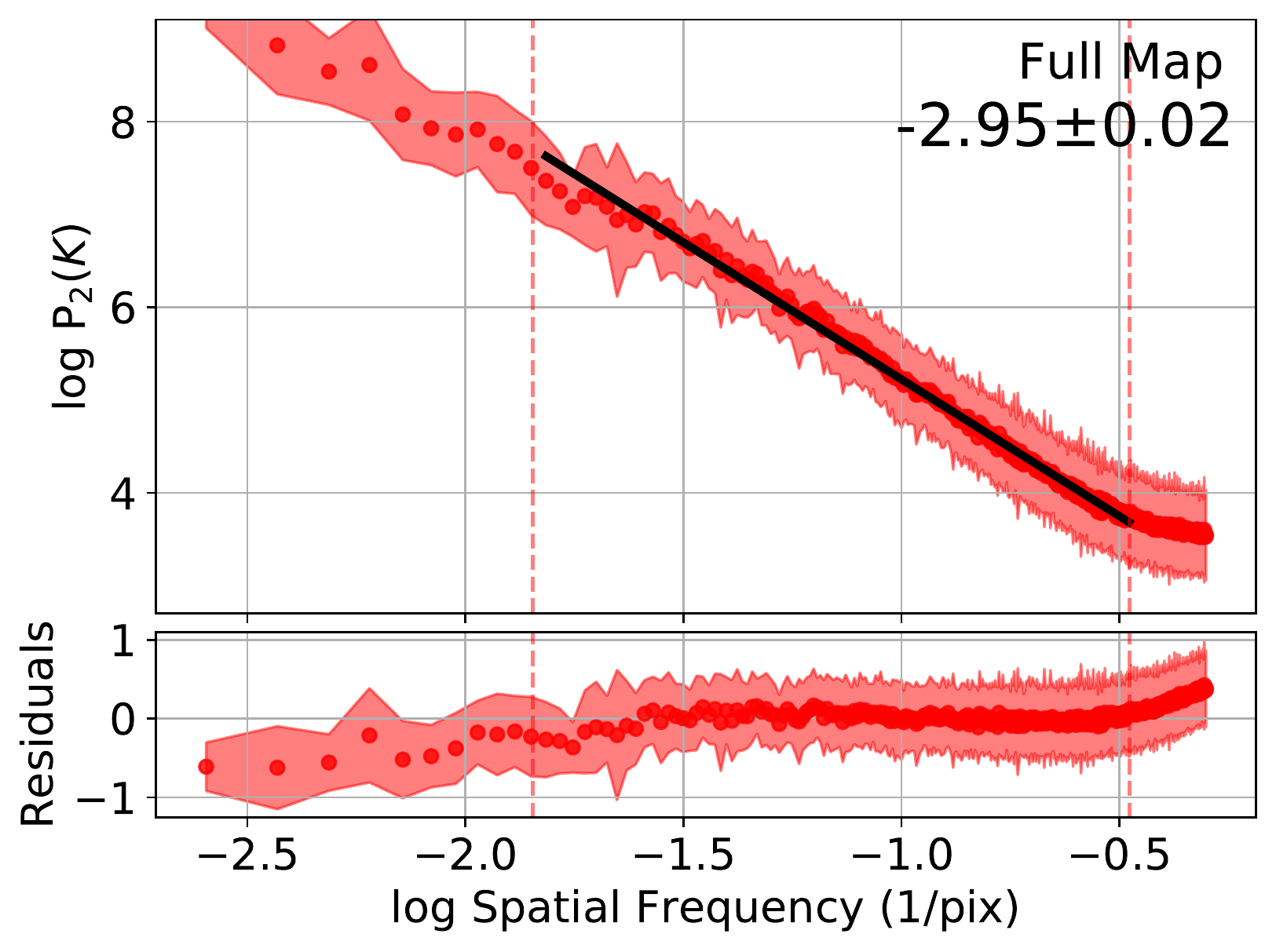}
\includegraphics[width=0.47\linewidth]{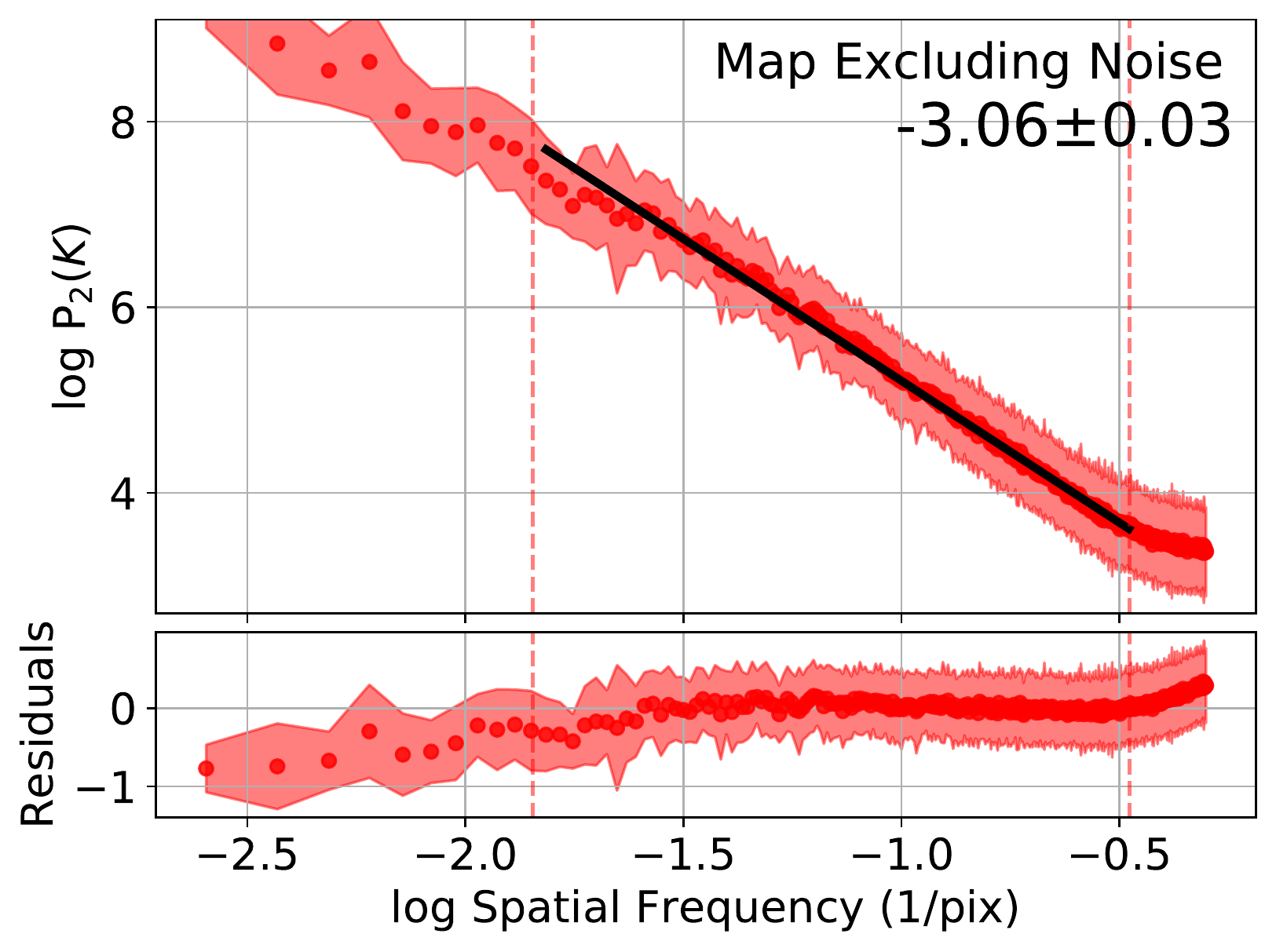}
\includegraphics[width=0.47\linewidth]{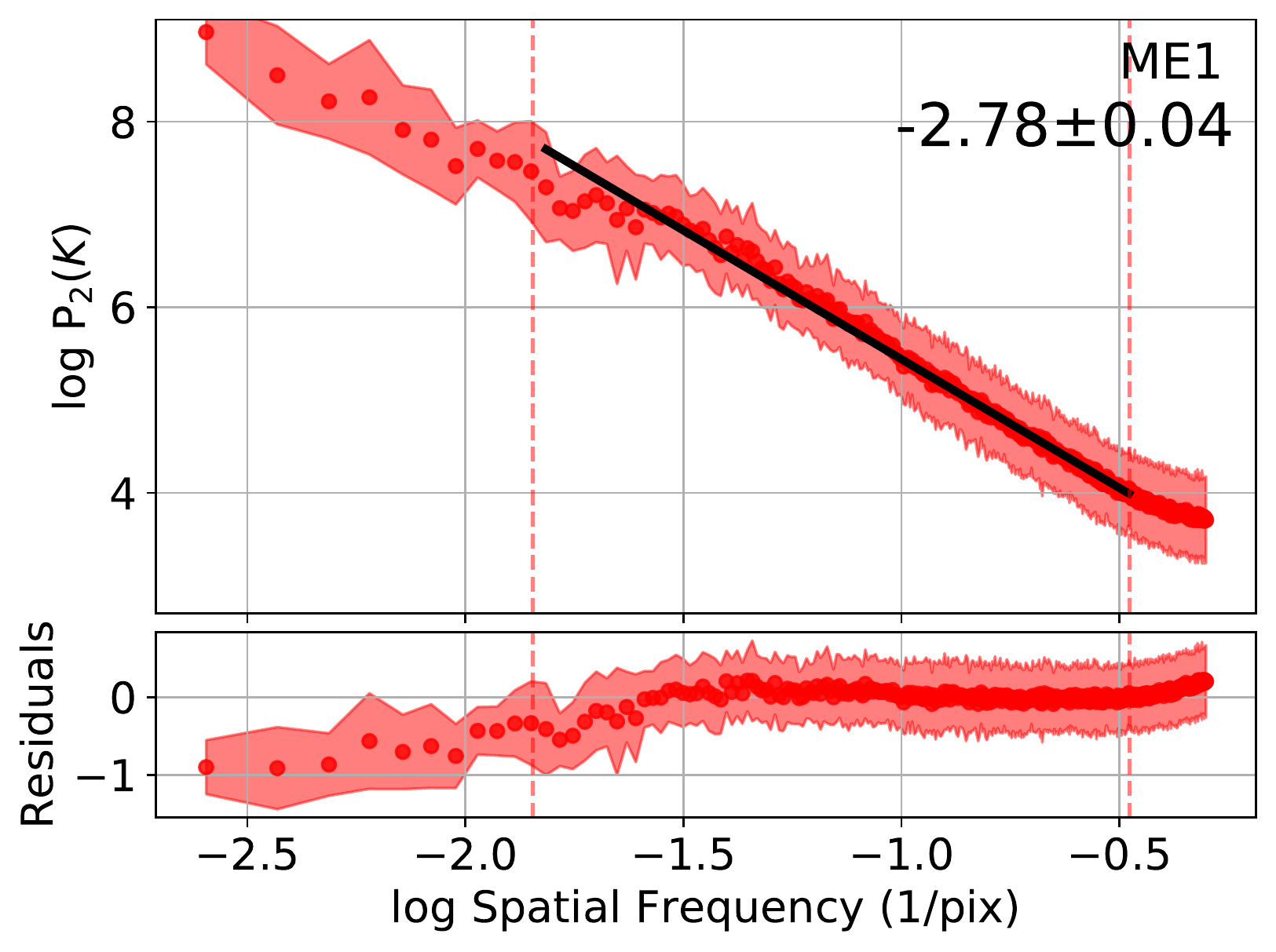}
\includegraphics[width=0.47\linewidth]{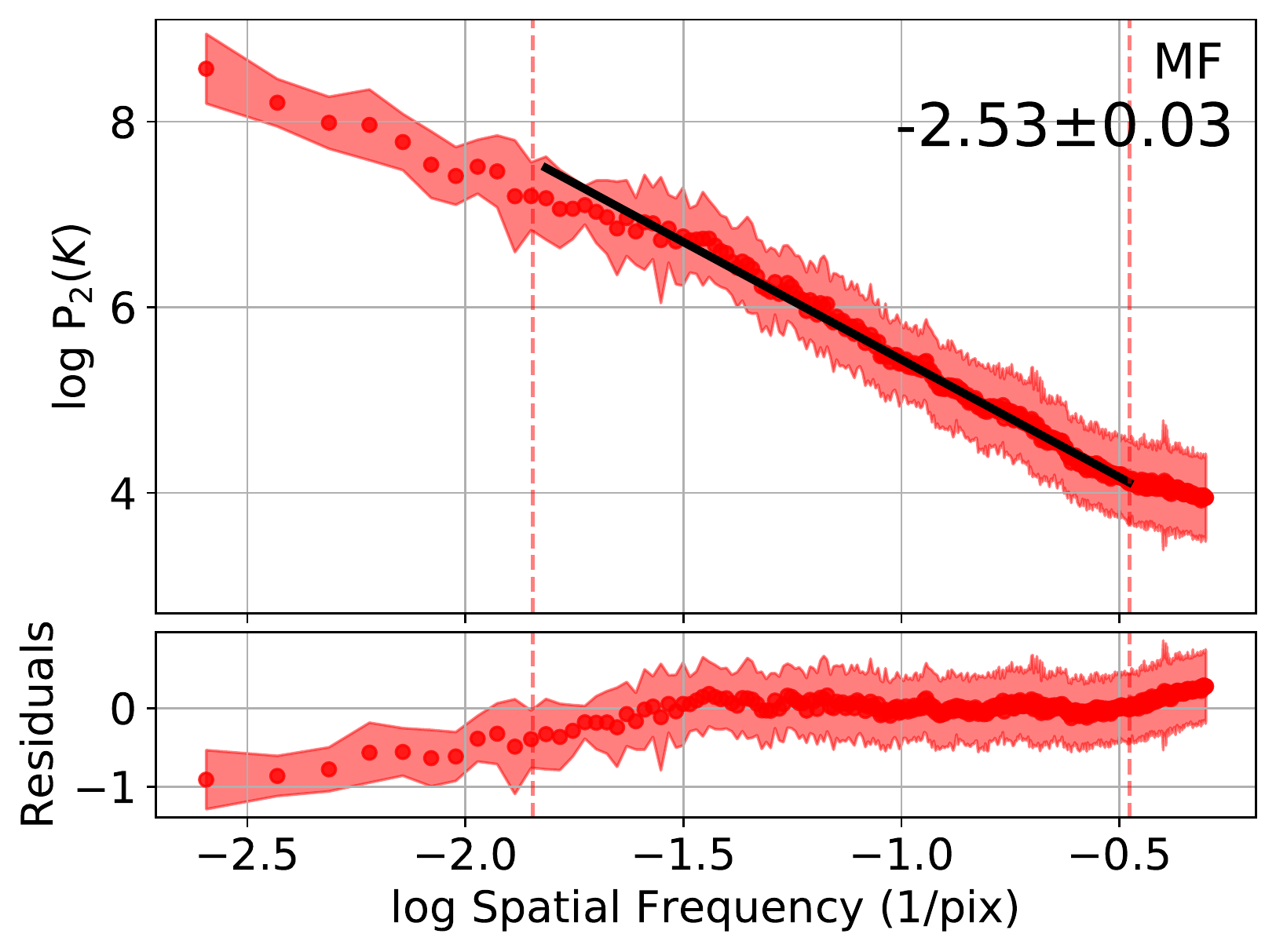}
\caption{The spatial power spectrum (SPS) of the full \13co\, integrated intensity map of Taurus and the SPS of the emission regions (excluding noise regions) where the emission is above 0.2 K and the SPS of the ME1 and MF predicted feedback regions }
\label{fig.ps-moment-0}
\end{figure*}

Figure~\ref{fig.ps-moment-0} shows the SPS of the full \13co\, integrated intensity map of Taurus, the SPS of the region where the emission is above 0.2 K (i.e., excluding noise) and the SPS of the model ME1 and MF predicted feedback regions. Figure~\ref{fig.ps-moment-0} shows that the slope of the SPS is flattened over the feedback injection region. If the emission is optically thin and the temperature is roughly constant, this indicates mass or energy has been injected into smaller scales by the feedback. Here, the \13co\ is mostly optically thin with the exception of dense cores.

\begin{figure*}[hbt!]
\centering
\includegraphics[width=0.46\linewidth]{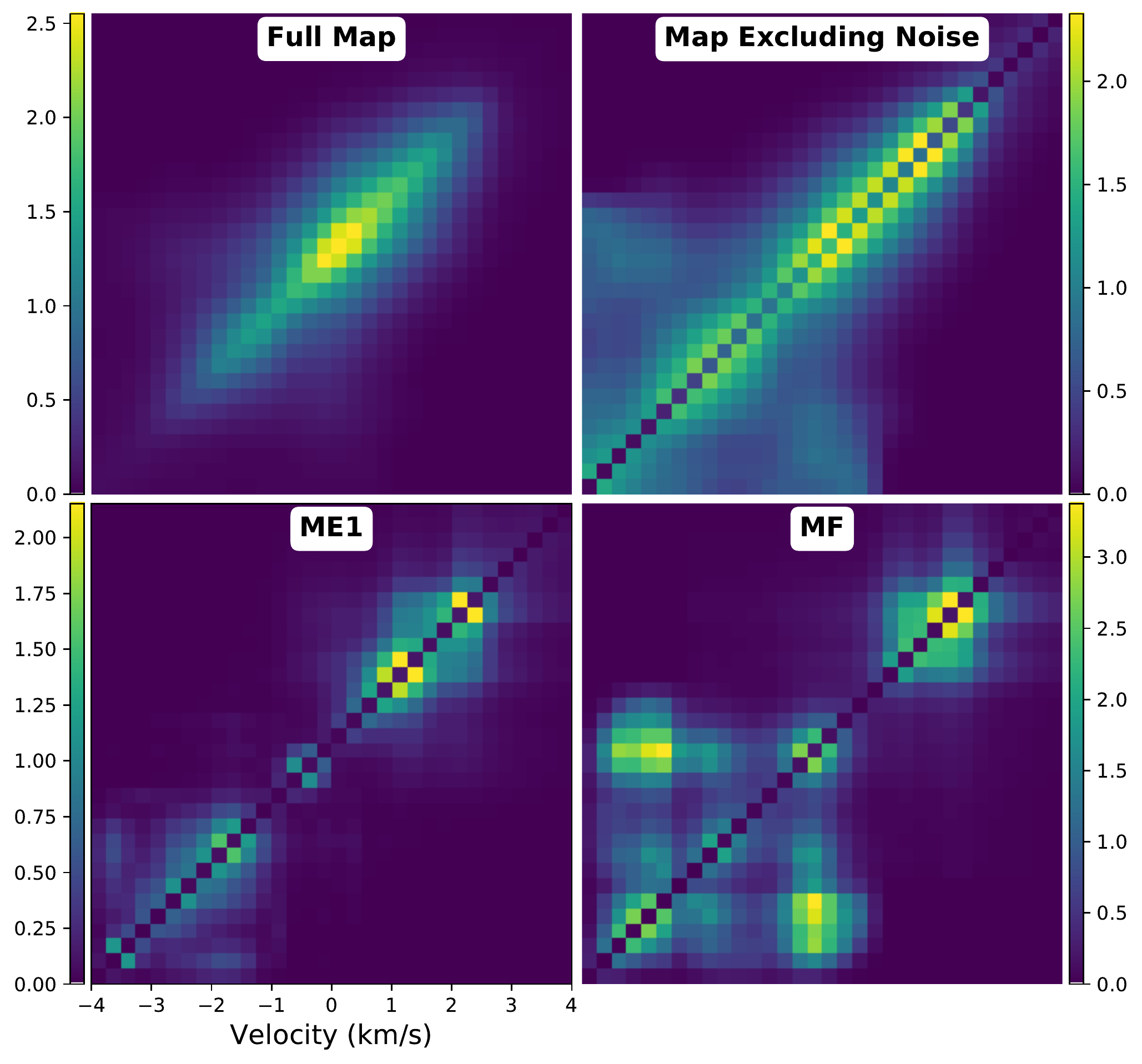}
\includegraphics[width=0.46\linewidth]{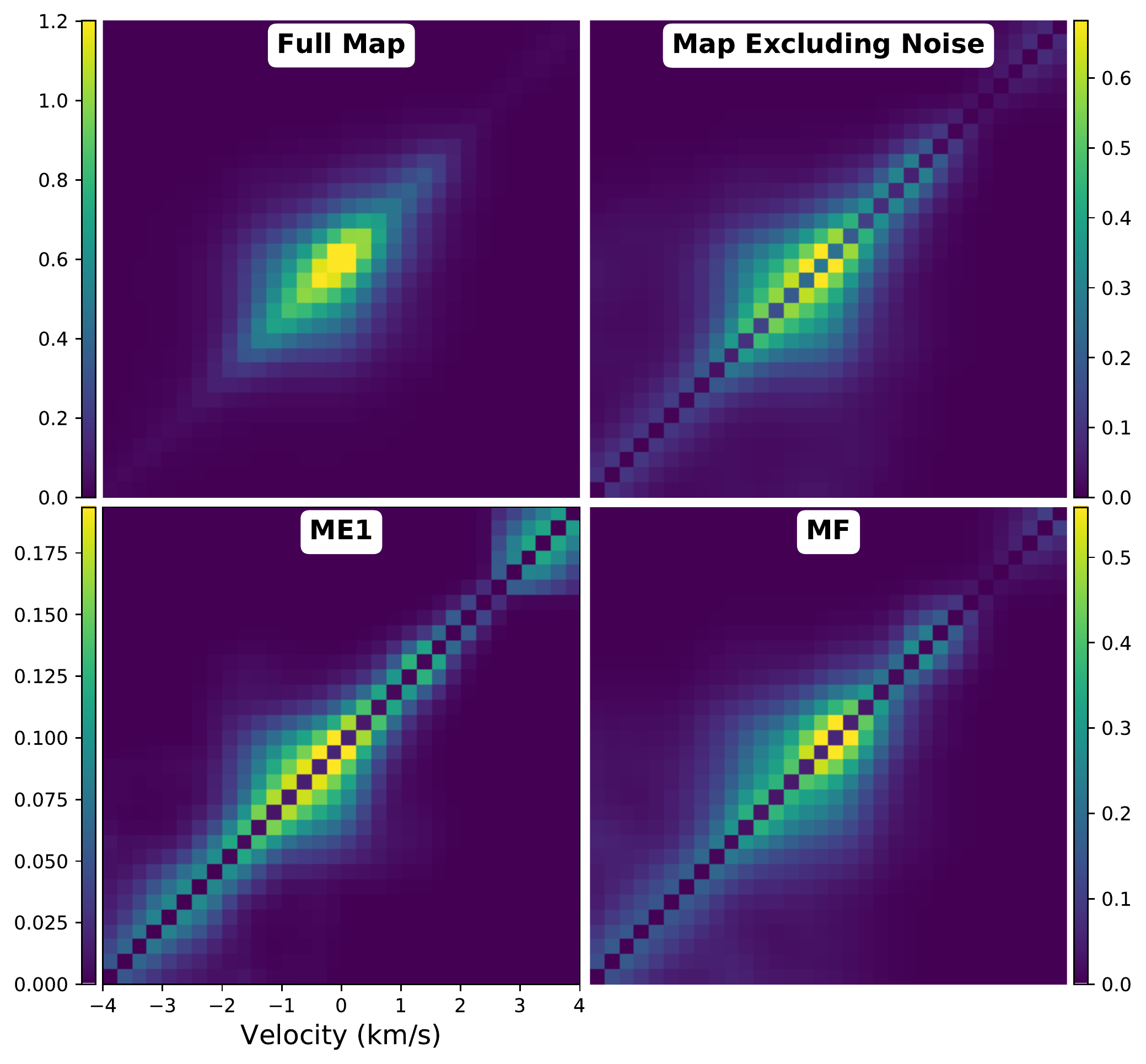}
\caption{The covariance matrices of the velocity channels on the full \13co\, integrated intensity map of Taurus and the covariance matrices of the emission regions and the covariance matrices of the ME1 and MF predicted feedback regions. Left panel: the covariance matrices of the velocity channels on synthetic data. Right panel: the covariance matrices of the velocity channels on Taurus \13co\ data.}
\label{fig.covariance-matrices}
\end{figure*}

Next, in Figure~\ref{fig.covariance-matrices} we present the covariance matrices of the velocity channels for the full \13co\, integrated intensity map of Taurus, the high signal-to-noise region and the prediction of the models ME1 and MF. 
For comparison, Figure~\ref{fig.covariance-matrices} also shows the covariance matrices calculated using the synthetic data. The covariance matrices of the predicted feedback regions clearly show off-diagonal velocity features, which indicate coherent motions at these velocities. These features can be characteristic of the expansion of bubbles or high-velocity gas \citep{2016ApJ...833..233B}, but it may also represent coherent cloud motions \citep[e.g.,][]{2019ApJ...875..162F}. In either case, 
the clear differences between the identified feedback and non-feedback gas underscore that {\sc casi-3d} is indeed identifying statistically distinct regions. 

\begin{figure}[hbt!]
\centering
\includegraphics[width=0.99\linewidth]{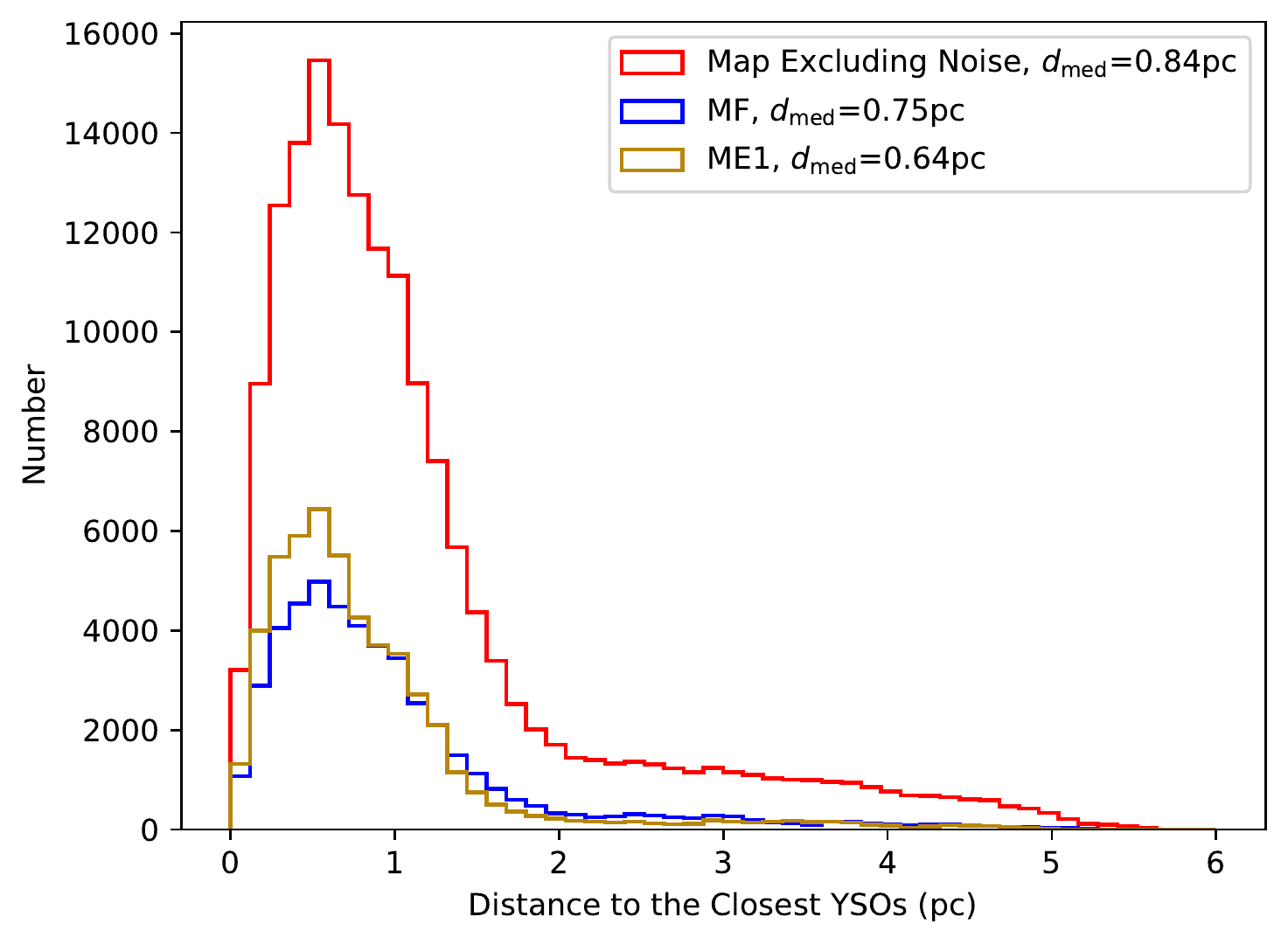}
\caption{The distribution of the projected distance between YSOs and the emitting gas, and the distribution of the projected distance between YSOs and the feedback gas predicted by ME1 and MF. }
\label{fig.YSO-distance-hist}
\end{figure}

Next, we assess the relative distance to the YSO locations, which provide additional evidence that our regions are associated with feedback.
Figure~\ref{fig.YSO-distance-hist} shows the distribution of the projected distances between the YSOs and the emitting gas, and the distribution of the projected distances between YSOs and the feedback gas predicted by ME1 and MF. The median value of the projected distance between the YSOs and the feedback gas is closer than that between the YSOs and all the emitting gas. The typical distance between the YSOs and the feedback gas is 0.7\,pc, which is also the typical size of the bubbles.

\begin{figure}[hbt!]
\centering
\includegraphics[width=0.99\linewidth]{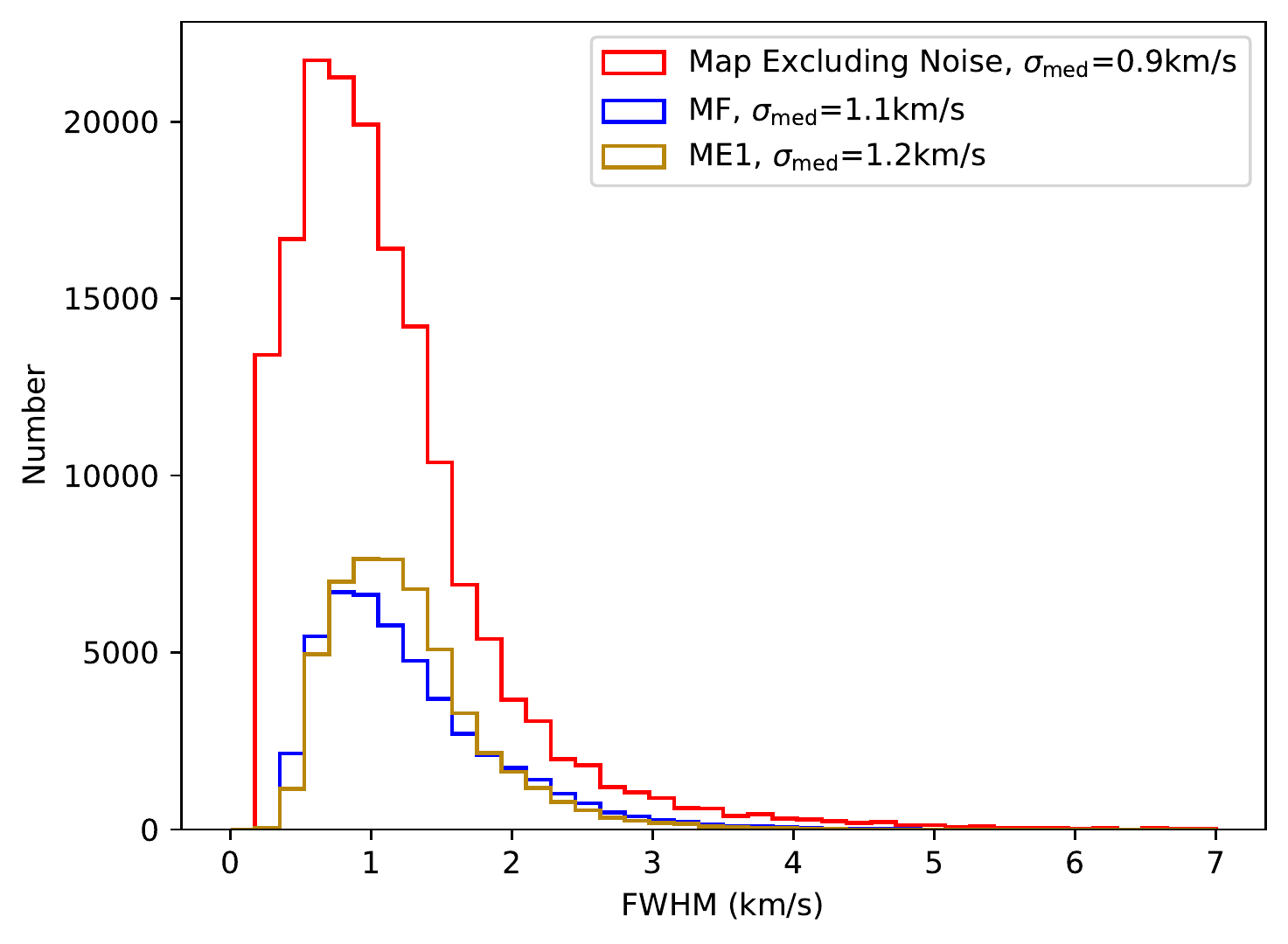}
\caption{The distribution of the FWHM of the emission regions and the FWHM of the ME1 and MF predicted feedback regions.}
\label{fig.fwhm-hist}
\end{figure}

Finally, we expect feedback regions to have larger velocity dispersions.
Figure~\ref{fig.fwhm-hist} shows the distribution of the full width at half maximum (FWHM) of the high signal-to-noise emission region and the FWHM of the ME1 and MF predicted feedback regions. The median values of the FWHM of the feedback regions are indeed larger than that of the FWHM of the full map. The higher FWHM indicates larger velocities in the spectrum associated with feedback.

\section{Conclusions}
\label{Conclusion}

We adopt a deep learning method, \CASI, and extend it to 3D (\CASItD) to identify stellar feedback features in 3D CO spectral cubes. By creating different training sets, we develop two deep machine learning tasks. Task I predicts the position of feedback. Task II predicts the fraction of the mass coming from feedback.  Our main findings are the
following:

\begin{enumerate}

\item \CASItD\ is a powerful method to identify bubbles. \CASItD\ performs well on synthetic test data and recovers feedback with an accuracy of 4\% on a pixel level.

\item  \CASItD\ successfully infers/predicts hidden information, e.g., the fraction of mass coming from feedback.

\item  We apply \CASItD\  to the \13co\ observations of the Taurus molecular cloud and show that \CASItD\ successfully identifies previously known, visually identified bubbles.

\item We find that training Task I reproduces the mass, momentum, and energy of individual bubbles inferred by human visual identifications. In contrast, Task II, which is trained on the feedback mass fraction, indicates that the true mass, momentum and energy are an order of magnitude lower.

\item \CASItD\ suggests previous studies overestimate feedback mass and energy in the Taurus molecular cloud. The feedback mass is overestimated by a factor of five. The feedback energy is overestimated by a factor of five compared to that calculated without subtracting the velocity gradient over the full map, and it is overestimated by a factor of ten compared to that calculated with subtracting the velocity gradient over the full map.

\item We carry out an analysis of the spatial power spectrum to quantify the turbulence properties in the feedback and non-feedback regions. We show that feedback flattens the slope of the spatial power spectrum of the full \13co\, integrated intensity map of Taurus, indicating that mass and/or energy has been injected at smaller scales by feedback. 

\item We calculate the covariance matrix and show that the presence of feedback appears as off-diagonal peaks in the covariance matrices.

\item The median value of the projected distance between YSOs and the feedback gas (0.64 pc predicted by model ME1 and 0.75 pc predicted by model MF) is closer than that between YSOs and all the emitting gas (0.84pc). The median value of the full-width at half maximum (FWHM) of the feedback regions (1.2 \kms\ predicted by model ME1 and 1.1 \kms\ predicted by model MF) is larger than that of the FWHM of the full emitting regions (0.9 \kms).

\end{enumerate}

In future work, we plan to apply \CASItD\ to other star-forming regions and other types of feedback, such as protostellar outflows \citep{2010ApJ...715.1170A}.

D.X., S.S.R.O., R.A.G. and C.V.O. were supported by NSF grant AST-1812747. S.S.R.O. also acknowledges support from NSF Career grant AST-1650486. The authors acknowledge the Texas Advanced Computing Center (TACC) at The University of Texas at Austin for providing HPC resources that have contributed to the research results reported within this paper.

\appendix
\section{CASI-3D Parameters}
\label{CASI-3D Parameters}

\subsection{Down-sampling Methods}
\label{Down-sampling Methods}

\begin{figure*}[hbt!]
\centering
\includegraphics[width=0.99\linewidth]{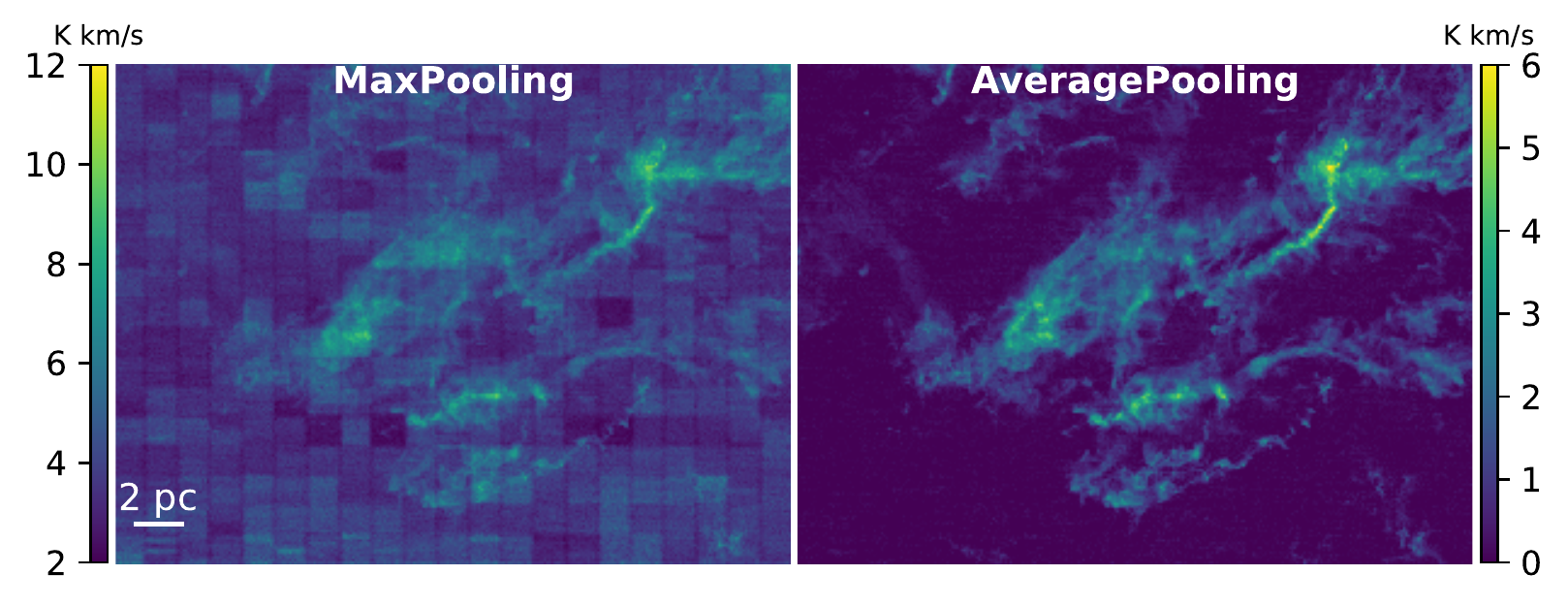}
\caption{Comparison of two different down-sampling methods tested on \13co\ Taurus molecular cloud data.}
\label{fig.pooling}
\end{figure*}

We test two widely used down-sampling methods to reduce the size of the data: max pooling and average pooling. Max pooling 
picks out the largest value to replace its adjacent pixels. Max pooling can extract the most important features, but it is not proficient in dealing with different noise backgrounds. Since all large-map sky surveys are conducted through substantial 
observing periods with different weather conditions and with different baselines, the noise level is different in different patches of the large map. When applying max pooling to down sample the data, the boundary between patches distinctly appears, which makes the data inconsistent 
across the map. On the other hand, average pooling extracts features smoothly and it preserves the overall value during down sampling., 
Figure~\ref{fig.pooling} shows an example of the two different down-sampling methods tested on \13co\ Taurus molecular cloud data. 

\subsection{Loss Function}
\label{Loss Function}

\begin{figure*}[hbt!]
\centering
\includegraphics[width=0.95\linewidth]{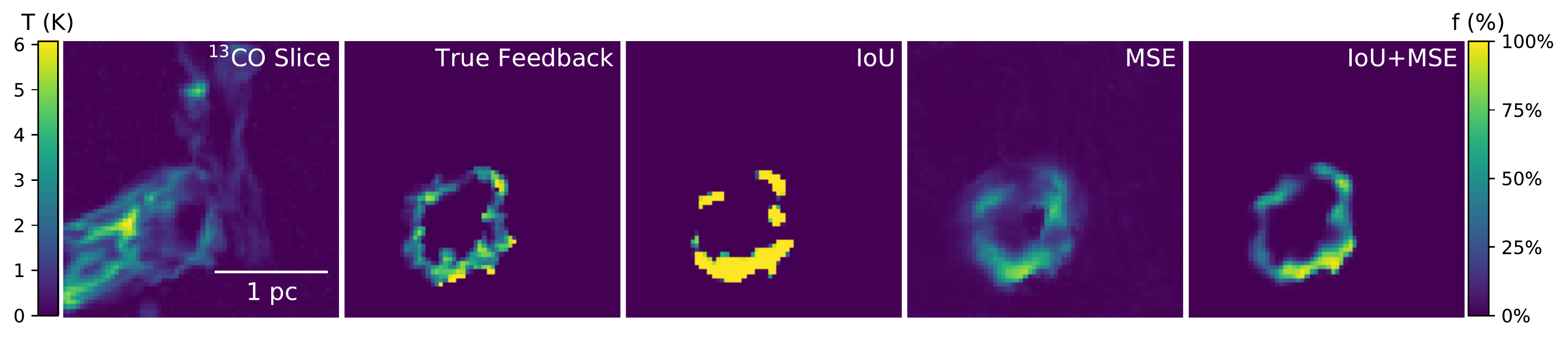}
\caption{The performance of model MF adopting different loss functions to predict the fraction of the mass that comes from stellar feedback on a test bubble. Left: integrated \13co\ intensity. Second from left: integrated true feedback fraction. Right panels: models using the IoU, MSE, and IoU+MSE loss functions, respectively. }
\label{fig.loss-function-test-img}
\end{figure*}

We test three types of loss functions -- mean squared error (MSE), intersection over union (IoU)  and a combination of MSE and IoU --
to predict the fraction of the mass that comes from stellar feedback. Figure~\ref{fig.loss-function-test-img} shows the performance of the model using different loss functions on a test bubble. The model adopting IoU as the loss function can capture the morphology of the bubble clearly but misses the value information. The IoU model predicts almost unity at the feedback position but does not reflect the actual fraction value. The MSE model is able to capture the position of larger feedback values but underestimate the smaller values which is useful to predict the emission but not the fraction. The model adopting both MSE and IoU as the loss function performs the best. This model not only captures the distinct bubble morphology but also returns reasonable fraction values.

\section{Training Sets}

\subsection{Comparison of \13co\ Emission with Different Cloud Thicknesses }
\label{12co and 13co Comparison}

\begin{figure*}[hbt!]
\centering
\includegraphics[width=0.98\linewidth]{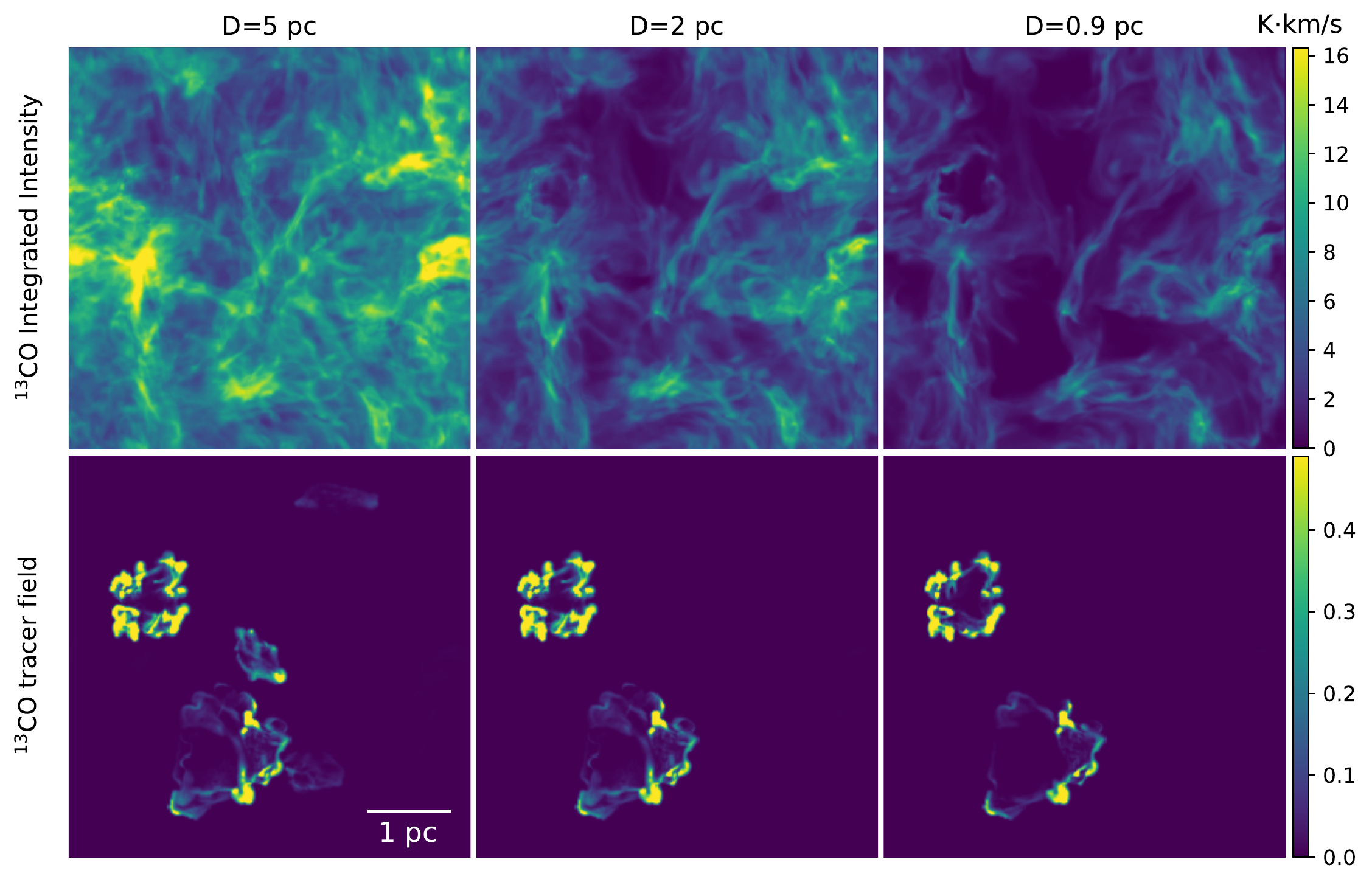}
\caption{Integrated intensity of \13co\ ($J$=1-0). Upper left: integrated intensity of \13co\ generated using the whole data cube. Upper middle and upper right: integrated intensity of \13co\ generated using the cropped data cube. Bottom left: integrated intensity of \13co\ wind tracer generated using the whole data cube. Bottom middle and bottom right: integrated intensity of \13co\ wind tracer generated using the cropped data cube. "D" is the line of sight thickness of the the data cube. }
\label{fig.bubbleco-inte-crop-co}
\end{figure*}

Figures~\ref{fig.bubbleco-inte-crop-co} shows the difference in the synthetic observations between the whole cube and the cropped data for \13co. The \13co\ bubble rim is embedded in the diffuse gas emission and the bubble cavity is not clear in the integrated intensity map when the thickness of the cloud is 5 pc. Since \co\ is even more optically thick, \co\ is not an appropriate proxy to trace stellar feedback winds \citep[e.g.,][]{2011ApJ...742..105A, 2015ApJS..219...20L}. When the thickness of the cloud becomes smaller, the bubble rim and its cavity are recognizable in the \13co\ integrated intensity map. Although some bubble rims or their cavities are not distinct in the integrated intensity map of \13co, these feedback features become recognizable in PPV space.

\subsection{Different Definitions {for the Bubble Extents}} 
\label{Different Definitions for the Bubble Extends}

\begin{figure*}[hbt!]
\centering
\includegraphics[width=0.98\linewidth]{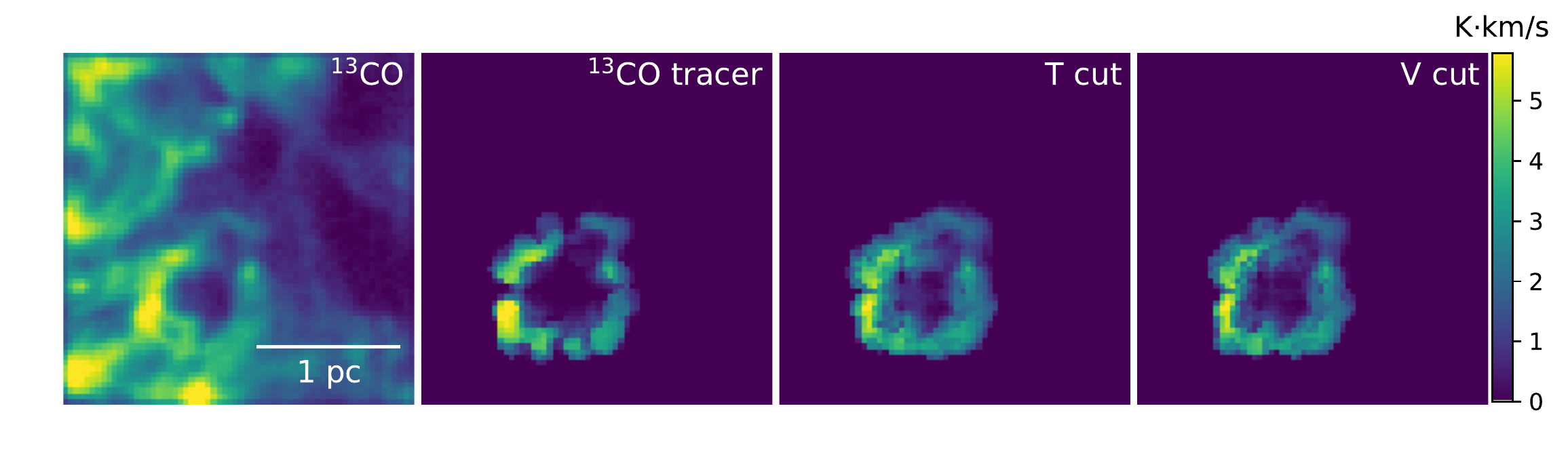}
\caption{ Different definitions for a synthetic bubble. First panel: the \13co\, integrated intensity map. Second panel: the original definition of the bubble using the {\sc orion} tracer field integrated over the velocity channels. Third panel: the temperature-based definition ($T> 12$ K plus the tracer field) integrated over the velocity channels. Fourth panel: the velocity-based definition (gas with expanding velocities plus the tracer field) integrated over the velocity channels. }
\label{fig.tacer-comp-map}
\end{figure*}

\begin{figure}[hbt!]
\centering
\includegraphics[width=0.48\linewidth]{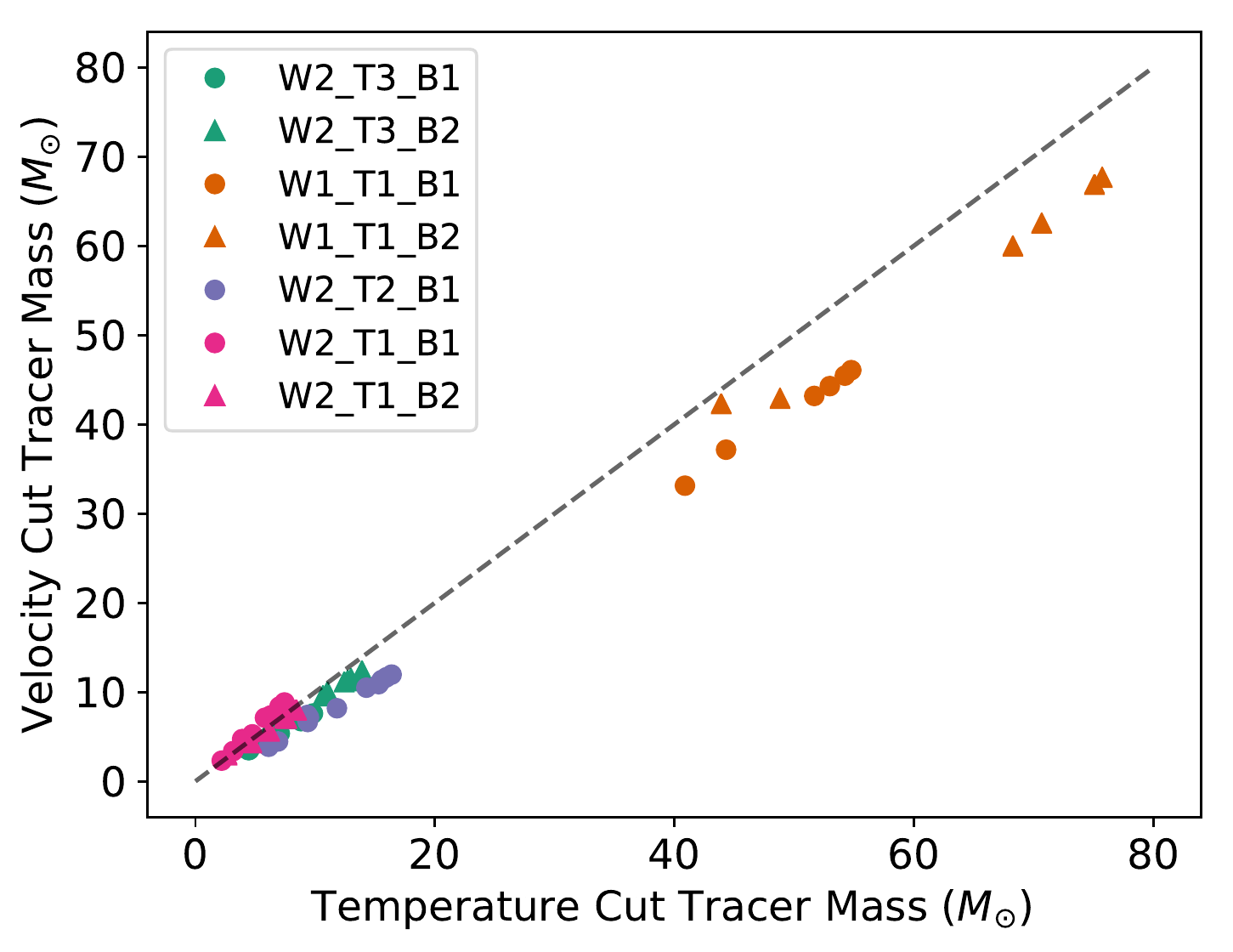}
\caption{Comparison between the velocity cut tracer mass and the temperature cut tracer mass. }
\label{fig.tacer-comp-T-V}
\end{figure}

{In this section we assess the impact of different choices for the bubble definition on the results. In addition to the bubble definition described in Section~\ref{Training Target: Tracer Field for Task I}, we }also examine the tracer field in the simulation data. The gas adjacent to the tracer gas has a velocity vector going outwards, which indicates the feedback gas compresses the ambient gas without direct contact. Although the fraction of feedback gas compared to the entire amount of gas contained in these voxels is almost zero, the adjacent layer contributes to the momentum and the energy of the cloud. Consequently, we define the tracer field with the velocity vector going outwards from the central stars, which increases the mass of the feedback bubble by a factor of 3. We furthermore test a temperature cut at $T\ge 12$ K near the tracer gas to calculate the bubble mass. The simulation data cubes have an average temperature of 10 K. The temperature drops quickly from the bubble rim to the ambient gas. We compare the different tracer definitions in Figure~\ref{fig.tacer-comp-map}. Both {the velocity cut plus the tracer field and the temperature cut plus the tracer field }are slightly larger in area than the original tracer field. {Both of these definitions yield bubbles that} are similar in shape. Since there are five individual stars in the simulation box, the bubbles generated by theses stars are easily connected to each other during the expansion. {This affects the gas velocities}, which makes it difficult for us to define the gas flow direction and determine which expansion is part of the shell. Under this circumstance, the temperature cut is a better option to define the {bubble boundary}. We compare the bubble mass calculated from the {velocity-based bubble definition} and the {the temperature-based bubble definition} in Figure~\ref{fig.tacer-comp-T-V}. Larger bubbles are more likely to overlap during the the expansion, which makes the {velocity-based bubble definition} mass slightly smaller than the {the temperature-based bubble definition} mass. Overall, {we conclude the temperature-based bubble definition} is the most appropriate definition of the {bubble boundary}.

{
\subsection{Comparison of the Training and Observed Bubble Mass Distributions}
\label{Bubble Mass Distribution in The Training Set}

In this section, we examine the distribution of the bubble masses in the training set. Figure~\ref{fig.mass-dis-train} shows the maximum bubble mass in the training set is $\sim$ 120 \msun, and it spans the range of the bubble masses in the test samples in Section~\ref{Assessing Model Accuracy Using Synthetic Observations}. Moreover, to extend the range of bubble masses, we have included ``zoomed-in” synthetic observations. In these 64$\times$64 postage-stamps, the original bubble is enlarged by a factor of two in both length and width, which indicates the bubble area and the mass both increase by a factor of 4. Since \CASItD\ takes postage-stamp cubes as inputs, regardless of the actual physical size of the cubes, this means the training set spans bubble masses up to $\sim 4\times120$ \msun, and it spans the range of the individual bubble masses in observations in Section~\ref{Physical Properties of the Individual Taurus Bubbles}. In some cases, a single $64\times 64$ postage-stamp cannot cover an entire bubble. Only part of the bubble appears within the input window, such as an arch or a half circle. These cases in the training set are consistent with the cases of larger bubbles that are contained in the full map prediction in Taurus. Thus, to obtain the masses of the largest bubbles in Taurus, we combine a stack of postage-stamps that cover different parts of each bubble to get the full prediction and then calculate the bubble mass as described in Section~\ref{Physical Properties of the Individual Taurus Bubbles}.

\begin{figure}[hbt!]
\centering
\includegraphics[trim={0 0 0 0.7cm},clip,width=0.46\linewidth]{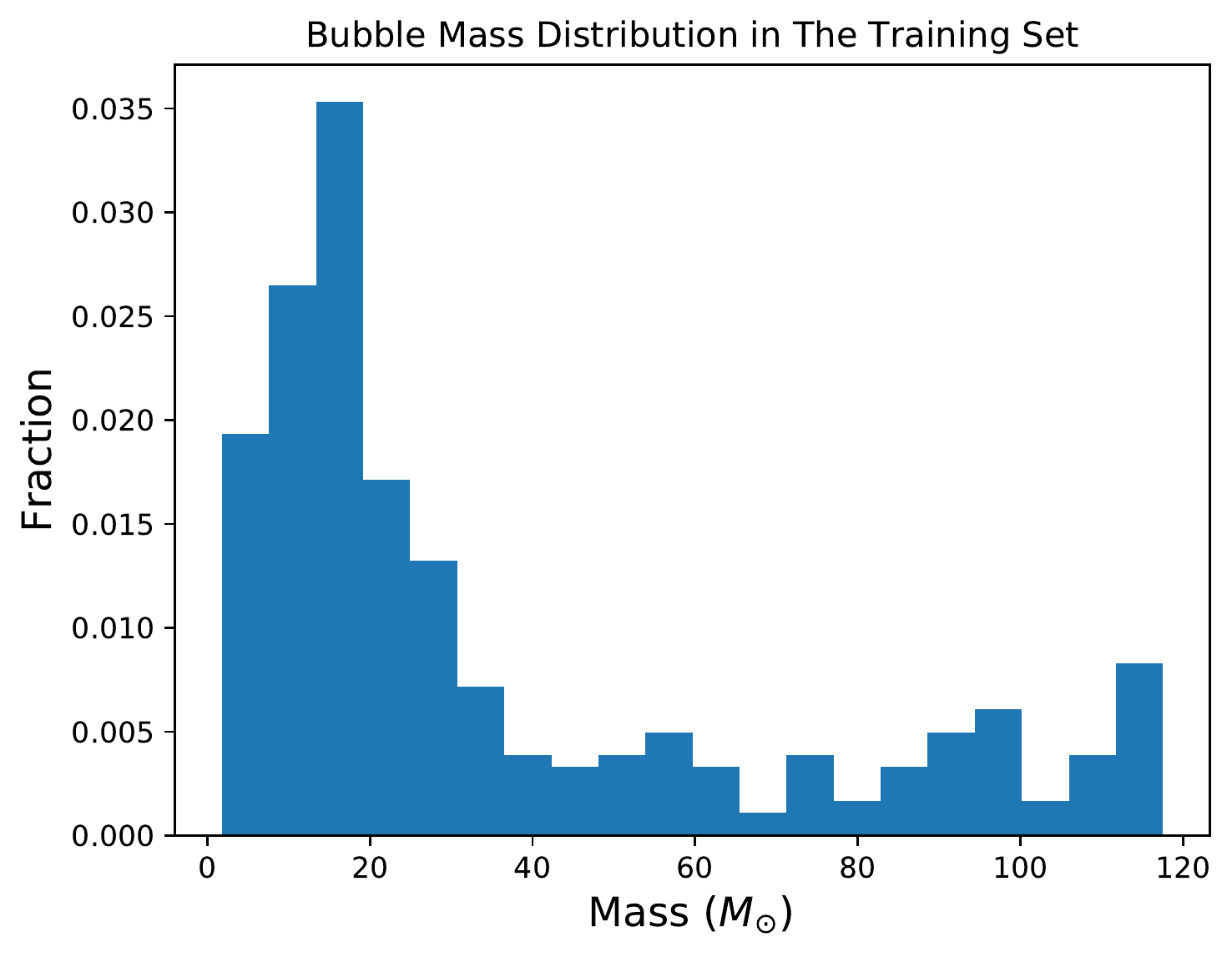}
\caption{The distribution of bubble masses in the training set. }
\label{fig.mass-dis-train}
\end{figure} 

}

{\bf
\section{Excitation Temperature Selection and Impact}
\label{Excitation Temperature Selection and Impact}

In this section, we explore the uncertainty in the bubble masses due to the choice of excitation temperature. We find that 25 K \citep[e.g.,][]{2012MNRAS.425.2641N,2015ApJS..219...20L} is the most appropriate choice to convert \13co\ emission to column density in the synthetic observations. We show the ratio between the mass estimated from \13co\ assuming a 25 K excitation temperature and the true mass calculated from the simulations in Figure~\ref{fig.temp-mass-ratio}. The ratio is within a factor of two of unity when assuming LTE and a 25 K excitation temperature, which in turn demonstrates that both LTE and the choice of 25 K are reasonable.

\begin{figure}[hbt!]
\centering
\includegraphics[width=0.48\linewidth]{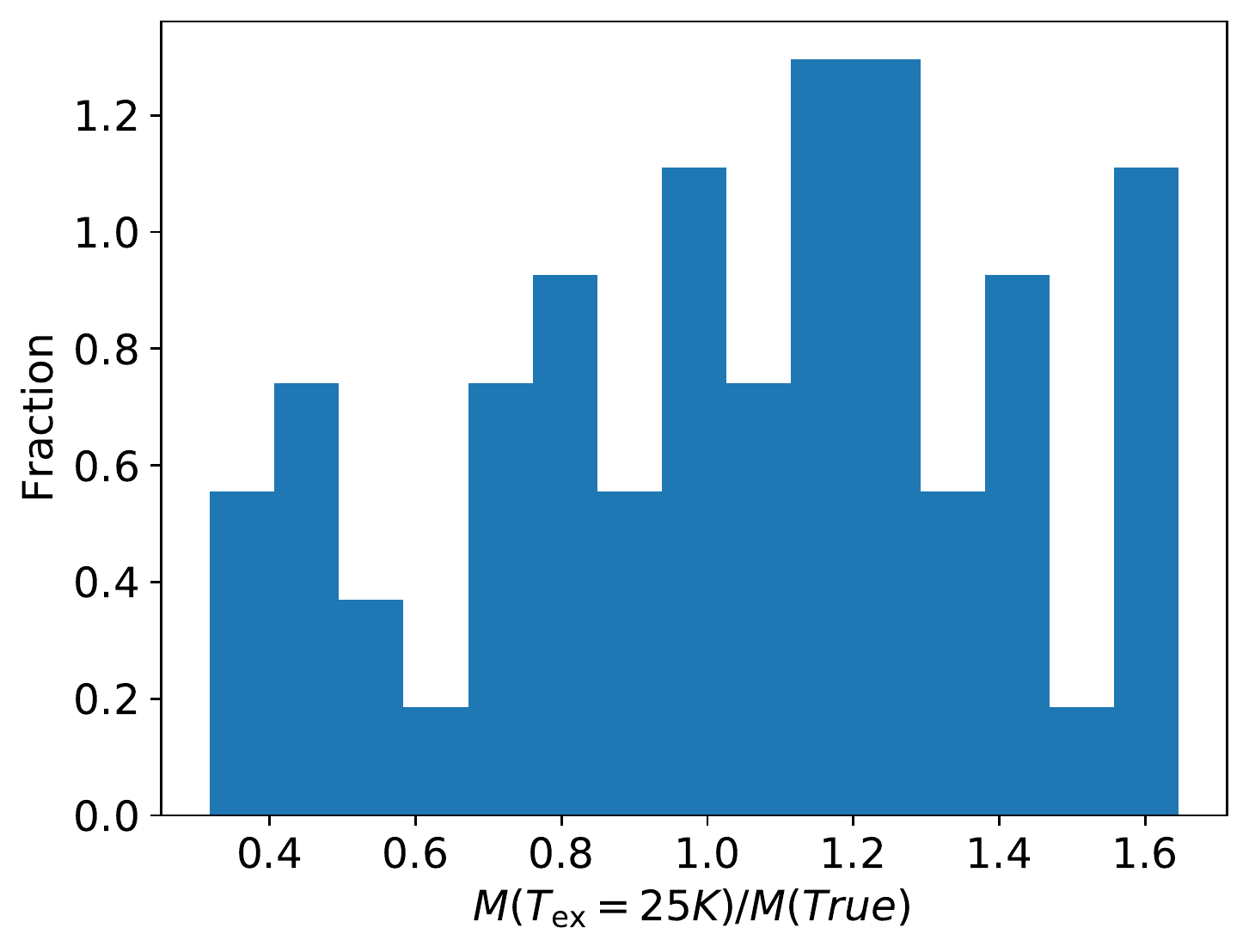}
\caption{The ratio between the mass estimated from \13co\ assuming a 25 K excitation temperature and the true mass calculated from the simulations.}
\label{fig.temp-mass-ratio}
\end{figure}

Under the assumption of LTE, the mass estimation goes linearly with the excitation temperature. From previous feedback mass estimates \citep[e.g.,][]{2011ApJ...742..105A,2015ApJS..219...20L}, the choice of excitation temperature ranges from 10 K to 50 K. This could introduce a factor of two uncertainty in the mass estimation, but it cannot account for a factor of ten. 

}

\section{Assessing the Sensitivity of the Data Window}
\label{Assessing the Sensitivity of the Data Window}

\begin{figure}[hbt!]
\centering
\includegraphics[width=0.48\linewidth]{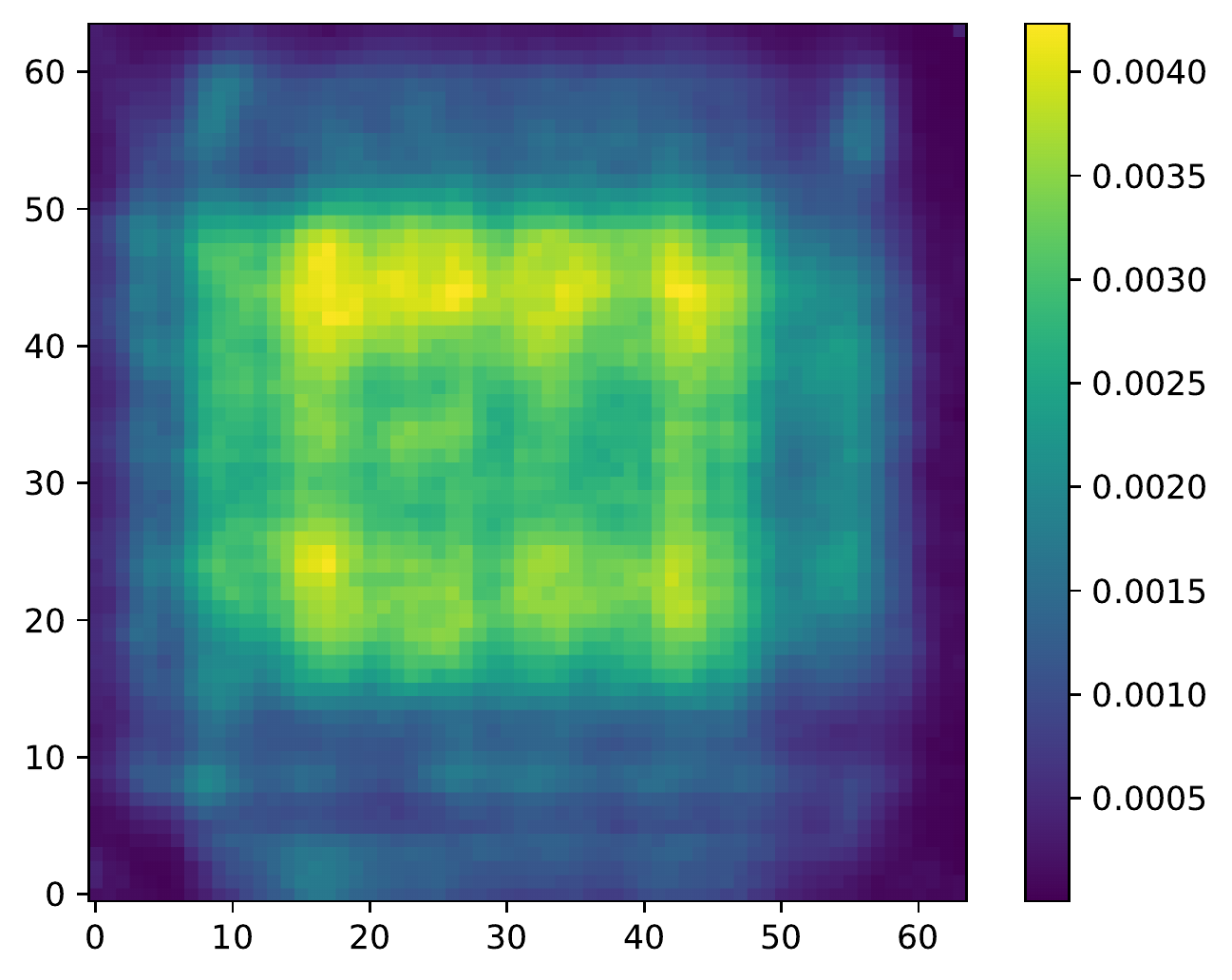}
\caption{The integrated response of a voxel in the cube, i.e., the fraction of voxels predicted to be associated with feedback in the same position in the stack of cubes summed over all velocity channels. }
\label{fig.fraction-ps-10}
\end{figure}

In this section we check the sensitivity of the data window to feedback as a function of voxel location. 
As discussed in Section~\ref{Predicting on Pixel by Pixel Position of Feedback}, the \CASItD\ models predict the full Taurus map feedback using a stack of $64\times64\times32$ cubes, in total 11,340 cubes. We examine the ``response" of each voxel in a $64\times64\times32$ cube. We define the 
response as the fraction of the stacked voxels over the 11,340 cubes that are detections in a $64\times64\times32$ cube. 
A detection is defined as a voxel above 90\% of the maximum prediction value for all overlapping voxels at the corresponding full map location. Figure~\ref{fig.fraction-ps-10} shows the response integrated over the velocity channels. The central region of the postage stamp is the most sensitive region, where a higher fraction of the stacked voxels fall above the prediction threshold. The boundary region of the postage stamp is less sensitive to features and detection is less efficient. This figure illustrates that choosing the appropriate cube offset is important to achieving the best sensitivity. 
To ensure that all data are 
covered by the highly sensitive part of the window, the maximum step size should not exceed 16 pixels. In our stacked prediction, we adopt a step size of 5 pixels, which is smaller than the required step size, to crop the full Taurus map. 

We note that the range of the window sensitivity likely depends on the training data and the target feature size. Here, we aim to find bubbles that have typical sizes greater than $\sim$16 pixels or a quarter of the the cube length. We recommend that other users of \CASItD\ check the window sensitivity for their problem to determine the appropriate offset size when applying  \CASItD\ to large data maps.

\section{YSOs in the Taurus molecular cloud}
\label{YSOs in the Taurus molecular cloud}

\begin{figure*}[hbt!]
\centering
\includegraphics[width=0.95\linewidth]{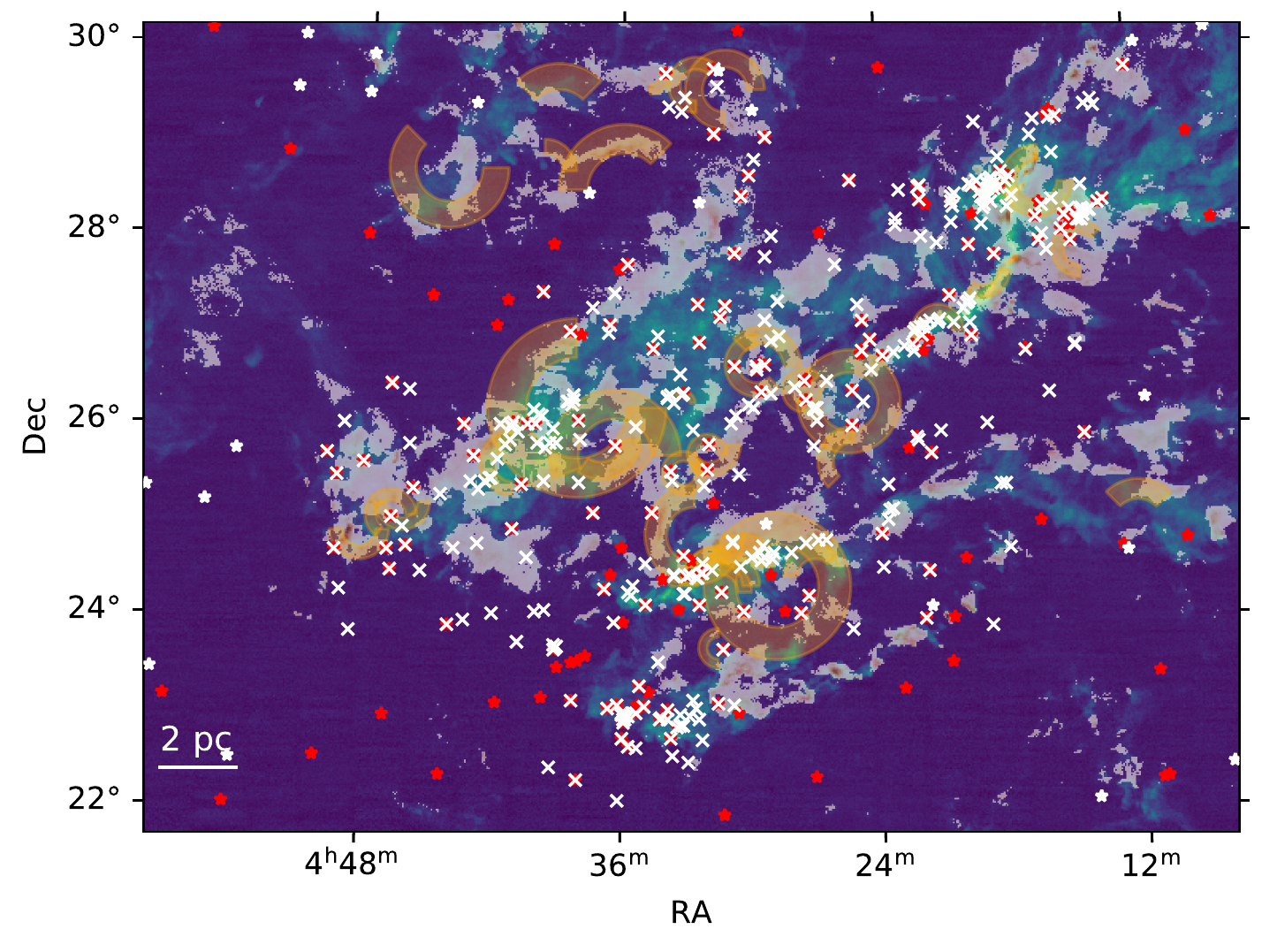}
\caption{The \13co\ integrated intensity of Taurus overlaid with the integrated prediction of feedback from models ME1 and MF (red). The arcs in yellow indicate the position of the previously identified bubbles from \citet{2015ApJS..219...20L}. The red star symbols indicate the locations of the Class III YSOs from \citet{2017ApJ...838..150K}. The white star symbols represent the locations of the YSOs from \citet{2011ApJS..196....4R}. The white cross symbols indicate the locations of the YSOs from \citet{2010ApJS..186..259R}.}
\label{fig.large-map-ME1-rebull}
\end{figure*}

\begin{figure*}[hbt!]
\centering
\includegraphics[width=0.95\linewidth]{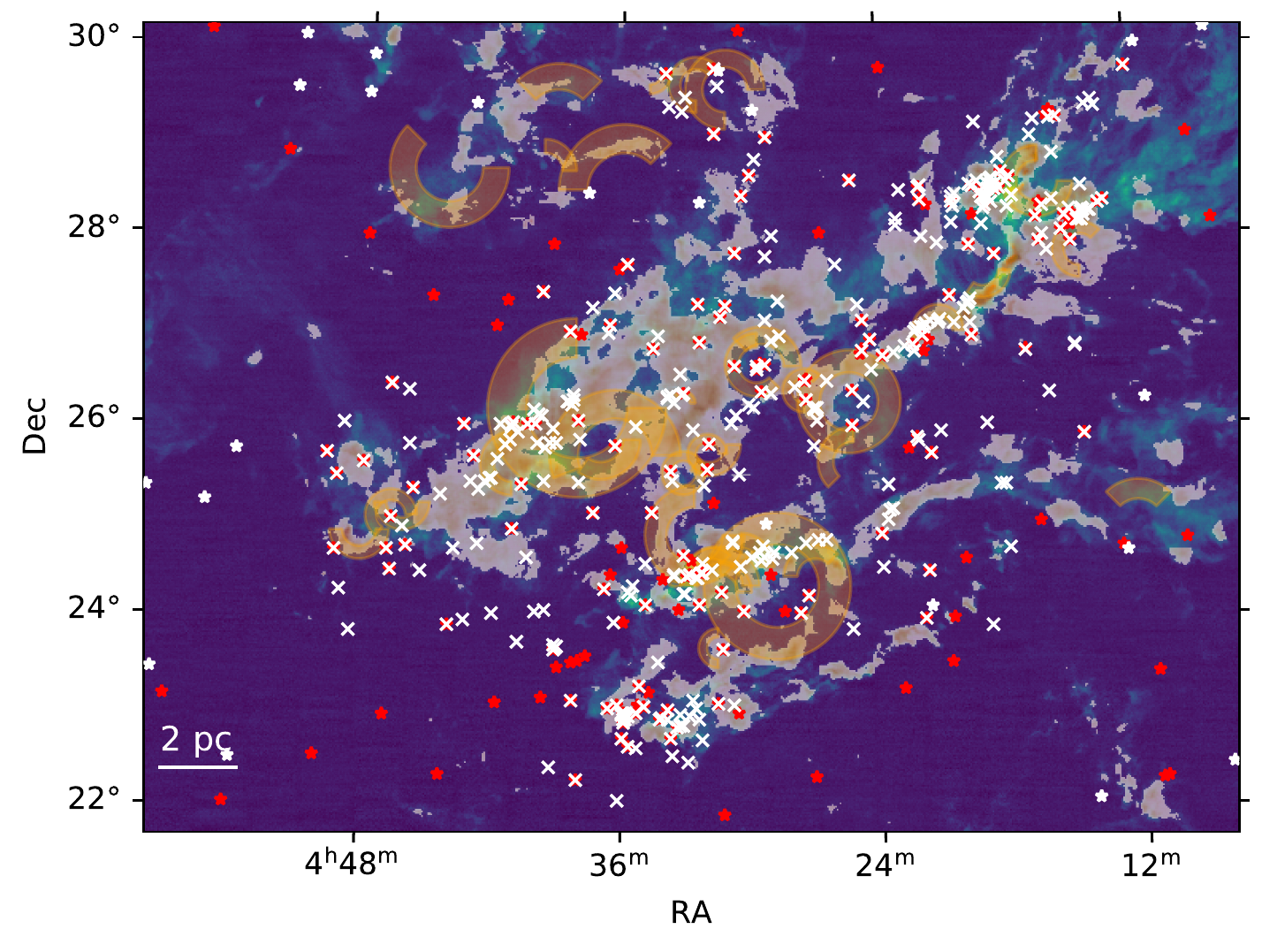}
\caption{The same as Figure~\ref{fig.large-map-ME1-rebull} but predicted by model MF. }
\label{fig.large-map-MF-rebull}
\end{figure*}

Figure~\ref{fig.large-map-ME1-rebull} and \ref{fig.large-map-MF-rebull} show the \13co\ integrated intensity of Taurus overlaid with the integrated prediction of feedback from models ME1 and MF (red). The arcs in yellow indicate the position of the previously identified bubbles from \citet{2015ApJS..219...20L}. The red star symbols indicate the locations of the Class III YSOs from \citet{2017ApJ...838..150K}. The white star symbols represent the locations of the YSOs from \citet{2011ApJS..196....4R}. The white cross symbols indicate the locations of the YSOs from \citet{2010ApJS..186..259R}.

\end{CJK*}

\end{document}